\documentclass[usegraphicx,usenatbib]{mn2e}

\usepackage[total={17.8cm,24.0cm},centering]{geometry}
\usepackage{times,deluxetable}

\newcommand{\apj}{ApJ}           
\newcommand{\apjl}{ApJ}           
\newcommand{\mnras}{MNRAS}       
\newcommand{\nat}{Nature}
\newcommand{\aap}{A\&A}
\newcommand{\araa}{ARA\&A}
\newcommand{\aj}{AJ}
\newcommand{\pasp}{PASP}
\newcommand{\apjs}{ApJS}           

\newcommand{\aaps}{A\&A Supp. Ser.}

\newcommand\ion[2]{#1$\;${\scshape{#2}}}
\newcommand{\hi}{{\sc H\,i}}
\newcommand{\mhi}{$M$(\hi)}
\newcommand{\sauron}{\texttt{SAURON}}
\newcommand{\atl}{ATLAS$^{\rm 3D}$}
\newcommand{\kms}{\hbox{km s$^{-1}$}}
\newcommand{\msun}{\hbox{$M_\odot$}}

\newcommand{\re}{\hbox{$R_{\rm e}$}}

\newcommand{\se}{\hbox{$\sigma_{\rm e}$}}

\newcommand{\vse}{\hbox{$(V/\sigma,\varepsilon)$}}
\newcommand{\plotone}[1]{\includegraphics[width=\columnwidth]{#1}}
\newcommand{\refsec}[1]{Section~\ref{#1}}
\newcommand{\reffig}[1]{Fig.~\ref{#1}}

\title[The \atl\ project -- I. The sample]
{The \atl\ project -- I. A volume-limited sample of 260 nearby early-type galaxies: science goals and selection criteria}

\author[M.~Cappellari et al.]
{Michele Cappellari$^1$\thanks{E-mail: cappellari@astro.ox.ac.uk},
Eric Emsellem$^{2,3}$,
Davor Krajnovi\'c$^2$,
Richard M. McDermid $^{4}$,\newauthor
Nicholas Scott$^1$,
G.~A. Verdoes Kleijn$^{5}$,
Lisa M. Young$^{6}$,
Katherine Alatalo$^7$,\newauthor
R.~Bacon$^3$,
Leo Blitz$^7$,
Maxime Bois$^{2,3}$,
Fr\'ed\'eric Bournaud$^{8}$,
M.~Bureau$^1$,\newauthor
Roger L. Davies$^1$,
Timothy A. Davis$^1$,
P. T. de Zeeuw$^{2,9}$,
Pierre-Alain Duc$^{10}$,\newauthor
Sadegh Khochfar$^{11}$,
Harald Kuntschner$^{12}$,
Pierre-Yves Lablanche$^{3}$,\newauthor
Raffaella Morganti$^{5,13}$,
Thorsten Naab$^{14}$,
Tom Oosterloo$^{5,13}$,
Marc Sarzi$^{15}$,\newauthor
Paolo Serra$^{13}$,
and Anne-Marie Weijmans$^{16}$\thanks{Dunlap Fellow}\\
$^1$Sub-department of Astrophysics, University of Oxford, Denys Wilkinson Building, Keble Road, Oxford OX1 3RH\\
$^2$European Southern Observatory, Karl-Schwarzschild-Str. 2, 85748 Garching, Germany\\
$^3$Universit\'e Lyon 1, Observatoire de Lyon, Centre de Recherche Astrophysique de Lyon\\ and Ecole Normale Sup\'erieure de Lyon, 9 avenue Charles Andr\'e, F-69230 Saint-Genis Laval, France\\
$^{4}$Gemini Observatory, Northern Operations Centre, 670 N. A`ohoku Place, Hilo, HI 96720, USA\\
$^{5}$Kapteyn Astronomical Institute, University of Groningen, Postbus 800, 9700 AV Groningen, The Netherlands\\
$^{6}$Physics Department, New Mexico Institute of Mining and Technology, Socorro, NM 87801, USA\\
$^7$Department of Astronomy, Campbell Hall, University of California, Berkeley, CA 94720, USA\\
$^{8}$Laboratoire AIM Paris-Saclay, CEA/IRFU/SAp – CNRS – Universit\'e Paris Diderot, 91191 Gif-sur-Yvette Cedex, France\\
$^{9}$Sterrewacht Leiden, Leiden University, Postbus 9513, 2300 RA Leiden, the Netherlands\\
$^{10}$Laboratoire AIM, CEA/DSM-CNRS-Université Paris Diderot, Dapnia/Service d'Astrophysique, CEA-Saclay, 91191 Gif-sur-Yvette Cedex, France\\
$^{11}$Max-Planck Institut f\"ur extraterrestrische Physik, PO Box 1312, D-85478 Garching, Germany\\
$^{12}$Space Telescope European Coordinating Facility, European Southern Observatory, Karl-Schwarzschild-Str. 2, 85748 Garching, Germany\\
$^{13}$Netherlands Institute for Radio Astronomy (ASTRON), Postbus 2, 7990 AA Dwingeloo, The Netherlands\\
$^{14}$Max-Planck Institut f\"ur Astrophysik, Karl-Schwarzschild-Str. 1, 85741 Garching, Germany\\
$^{15}$Centre for Astrophysics Research, University of Hertfordshire, Hatfield, Herts AL1 9AB, UK\\
$^{16}$Dunlap Institute for Astronomy \& Astrophysics, University of Toronto, 50 St. George Street, Toronto, ON M5S 3H4, Canada
}

\date{Accepted 2010 December 3. Received 2010 November 2}

\pagerange{\pageref{firstpage}--\pageref{lastpage}} \pubyear{2011}

\begin{document}
\label{firstpage}
\maketitle

\clearpage
\begin{abstract}
The \atl\ project is a multi-wavelength survey combined with a theoretical modelling effort. The observations span from the radio to the millimeter and optical, and provide multi-colour imaging, two-dimensional kinematics of the atomic (\hi), molecular (CO) and ionized gas (H$\beta$, [\ion{O}{iii}] and [\ion{N}{i}]), together with the kinematics and population of the stars (H$\beta$, Fe5015 and Mg {\em b}), for a carefully selected, volume-limited ($1.16\times10^5$ Mpc$^3$) sample of 260 early-type (elliptical E and lenticular S0) galaxies (ETGs). The models include semi-analytic, N-body binary mergers and cosmological simulations of galaxy formation. Here we present the science goals for the project and introduce the galaxy sample and the selection criteria.  The sample consists of nearby ($D<42$ Mpc) morphologically-selected ETGs extracted from a {\em parent} sample of 871 galaxies (8\% E, 22\% S0 and 70\% spirals) brighter than $M_K<-21.5$ mag (stellar mass $M_\star\ga6\times10^9$ $M_\odot$). We analyze possible selection biases and we conclude that the parent sample is essentially complete and statistically representative of the nearby galaxy population. We present the size-luminosity relation for the spirals and ETGs and show that the ETGs in the \atl\ sample define a tight red sequence in a colour-magnitude diagram, with few objects in the transition from the blue cloud. We describe the strategy of the \sauron\ integral-field observations and the extraction of the stellar kinematics with the pPXF method. We find typical 1$\sigma$ errors of $\Delta V\approx6$ \kms, $\Delta\sigma\approx7$ \kms, $\Delta h_3\approx\Delta h_4\approx0.03$ in the mean velocity, the velocity dispersion and Gauss-Hermite (GH) moments for galaxies with effective dispersion $\sigma_e\ga120$ \kms. For galaxies with lower $\sigma_e$ ($\approx40$\% of the sample) the GH moments are gradually penalized by pPXF towards zero to suppress the noise produced by the spectral under-sampling and only $V$ and $\sigma$ can be measured. We give an overview of the characteristics of the other main datasets already available for our sample and of the ongoing modeling projects.
\end{abstract}


\section{Introduction}

\subsection{Scientific background}

Observations of high-redshift galaxies and the cosmic microwave background \citep{Spergel2007} have revealed the Universe to be dominated by dark matter and dark energy \citep{Riess1998,Perlmutter1999}, providing a working paradigm for the formation of structure \citep[e.g.][]{Springel2005nat}. However, the mechanisms that form the luminous content of the dark-matter potential (i.e.\ the stars and galaxies that we observe) remain the key unknowns of modern extra-galactic astronomy. These processes are driven by the hydrodynamics and chemistry of the gas, combined with complex radiative feedback processes. High-redshift observations alone are not sufficient to constrain these processes, lacking spectral information and spatial resolution \citep{Faber2007}. It is therefore necessary to complement these studies with detailed analysis of nearby objects, tracing the {\it fossil record} of the formation process. Early-type (elliptical E and lenticular S0) galaxies (ETGs) are especially useful as they are old, have smaller levels of star formation and limited amount of dust, which simplifies the interpretation of the observations. Significant progress has been made in this direction in the past few decades, building on the classic observational works that still capture much of our understanding of the structure of local ETGs \citep[e.g.][]{Hubble1936,faber76,Davies1983,Dressler1987,Djorgovski1987,
bender92,kormendy95}.

A major step forward was brought by the era of large galaxy surveys. Thanks to the unprecedented sample size, one of the most important contributions of the Sloan Digital Sky Survey \citep[SDSS,][]{York2000} was to firmly establish a statistically significant bimodality in the colour distribution of local galaxies, such that they can be clearly separated in a so-called `blue cloud', generally consisting of star-forming spiral galaxies, and a `red sequence', mostly of non-star-forming ETGs \citep{Strateva2001,Baldry2004}. Accurately quantifying this bimodality, and the realization that it can be traced back in time to higher redshift \citep{Bell2004,Faber2007}, allowed a dramatic improvement in the detailed testing of galaxy formation scenarios.

The bimodality can only be explained with the existence of a mechanism, which suppresses episodes of intense star formation by evacuating gas from the system, resulting in a rapid transition of galaxies from the blue cloud to the red sequence \citep{Springel2005,Faber2007}. Many simulation groups have reproduced the bimodality qualitatively, though with rather different assumptions for the star formation and feedback processes \citep{Granato2004,DiMatteo2005,Bower2006,Cattaneo2006,Croton2006}. A generic feature of these models is that red-sequence galaxies form by dissipational `wet mergers' of gas-rich blue-cloud galaxies, followed by quenching of the resulting intense star-formation by rapid ejection of the gas, caused by the feedback from a central supermassive black hole, supernovae winds, by shock heating of the gas in the most massive halos \citep{Keres2005,Dekel2006} or gravitational gas heating \citep{Naab2007,Khochfar2008,Johansson2009}. The merging of the most massive blue galaxies, however, is not sufficient to explain the population of most-massive red-sequence galaxies. Dissipationless `dry mergers' of gas-poor, red-sequence galaxies is therefore also required, evolving galaxies {\em along} the red-sequence as they increase in mass \citep{Khochfar2003,Naab2006kb,Hopkins2009,Khochfar2009,Oser2010}.

Both wet and dry major mergers generally produce red, bulge-dominated galaxies when feedback is included in the models. The kinematic structure of the remnants is however very different. In a major (1:1) merger between blue gas-rich galaxies, the gas tends to form a disk, so that the end result of the merger, after the gas has been removed from the system by ejection, heating or conversion to stars, will be a red stellar system dominated by rotation \citep{Cox2006,Naab2006,Robertson2006,Jesseit2009}. In major mergers between red gas-poor galaxies, dissipationless processes dominate, resulting in a red galaxy with little or no net rotation \citep{Barnes1992,Hernquist1992,Naab1999,Naab2003,Cox2006}. Unlike major mergers, minor mergers (1:3 or less) retain more closely the structure of the progenitor, to an extent that depends on the amount of mass and gas accreted, so that the remnant of a spiral galaxy will always display significant rotation \citep{Naab2006,Robertson2006,Bournaud2007,Jesseit2009}. These simulations demonstrate that if galaxies assemble by mergers, the existence of the red/blue galaxies dichotomy therefore also suggests the existence of a {\em kinematical} differentiation {\em within} the red sequence between fast and slow rotating galaxies.

Various classic observational indicators of an ETGs dichotomy have been proposed in the past two decades. ETGs have been found to exhibit trends as a function of luminosity in terms of (i) their distribution on the \vse\ diagram, which relates the ratio of ordered $V$ and random $\sigma$ stellar motion to the galaxy ellipticity $\varepsilon$ \citep[e.g.][]{Illingworth1977,Binney1978,Davies1983}, (ii) their isophote shape (disky or boxy) \citep{Bender1989,Kormendy1996}, (iii) the inner slope of their photometric profiles: cored/cuspy \citep{Ferrarese1994,Lauer1995,faber97} or excess/deficit of core light \citep{Graham2004,Ferrarese2006,Kormendy2009}. However, none of these signatures have been able to give clear evidence for a distinction between the two classes of red-sequence galaxies, primarily because they are all essentially secondary indicators of the galaxies' internal kinematic structure.

By the application of integral-field spectroscopy to a representative sample of nearby ETGs, the \sauron\ survey \citep{deZeeuw2002} has revealed the full richness of the kinematics of these objects \citep{Emsellem2004,mcdermid06,Krajnovic08}. From the two-dimensional nature of this unique data set, two distinct morphologies of stellar rotation fields are clearly evident, corresponding to the above described fast- and slow-rotators. In two companion papers of that survey a global quantitative measure of this morphology was defined, termed $\lambda_R$, that can be used to kinematically classify these galaxies in a way that is more robust than the \vse\ diagram, is nearly insensitive to projection effects \citep{Emsellem2007,Cappellari2007}. $\lambda_R$ relates directly to their formation, and is precisely reproducible in current cosmological simulations \citep{Jesseit2009,Bois2010}. This is the basic new finding we plan to exploit in the present project to improve our understanding of the structure and formation of ETGs. Additional results of the \sauron\ survey on ETGs include the robustness and empirical `calibration' of the simple virial mass estimator to measure mass in the central parts of ETGs and a determination of their dark matter fraction \citep{Cappellari2006}. The survey found a high incidence of ionized gas in ETGs \citep{Sarzi2006} and explained their ionization mechanism as mainly due to the evolved stellar population \citep{Sarzi2010}. It was shown that the stellar population gradients correlate well with the escape velocity, both locally within galaxies and globally among different ETGs \citep{Scott2009}. Star formation in ETGs only happens in fast rotators and follows two distinct modes: in disks or widespread \citep{Shapiro2010}, where the latter cases are in low-mass systems \citep{Jeong2009,Kuntschner2010}. Disks in fast rotators have enhanced metallicity, while kinematically distinct cores in slow rotators show no stellar population signatures \citep{Kuntschner2006,Kuntschner2010}.

\subsection{Goals of the Project}

Due to the exploratory character of the \sauron\ survey \citep{deZeeuw2002}, the ETGs were selected to sample, with a relatively small number of objects, a wide range of masses, shapes and morphologies. This was done by selecting galaxies brighter than a total magnitude $M_B<-18$ mag equally subdivided into 24 E and 24 S0. Within each E/S0 subclass the selected objects sample uniformly a grid in the $(M_B,\varepsilon)$ plane. Although that approach was crucial in bringing the fast/slow rotator dichotomy to light and in most of the findings mentioned in the previous section, the selection criteria impose complex biases and do not allow for a quantitative statistical comparisons of galaxy properties with simulations, which is a main goal of the \atl\ project. Moreover, with only 48 galaxies, the statistical uncertainties are large.

The power of the kinematic classification based on $\lambda_R$ is to be able to study differences in the formation process along the red-sequence galaxy population.
The $\lambda_R$ parameter describes in a compact way the present status of the galaxies, however it is essential to obtain information on the formation history and the detailed dynamical structure as well. The stellar population contains a record of the more distant history (a few Gyr). Recent gas accretion is recorded in the cold atomic gas components, generally detected on galaxy scales with radio observations of \hi, while the ongoing accretion and star formation activity is traced by cold molecular gas (e.g.\ CO), often detected in regular disks in the central regions. For comparison with theoretical predictions one needs to observe all these quantities for a statistically significant, volume-limited sample of galaxies complete to some useful lower limit in mass. With these ideas in mind we carefully selected the \atl\ sample of ETGs and we systematically observed all the above quantities. The \atl\ dataset now provides a complete inventory of the baryon budget and a detailed two-dimensional description of stellar and gaseous kinematics, together with resolved stellar population within the main body of a complete and statistically significant sample of ETGs. Our goal is to use this dataset to perform archeological cosmology by specifically answering the following questions:
\begin{enumerate}
\item How do slow rotators form? What are the physical processes that determine their kinematic and photometric features?   What is the role of major and minor mergers in their formation history? This will be reflected in the kinematics, gas content and stellar population.

\item Why are most ETGs fast rotators? There seems to be a dominant formation mechanism that delivers galaxies with quite homogenous rotation properties. Can this be merging? Can significant major merging be excluded?

\item How is star formation in ETGs quenched? Is it different for fast and slow rotators ETGs? How does it depend on environment? Can we infer the quenching mechanism from the amount and distribution of the left-over gas, the presence of AGNs or metallicity gradients? The distribution of stellar population and gas properties constitute a stringent test for future galaxy formation models.

\item Most past studies have focused on single stellar population models of ETGs, but cosmological models predict more complex histories. Can we infer the star formation history in ETGs for detailed comparison with simulations?

\item How do counter rotating cores in massive and old ETGs form and survive to the present time? Are these relics of the very early Universe?

\item Can we link the present day properties of ETGs to results form existing and upcoming surveys at higher redshift with respect to e.g. masses, sizes, stellar populations, gas fractions, star formation? Our study will constitute a $z=0$ redshift benchmark to trace the time evolution of galaxies.

\end{enumerate}

The \atl\ sample includes all nearby ETGs observable from the northern Earth hemisphere, and for this reason we hope its homogeneous dataset will ultimately constitute a legacy for future studies. We trust that our and other groups will exploit our data and sample well beyond what we had originally envisioned. Our first steps in the directions outlined above are presented in the following papers, while the other aspects will be presented in subsequent papers of this series:
\begin{enumerate}
\item \citet[hereafter Paper II]{Krajnovic2010}, which describes the morphology of the kinematics and the kinematical misalignment in ETGs;
\item  \citet[hereafter Paper III]{Emsellem2010}, which presents a census of the stellar angular momentum in the central region of ETGs;
\item \citet[hereafter Paper IV]{Young2010}, which quantifies the distribution of molecular gas content in ETGs;
\item \citet[hereafter Paper V]{Davis2010}, which studies the \citet{Tully1977} relation from the width of the molecular lines in ETGs;
\item \citet[hereafter Paper VI]{Bois2011}, which studies the formation of the fast and slow-rotator galaxies via numerical simulations of binary mergers;
\item \citet[hereafter Paper VII]{Cappellari2011}, which revisits the morphology of nearby galaxies and presents the {\em kinematic} morphology-density relation.
\end{enumerate}

Here in Section~2 we discuss the selection criteria for the {\em parent} sample of galaxies, from which the \atl\ sample of ETGs was extracted (Section~3). In Section~4 we present the \sauron\ observing strategy for the survey, the integral-field data, and the kinematic extraction, while other additional datasets and simulations from our project are listed in Section~5. We give a summary in Section~6. In the paper we assume $H_0=72$ \kms\ Mpc$^{-1}$.

\section{The Parent Sample}

\subsection{Selection criteria}

Our final \atl\ sample will focus on ETGs only, however before any morphological classification, we want to select all galaxies in the nearby volume above a certain total stellar mass. As we did not have dynamical information for all galaxies in the local volume at the beginning of our survey, the best proxy for mass available was the near-infrared ($\sim2.2$ \micron) $K_s$-band luminosity provided by the two-micron all sky redshift survey (2MASS; \citealt{Skrutskie2006}), which is unique for its  full sky completeness and excellent photometric homogeneity. The $K_s$-band is 5--10 times less sensitive to dust absorption than optical wavelengths and therefore can be used to select both dust-rich spirals and dust-poor ETGs to a similar mass level. Moreover the mass-to-light ratio of the stars in the near infrared varies only about a factor $\approx2$, which is $\sim3\times$ less than at optical wavelengths \citep{Bell2001,Maraston2005}, thus providing a better approximation to the stellar mass than an optical selection.

To derive luminosities from the observed apparent magnitudes we need distances. Numerous accurate determinations have been accumulated in the literature in the past few years. But we will resort to redshift distances when a more accurate distance is not available. In addition we enforce obvious observability criteria. This leads to the following selection steps:
\begin{enumerate}
\item Choose a representative local volume with radius $D=42$ Mpc. It approximates the redshift selection $c z<3000$ \kms\ of the \sauron\ survey, for an adopted $H_0=72$ \kms\ Mpc$^{-1}$ \citep{Dunkley2009}. It makes sure that key spectral features, such as H$\beta$, [\ion{O}{iii}] and Mg~{\it b}, fall within the \sauron\ wavelength range and allows for a significant overlap with previous observations;
\item Specify the observability criterion from the William Herschel Telescope on La Palma $|\delta-29^\circ|<35^\circ$, where $\delta$ is the sky declination;
\item Exclude the dusty region near the Galaxy equatorial plane $|b|<15^\circ$, with $b$ the galactic latitude;
\item Select all galaxies from the 2MASS extended source catalog (XSC; \citealt{Jarrett2000}) with apparent total magnitude $K_T<11.6$ mag (defined by the XSC parameter \texttt{k\_m\_ext}) and satisfying the observability criteria (ii)-(iii). Given the near completeness of the XSC down to $K_T\approx13.5$, this selection is essentially complete. It ensures that all candidate galaxies brighter than an absolute total magnitude $M_K=K_T-5\log D-25=-21.5$ mag are selected. This $K_s$-band luminosity limit roughly corresponds to a $B$-band selection $M_B\la-18$, for the typical $B-K_s\approx3.5$ mag colour of ETGs, at the faint end of our selection. This criterion is thus again similar to the one in the \sauron\ survey and allows for a significant overlap in the samples, reducing the required observing time. This step provides a sample of $\sim20,000$ extended objects classified as galaxies;
\item Assign a distance to as many galaxies as possible in the selection and include in the \atl\ parent sample the ones with $D<42$ Mpc and $M_K<-21.5$ mag; The distance selection requires some further explanation and may introduce incompleteness biases which are discussed in the next section.
\end{enumerate}

A summary of the selection criteria is given in Table~\ref{tab:parent_criteria}, while some of the main characteristics of the resulting galaxy sample are given in Table~\ref{tab:parent_specs}. This is the sample of galaxies, which includes both spiral and ETGs, from which the \atl\ sample of ETGs will be extracted. The names and the characteristics of the resulting 871 galaxies in the \atl\ parent sample are given in Table~\ref{tab:atlas3d_sample} (for the ETGs) and Table~\ref{tab:atlas3d_spirals} (for the spirals). As the evolution of spirals and ETGs are closely related, the spirals of the parent sample are critical to properly interpret the \atl\ results on ETGs.

\begin{table}
\caption{Selection criteria for the galaxies in the \atl\ parent sample}
\centering
\begin{tabular}{rl}
\hline
Distance: & $D<42$ Mpc \\
Galaxies total mag: & $M_K<-21.5$ mag \\
Observability: & $|\delta-29^\circ|<35^\circ$ \\
Galaxy zone of avoidance: & $|b|>15^\circ$ \\
\hline
\label{tab:parent_criteria}
\end{tabular}
\end{table}

\begin{table}
\caption{Main characteristics of the \atl\ parent sample}
\centering
\begin{tabular}{rl}
\hline
Survey Volume: & ${\rm Vol}=1.16\times10^5$ Mpc$^3$ \\
Galaxy $K$-band luminosity: & $L>8.2\times10^9$ $L_{K,\odot}$ \\
Galaxy stellar mass: & $M_\star\ga6\times10^9$ \msun \\
Galaxy $B$-band total mag: & $M_B\la-18.0$ mag\\
Galaxy SDSS $r$-band total mag: & $M_r\la-18.9$ mag\\
Total number of galaxies: & $N_{\rm gal}=871$ \\
Spiral and irregular galaxies: & $N_{\rm Sp}=611$ (70\%) \\
S0 galaxies in \atl\ ($T>-3.5$): & $N_{\rm S0}=192$ (22\%) \\
E galaxies in \atl\ ($T\le-3.5$): & $N_{\rm E}=68$ (8\%) \\
\hline
\label{tab:parent_specs}
\end{tabular}
\end{table}

\subsection{Sources of distances and errors}
\label{sec:distances}

\begin{figure}
\plotone{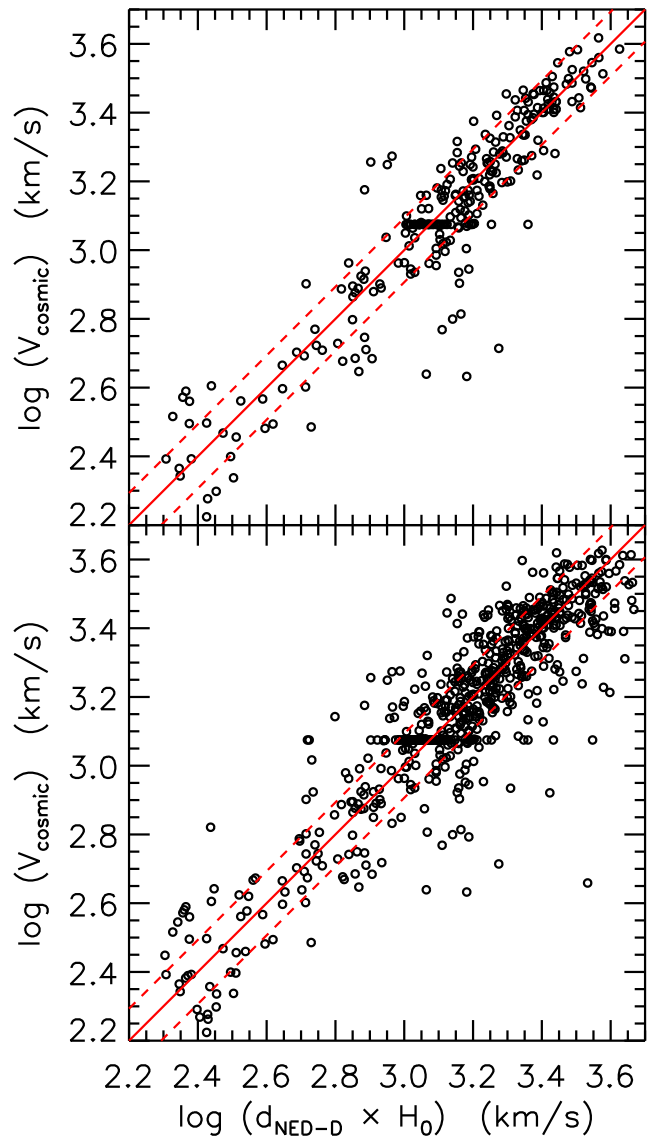}
\caption{Accuracy of redshift distances. {\em Top Panel:} Recession velocities, converted into velocities $V_{\rm cosmic}$ characteristic of the expansion of the universe, against the median of the NED-D  redshift-independent distances for 291 galaxies with at least 3 independent determination (in most cases including SBF distances). The solid line is a one-to-one relation, while the dashed lines indicate $\pm24\%$ deviations. {\em Bottom Panel:} Same as in the top panel for 705 galaxies with at least one distance determination in NED-D.}
\label{fig:redshift_test}
\end{figure}

Numerous sources of distances for nearby galaxies have been accumulated over the past decades. In most cases the distances are based on redshift as provided by large redshift surveys, but a number of more accurate distances are available based on other methods (see e.g.\ the recent compilations of \citealt{Tully2008,Tully2009,Karachentsev2010}). For the $\sim20,000$ 2MASS galaxies with $M_K<-21.5$ mag we tried to automatically assign the most accurate available distance according to the following order of priority:
\begin{enumerate}
\item Distance obtained with the surface brightness fluctuation (SBF) method for the ACS Virgo Cluster Survey \citep{Mei2007,Blakeslee2009} (91 values). These are claimed to be accurate to about 3\% in distance;
\item SBF distance from ground-based observation from \citet{Tonry2001} (300 values), which have a median error of 10\% in distance. These have been converted to the same zeropoint of \citet{Mei2007} by subtracting 0.06 mag to the distance moduli (see discussion in \citealt{Blakeslee2010});
\item Distances from the NED-D compilation\footnote{http://nedwww.ipac.caltech.edu/Library/Distances/} by Madore, Steer and the NED team (V3.0 June 2010, about 2000 values). The list includes accurate determinations using (1) SBF, (2) the tip of the red giant branch (TRGB), (3) Cepheids variables, with a claimed comparable accuracy of $\sim10\%$, but the list also includes various other methods like the ones based on the (4) \citet{Tully1977} or (5) Fundamental Plane \citep{Djorgovski1987,Dressler1987} relations, on the (6) luminosity of Type Ia Supernovae, or on the luminosity functions of (7) globular clusters and (8) planetary nebulae. The latter methods are expected to be accurate to better than $\la20\%$ \citep{Tully2008}. For a number of galaxies more than one independent distance was available and we adopted the median of the values.
\item When no better distance was available, for galaxies within 12$^\circ$ of the projected center of the Virgo cluster (defined by the galaxy M87) with heliocentric radial velocities $V_{\rm hel}<2300$ \kms, we assigned the distance of the cluster (assumed to be 16.5 Mpc from \citealt{Mei2007}). These distances should be accurate to $\sim7\%$, considering the intrinsic depth of the cluster. Two galaxies were later removed from Virgo as that distance implied a too high and non-physical dynamical $M/L$ as determined in \citealt{Cappellari2010iau}.
\item Finally if none of the above criteria was met, we assigned a distance based on the observed heliocentric radial velocities $V_{\rm hel}$, which we converted to velocities $V_{\rm cosmic}$ characteristic of the expansion of the universe following \citet{Mould2000},\footnote{Corrected as described in the corresponding Erratum.} but only including the contribution of the Virgo attractor, and assuming $H_0=72$ \kms\ Mpc$^{-1}$ \citep{Dunkley2009}. We repeated our sample selection using heliocentric velocities extracted from either the HyperLeda\footnote{http://leda.univ-lyon1.fr/} \citep{Paturel2003} or NED \footnote{http://nedwww.ipac.caltech.edu/} databases and obtained identical final \atl\ samples. We compared the $V_{\rm hel}$ of galaxies from NED and HyperLeda and found a general very good agreement in the two databases, with a biweight rms scatter of 0.4\%, as expected given that in most cases the redshifts come from the same source. However in some cases significant differences exists: we found $83/5398$ galaxies (1.5\%) with $V_{\rm hel}$ differences larger than 20\%. For our $V_{\rm hel}$ in the parent sample we adopted NED (June 2010), which currently includes as major sources the Center for Astrophysics Redshift Catalog \citep{Huchra1983}, the RC3 \citep{deVaucouleurs1991}, the ZCAT compilation \citep{Huchra1992} and the Sloan Digital Sky Survey DR7 \citep{Abazajian2009}. After obtaining our new accurate \sauron\ redshifts (\refsec{sec:kinematics}) we updated the redshift-based distances and this lead to the removal of one observed galaxy. However we retained in the sample two observed galaxies formally with $D>42$ Mpc, but still inside our volume within the distance errors.
\end{enumerate}

To estimate the representative errors of the redshift distances we correlated them against the direct distance estimates in the NED-D compilation. Specifically we selected the 285 galaxies in common with our sample with at least 3 independent distance determinations (in most cases including SBF distances), and we correlated the median of their $d_{\rm NED-D}\times H_0$ values against the $V_{\rm cosmic}$ (\reffig{fig:redshift_test}). We found a biweight dispersion of 24\% in the residuals. The minimum value in the median residual was obtained with $H_0=72$ \kms\ Mpc$^{-1}$, consistent with the adopted WMAP estimate. Assuming a typical rms error of 10\% for the best set of NED-D distances, this implies an intrinsic rms error of $\sim21\%$ in the redshift distances. However a significant number of outliers exist. If we repeat the comparison for all the 692 galaxies in common, with a NED-D distance, the biweight dispersion increases to 29\%, which would imply a redshift error of $\sim27\%$. If we only consider the Local Group's peculiar velocity into Virgo in the calculation of $V_{\rm cosmic}$, the scatter increases significantly and systematic deviations appear. Including the infall of galaxies into the Virgo attractor improves the agreement. However including other attractors as done by \citet{Mould2000} does not appear to further reduce the scatter. For this reason we only included the more secure Virgo attractor contribution in our redshift distances corrections.

\subsection{Estimating redshift incompleteness}
\label{sec:redshift_incompleteness}

\begin{figure}
\plotone{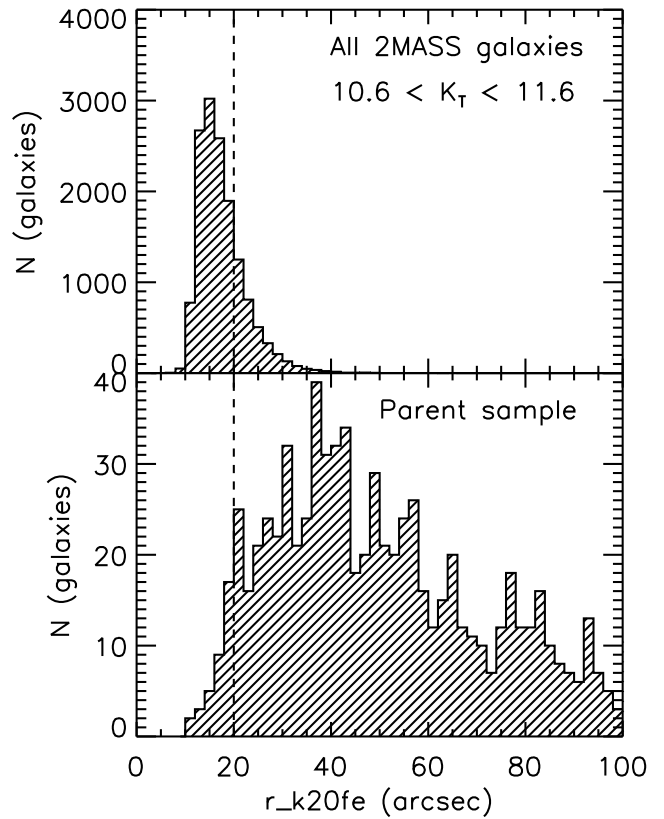}
\caption{Apparent size distribution of 2MASS galaxies. {\em Top Panel:} The apparent size, as described by the 2MASS XSC parameter \texttt{r\_k20fe}, for all the galaxies satisfying our observability criteria and with $10.6<K_T<11.6$. {\em Bottom Panel:} Same as in the top panel for the galaxies in the parent sample ($D<42$ Mpc and $M_K<-21.5$ mag).}
\label{fig:size_distribution}
\end{figure}

Not all the $\sim20,000$ 2MASS galaxies satisfying our observability criteria and with $K_T<11.6$ have a redshift measurement. This may introduce biases in our volume limited sample selection. Specifically $4146/14461$ (29\%) of the galaxies in the faintest magnitude bin $10.6<K_T<11.6$ do not have a redshift in NED.\footnote{The situation will change in the near future when the 2MASS Redshift Survey will become available, which is already complete down to $K_s=11.25$ \citep{Crook2007} and ultimately aims for a redshift completeness down to $K_s=13.0$ \citep{Huchra2005}.} The redshift completeness quickly improves for brighter apparent magnitudes, and in fact only 4\% of the galaxies with $K_T<10.6$ have no redshift. One way to estimate how many of these galaxies are likely to be within our selection criteria, is to look at their size distribution as measured by 2MASS. In fact one expects many of the apparently faint and small galaxies to be intrinsically brighter and larger, but to appear faint and small due to the large distance.

To quantify the galaxy angular sizes we use the 2MASS XSC parameter \texttt{r\_k20fe}, which gives the semi-major axis of the 20 mag arcsec$^{-2}$ surface brightness isophote at $K_s$. The size distribution for the galaxies in the faintest magnitude bin, according to this size parameter is shown in the top panel of \reffig{fig:size_distribution}. As expected the distribution presents a dramatic increase in the numbers for very small objects. For comparison we show in the bottom panel the size distribution of the 2MASS galaxies which satisfy our selection criteria $D<42$ Mpc and $M_K<-21.5$ mag. The latter sample has a peak in the size distribution around $\texttt{r\_k20fe}\approx40\arcsec$, while the number of objects sharply drops for $\texttt{r\_k20fe}\la20\arcsec$ (only 4\% of the objects have sizes below that limit). This lack of apparently small objects is not due to any redshift selection criteria. In fact among all the galaxies with measured redshift, about equal numbers have size larger or smaller than $\texttt{r\_k20fe}=20\arcsec$. The apparent galaxy size is just an efficient way to select, without redshift information, galaxies unlikely to belong to our volume-limited sample.

Excluding all objects apparently smaller than $\texttt{r\_k20fe}<20\arcsec$, likely outside the limits of our local volume, we find a redshift incompleteness of $478/3383$ (14\%) in the faintest magnitude bin. Among the galaxies that {\em do} have redshift in this set, only $68/2905$ (2\%) satisfy the selection criteria for the parent sample (most of the others are still outside the local volume). This implies that statistically we may expect $\sim11$ galaxies (1\% of the parent sample) to be possibly missed from the parent sample due to redshift incompleteness in the faintest magnitude bin. We conclude that we can safely ignore this possible bias from any conclusion derived from the sample.

\subsection{Luminosity function}

\begin{figure}
\plotone{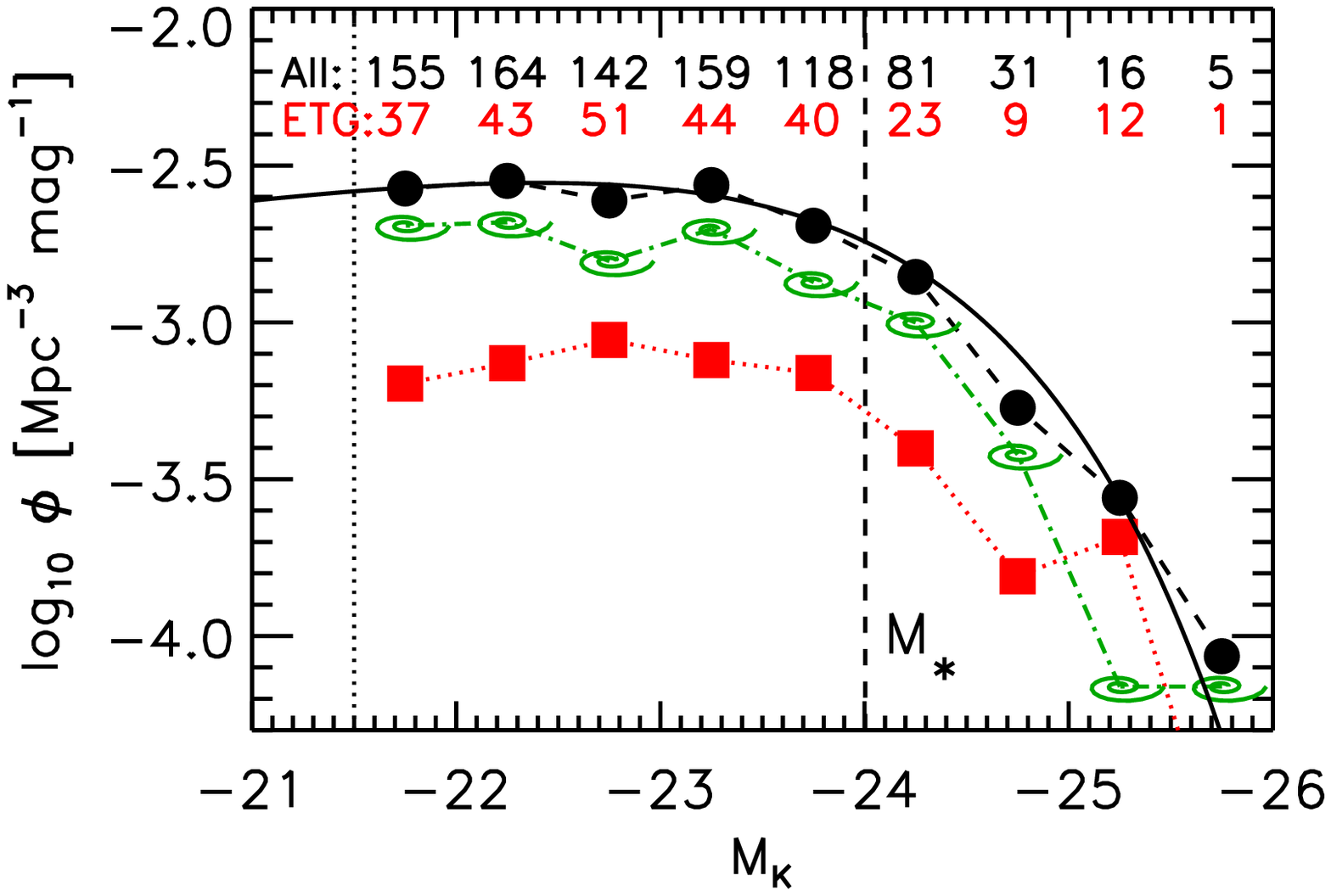}
\caption{$K_s$-band luminosity function (LF) of the \atl\ {\em parent} sample of 871 galaxies (black filled circles). The LF for the spiral galaxies (green spirals) and the 260 ETGs which constitute the \atl\ sample (red squares) separately are also shown. The solid curve shows the \citet{Schechter1976} function derived by \citet{Bell2003} from a fit to 6,282 galaxies. It was not fitted to our data! The black numbers above the symbols indicate the total number of galaxies included in each 0.5 mag bin, while the red ones are the corresponding numbers for the ETGs of the \atl\ sample. There is no evidence of incompleteness down to the magnitude limit of the survey (vertical dotted line), which is $\approx2.5$ mag below $M_\star$ (vertical dashed line).}
\label{fig:luminosity_function}
\end{figure}

The \atl\ parent sample was carefully selected to provide a volume-limited sample of galaxies in the nearby universe. It should be representative of the galaxy population at low redshift, apart from the unavoidable cosmic variance, due to the relatively limited size of the volume. A first test of the representativeness of our parent sample is to compare its $K_s$-band luminosity function against that measured from larger volumes. For this we compare in \reffig{fig:luminosity_function} the luminosity function of the parent sample against that derived from a much larger sample, at a mean redshift $\langle z\rangle\approx0.08$, by \citet{Bell2003}, using 2MASS $K_s$-band photometry as we do, and SDSS redshifts. It agrees well with the ones by \citet{Kochanek2001} and \citet{Cole2001}. The comparison shows excellent agreement between the two luminosity functions, both in shape and normalization, and indicates that our parent sample is representative of the general galaxy population. In particular this test shows no sign of incompleteness at the faint end, in agreement with the discussion of \refsec{sec:redshift_incompleteness}.

\subsection{Size - Luminosity relations for spirals and ETGs}

\begin{figure}
\plotone{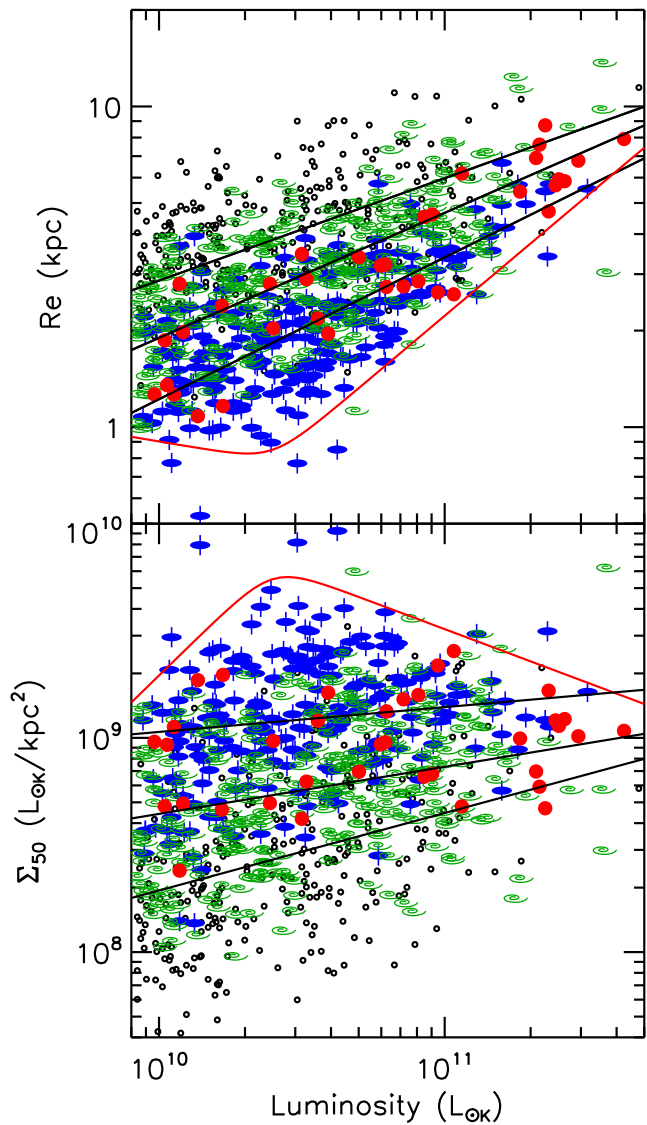}
\caption{{\em Top Panel:} Size-Luminosity relation for the parent sample. The $K$-band luminosity of all the galaxies in the parent sample is plotted against the effective radius \re\ (\refsec{sec:observing}). Red filled circles are slow rotators, blue ellipses with vertical axis are fast rotators. These 260 objects constitute the \atl\ sample. The green spirals represent spiral galaxies of type Sa--Sb ($-0.5<T\le4$), while later spiral types ($T>4$) are plotted as small open circles. The solid lines are the best fitting relations described in the text. From the bottom to the top they are fit to the fast rotators ETGs, to the Sa--Sb and to the later spiral types. The red curve approximates the boundary of the zone of avoidance in the observed galaxy distribution. {\em Bottom Panel:} Effective surface brightness versus luminosity. The symbols and the lines are the same as in the top panel.}
\label{fig:luminosity_size}
\end{figure}

To illustrate the general characteristic of the galaxies in the parent sample \reffig{fig:luminosity_size} shows the $K$-band size-luminosity relation and the effective surface brightness $\Sigma_{50}\equiv L_{\odot K}/(2\pi\re^2)$ for different morphological types. This plot shows a similar distribution as the one inferred from much larger galaxy samples based on SDSS photometry, further confirming the representativeness of our sample (compare \reffig{fig:luminosity_size} with fig.~2 of \citealt{vanDokkum2008}). In our plot we show the fast/slow rotator classification for 260 ETGs of the \atl\ sample (Paper III) together with the early spirals (Sa--Sb) and later spiral types (Sc--Irr) of the parent sample.
There is a clear trend in the $\re-L_K$ diagram as a function of galaxy morphology. To quantify this trend we fitted linear relations to the logarithmic coordinates assuming the same fractional errors for both axes and requiring $\chi^2/{\rm DOF}=1$. The fit was performed using the \texttt{FITEXY} routine which is based on the algorithm by \citet{press92} and is part of the IDL Astronomy User's Library\footnote{Available from http://idlastro.gsfc.nasa.gov/} \citep{Landsman1993}. The best fitting power-law $\re-L_K$ relations are
\begin{equation}
\log \left[\frac{\re(\rm FR)}{\rm kpc}\right] = 0.53 + 0.44 \log \left[\frac{L_K(\rm FR)}{10^{11} L_{\odot,K}}\right],
\label{eq:lum_size_fr}
\end{equation}
\begin{equation}
\log \left[\frac{\re(\rm Sa-Sb)}{\rm kpc}\right] = 0.67 + 0.39 \log \left[\frac{L_K(\rm Sa-Sb)}{10^{11} L_{\odot,K}}\right],
\end{equation}
\begin{equation}
\log \left[\frac{\re(\rm Sc-Irr)}{\rm kpc}\right] = 0.78 + 0.32 \log \left[\frac{L_K(\rm Sc-Irr)}{10^{11} L_{\odot,K}}\right],
\end{equation}
for fast rotators, Sa--Sb spirals and Sc or later spiral types respectively. There is a clear know zone of avoidance at small sizes and large luminosities, which is also well defined in our parent sample and approximated within the region where we have data by a double power-law (cfr.\ \citealt{Lauer1995}):
\begin{equation}
\re = 2^{(\gamma-\beta)/\alpha}R_{{\rm e},b}\left(\frac{L_K}{L_{K,b}}\right)^\gamma\left[1 + \left(\frac{L_K}{L_{K,b}}\right)^\alpha\right]^{(\beta-\gamma)/\alpha}.
\end{equation}
This equation defines a minimum effective radius $R_{{\rm e},b}=0.85$ kpc and a corresponding maximum effective surface brightness at a characteristic galaxy luminosity $L_{K,b} = 2.5\times10^{10}$ L$_{\odot,K}$. The logarithmic power slope for $L_K\ll L_{K,b}$ is $\gamma=-0.15$, while for $L_K\gg L_{K,b}$ it is $\beta=0.75$, so that for large luminosities $\re\propto L_K^{0.75}$. A sharp transition between the two regimes is set by $\alpha=8$.

Consistently with other larger local galaxy samples \citep{vanDokkum2008,Trujillo2009,Taylor2010}, we find no massive and superdense ($L_K\ga10^{11}$ $L_{\odot,K}$ and $\re\la2$ kpc) ETGs in our parent sample, contrary to what is found from photometry of ETGs at redshift $z\ga1.5$ \citep[e.g.][]{Daddi2005,Trujillo2006,Cimatti2008,vanDokkum2008}. Similar $\re-L$ relations were derived by \citet{Shen2003} for early-type and late-type galaxies, defined as those having a \citet{sersic68} index larger or smaller than $n=2.5$ respectively. Their relation also showed a trend for late-types to have larger sizes (by definition, due to the smaller $n$) at given luminosity (or mass) and a more shallow $\re-L$ relations, as we find using the morphological selection. A trend in the $\re-M$ relation involving colours, with red-sequence galaxies having smaller sizes, can be seen in \citet{vanDokkum2008}. While a trend involving age was presented by \citet{vanderWel2009} and \citet{Shankar2009}, and confirmed by \citet{Valentinuzzi2010}, who find smaller sizes for older objects, at given stellar mass. All these trends are consistent with our finding using fast/slow rotators ETGs, in combination with traditional morphology of spiral galaxies (see also \citealt{Bernardi2010}), when one considers that later galaxy types tend to be more gas rich and have a younger stellar population. However here we interpret the observed trends as due to a variation in the bulge fraction, with bulges progressively increasing (by definition) from Sd-Sc to Sb-Sa and to fast rotators ETGs (see Paper~VII). A detailed study of scaling relations in our sample will be presented in a subsequent paper.

\section{The \atl\ Sample}

\subsection{Morphological selection}

\begin{figure*}
\includegraphics[width=0.95\textwidth]{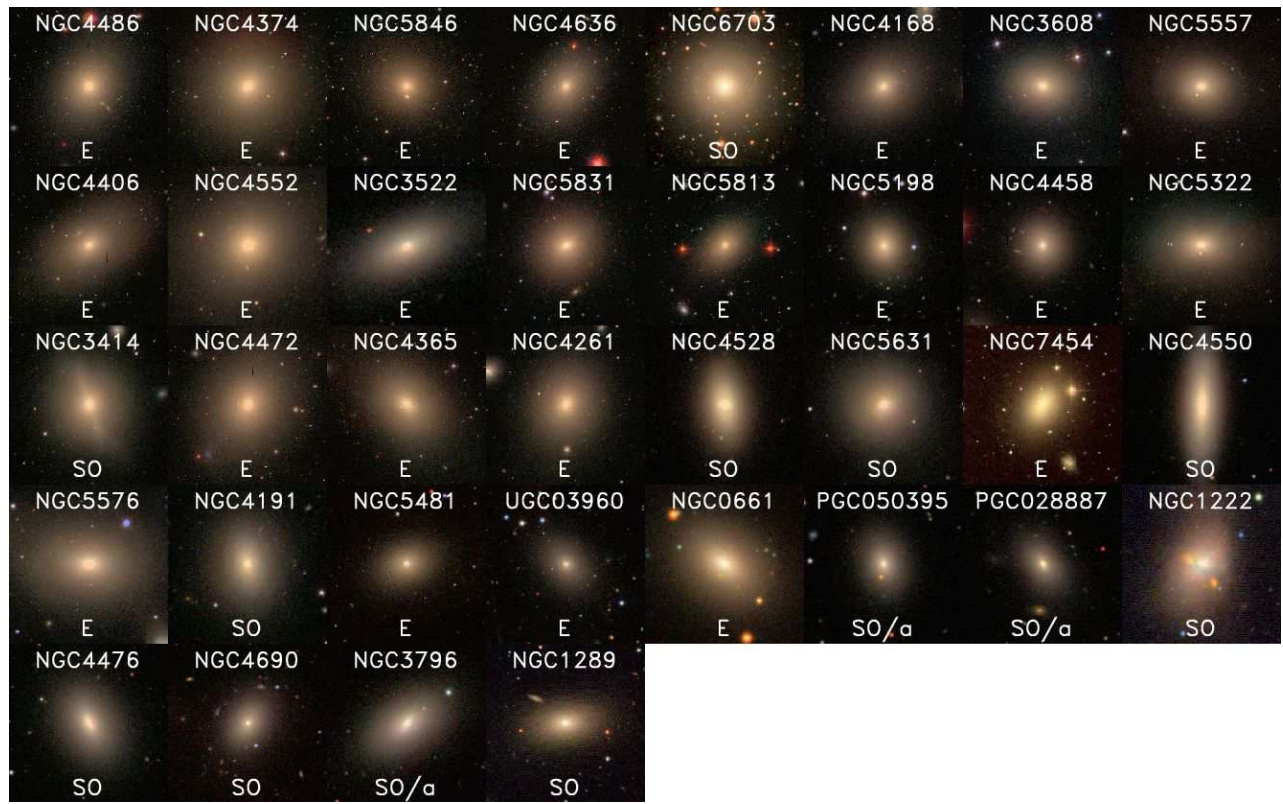}
\caption{Morphology of slow rotators ETGs sorted by increasing $\lambda_R$. Postage stamps of the SDSS DR7 and INT red-green-blue composite images of slow rotators in the \atl\ sample. The image of each galaxy was scaled so that the plot side is equal to 10\re, where \re\ is the projected half-light radius given in Table~\ref{tab:atlas3d_sample}. From left to right and from top to bottom the panels are sorted according to their specific stellar angular momentum, as measured by the parameter $\lambda_R$ given in Paper III. The galaxy name is given at the top of each panel and the morphological classification from HyperLeda at the bottom. At this scale slow-rotator ETGs appear generally featureless except for the synchrotron jet in NGC~4486 and obvious signs of interactions in NGC~1222, NGC~3414 and NGC~5557. The only significant flat galaxy in this class is NGC~4550, while two other galaxies NGC~3796 and NGC~4528 show evidence of bar perturbations, which is typically associated to stellar disks. All these three objects contain counter-rotating stellar disks (Paper II). (This figure is better appreciated on a computer screen rather than on printed paper.)}
\label{fig:postage_stamps_sr}
\end{figure*}

\begin{figure*}
\includegraphics[width=0.93\textwidth]{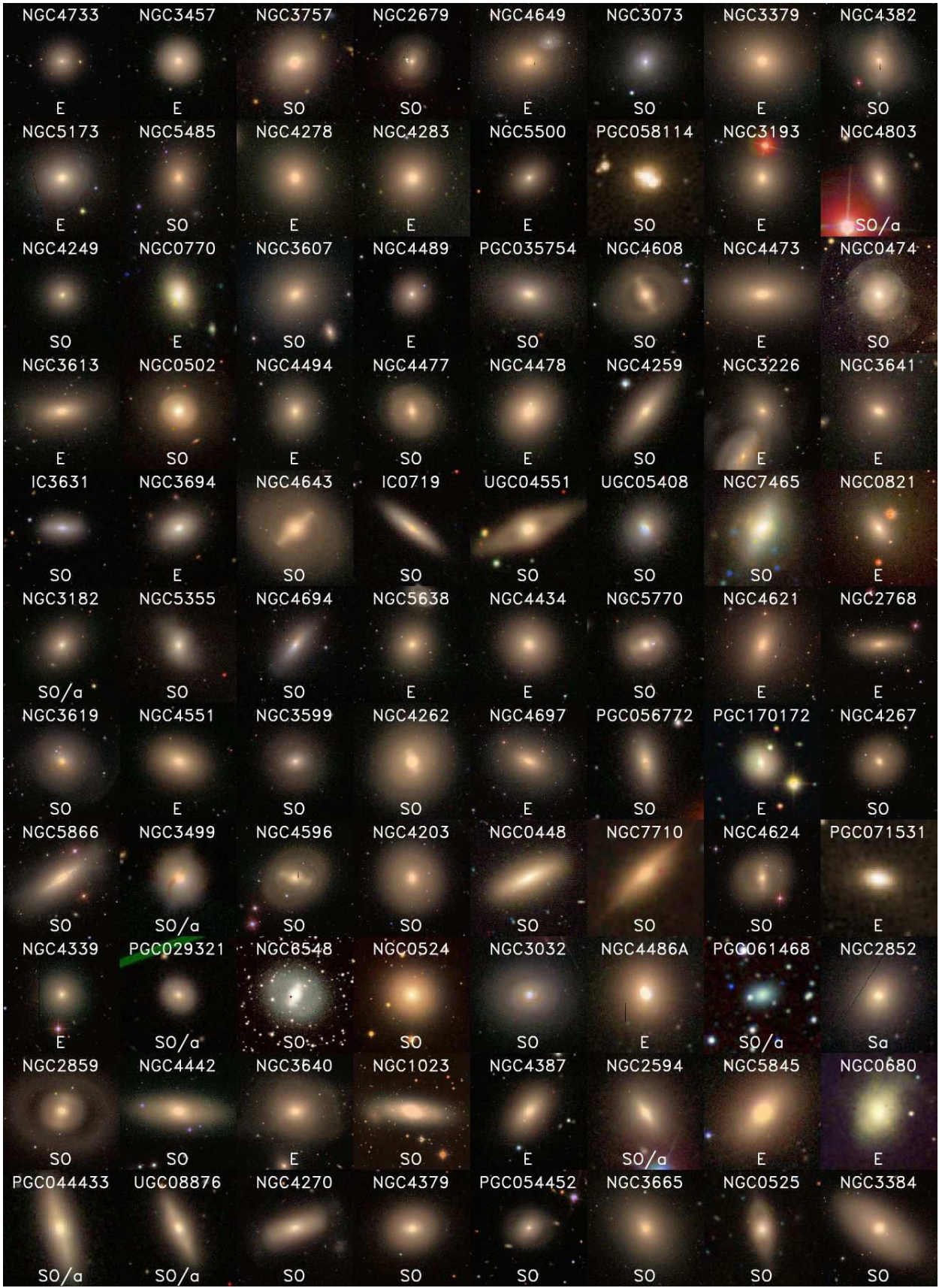}
\caption{Same as in \reffig{fig:postage_stamps_sr} for the fast rotators ETGs in the \atl\ sample, sorted by increasing $\lambda_R$. The first panel contains mostly round objects. Many of them are barred (Paper II), nearly face-on, S0 galaxies and often contain stellar rings (e.g.\ NGC4608), while others appears face-on from the geometry of their dust. This suggests that the round shape and low $\lambda_R$ of these objects is not intrinsic, but due to their low inclination ($i=90^\circ$ being edge on). On the contrary the last panel is dominated by nearly edge-on disks, which explains their high $\lambda_R$.}
\label{fig:postage_stamps_fr}
\end{figure*}

\begin{figure*}
\includegraphics[width=0.93\textwidth]{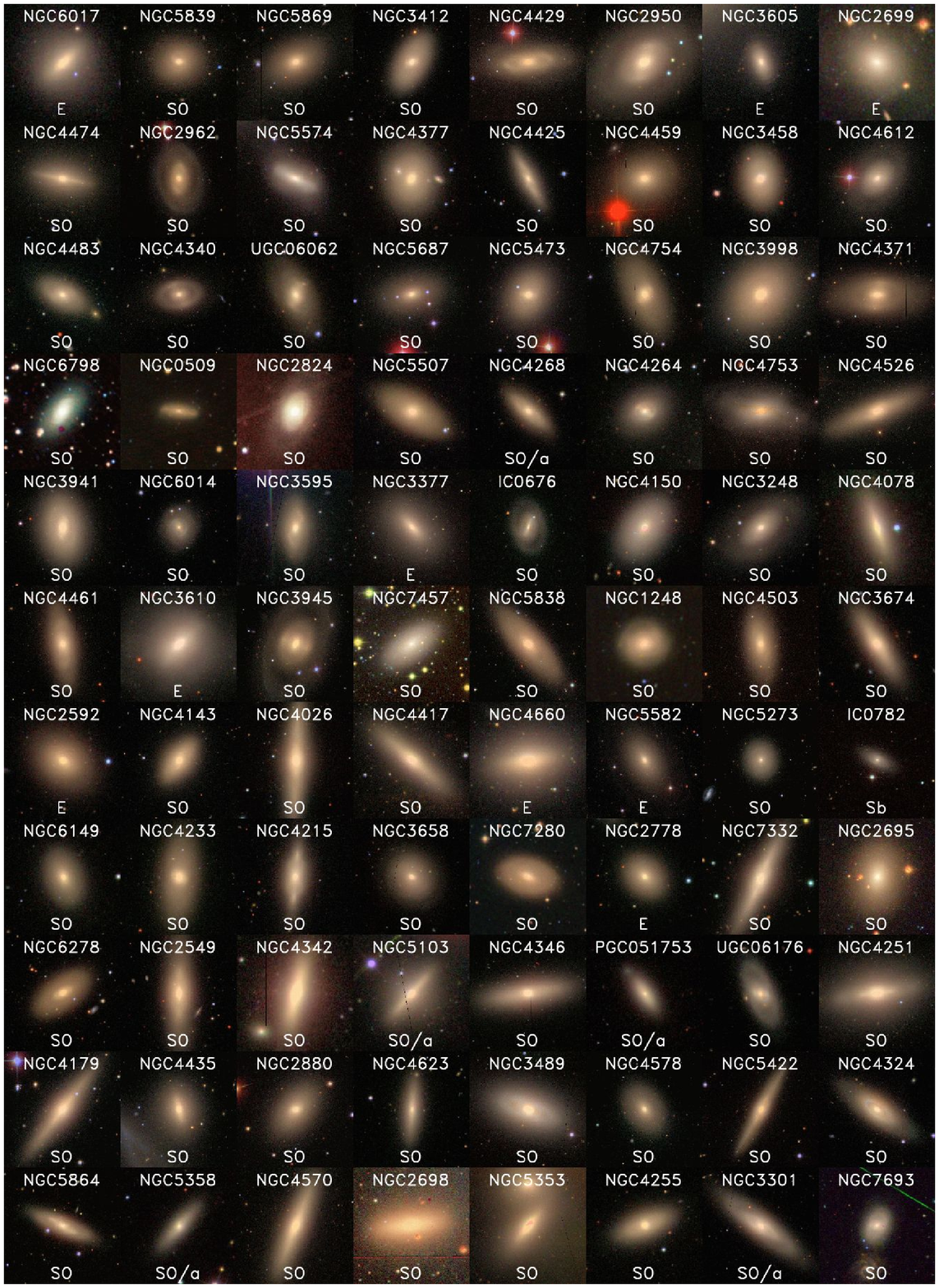}
\addtocounter{figure}{-1}
\caption{ --- continued}
\end{figure*}

\begin{figure*}
\includegraphics[width=0.93\textwidth]{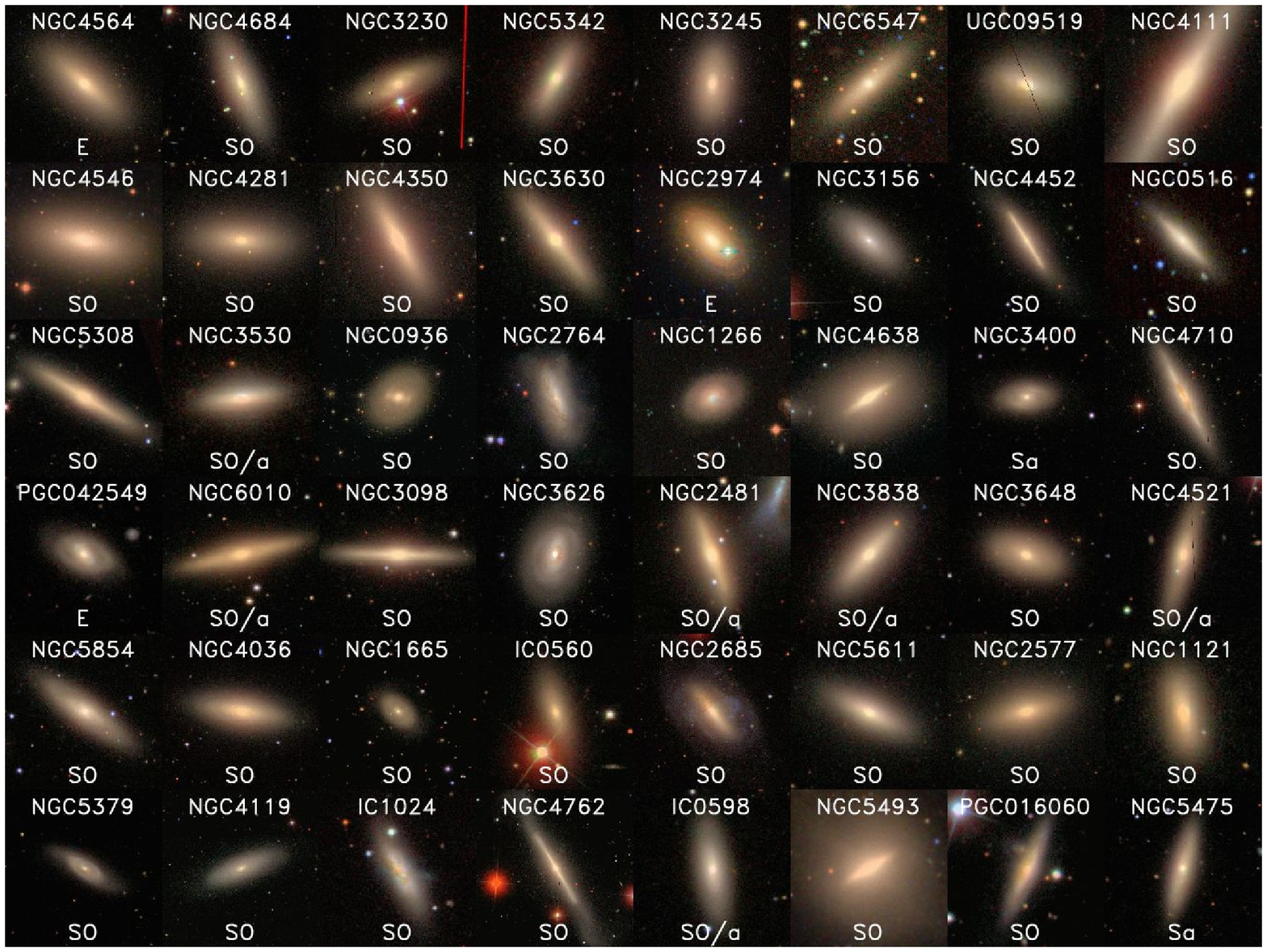}
\addtocounter{figure}{-1}
\caption{ --- continued}
\end{figure*}

We established that the \atl\ parent sample is essentially complete within the selection criteria and representative of the nearby galaxy population. We also verified that its luminosity function agrees with the one derived from much larger volumes of the Universe. The \atl\ survey however is focused on the study of the fossil record of galaxy formation as recorded in the structure of ETGs. The \atl\ sample is a subset of the parent sample, consisting of all the ETGs in that sample.

The distinction between red-sequence and blue-cloud galaxies is related, but different from the early-type versus spiral morphological separation. E and S0 galaxies invariably lie on the red sequence, while late type spirals are generally on the blue cloud. However large fractions of spirals populate the red sequence as well and overlap with early-type galaxies \citep{Strateva2001,Conselice2006,vandenBergh2007,Bernardi2010}. An accurate morphology is easier to obtain for nearby galaxies and it is more robust than colour to dust and inclination effects. For this reasons a morphological classification is our preferred selection criterion.

To perform the morphological selection we considered using the morphological classification provided in available catalogues like the RC3 or its ongoing evolution HyperLeda. A problem with those classifications is the possible non-homogeneity of the classification process. Moreover the classification in those catalogues was performed using photometry in a single band, often from photographic plates. Given that for the majority of the galaxies in our parent sample excellent quality multi-band photometry is available from the SDSS DR7 \citep{Abazajian2009}, we decided to revisit the classification of the whole parent sample using the best available imaging.

The morphological classification of a given galaxy using multi-colour imaging may differ from the one obtained from photographic plates of the same object. Nonetheless we tried as much as possible to be consistent with the currently accepted morphological criteria. We just need to separate the parent sample into two classes: ETGs and spirals. This makes our task much simpler and reproducible than a more detailed morphological classification into E, S0 and spiral subclasses Sa--Sd.

Since the introduction of the classic tuning-fork diagram by \citet{Hubble1936}, for the past half century, essentially all authors have converged on a simple criterion to differentiate ETGs from spirals. The criterion, which defines the {\em revised} Hubble classification scheme, is outlined by \citet{Sandage1961} in the Hubble Atlas and is based entirely on the presence of spiral arms (or dust lanes when seen edge-on): ``The transition stages, S0 and SB0, are firmly established. In both sequences, the nebulae may be described as systems definitely later than E7 but showing no spiral structure''. This same criterion was adopted unchanged in the extension to the classification scheme by \citet{deVaucouleurs1959,deVaucouleurs1963}, which was applied to the widely used RC2 and RC3 catalogues \citep{deVaucouleurs1976,deVaucouleurs1991} and HyperLeda \citep{Paturel2003}. Although other characteristics of galaxies change with morphological classification (e.g.\ the bulge/disk ratio), they are ignored in the separation between early-types and spiral galaxies \citep[see review by][]{Sandage1975}. We adopted the same criterion here to select the ETGs belonging to the \atl\ sample from the parent sample.

Our morphological selection was done by visual inspection of the true-colour red-green-blue images \citep{Lupton2004} provided by the SDSS DR7 which are available for 82\% of the galaxies in the parent sample. For the remaining objects we used the $B$-band DSS2-blue images in the Online Digitized Sky Surveys\footnote{http://archive.eso.org/dss}. We revisited the classification for the galaxies without SDSS DR7 data after obtaining our own INT imaging  (\refsec{sec:data}) and this lead to the removal of a couple of galaxies from the ETGs sample.
At the end of our classification we compared the agreement between our separation into early-types and spirals and the one provided by the $T$-type given by HyperLeda, which defines as ETGs (E and S0) those having $T\le-0.5$. We found agreement in 97\% of the cases, confirming the robustness and reproducibility of the morphological selection. The few disagreements with HyperLeda could all be easily explained by the high quality of the multi-colour SDSS images, which allowed for a better detection of faint spiral structures. The \atl\ sample of 260 ETGs obtained from this selection is given in Table~\ref{tab:atlas3d_sample} and illustrated in \reffig{fig:postage_stamps_sr} and in \reffig{fig:postage_stamps_fr}, together with the HyperLeda morphological classification as provided by their $T$-type (E: $T\le-3.5$; S0: $-3.5<T\le-0.5$; S0/a: $-0.5<T\le0.5$).

\subsection{Colour-magnitude diagram}

\begin{figure}
\plotone{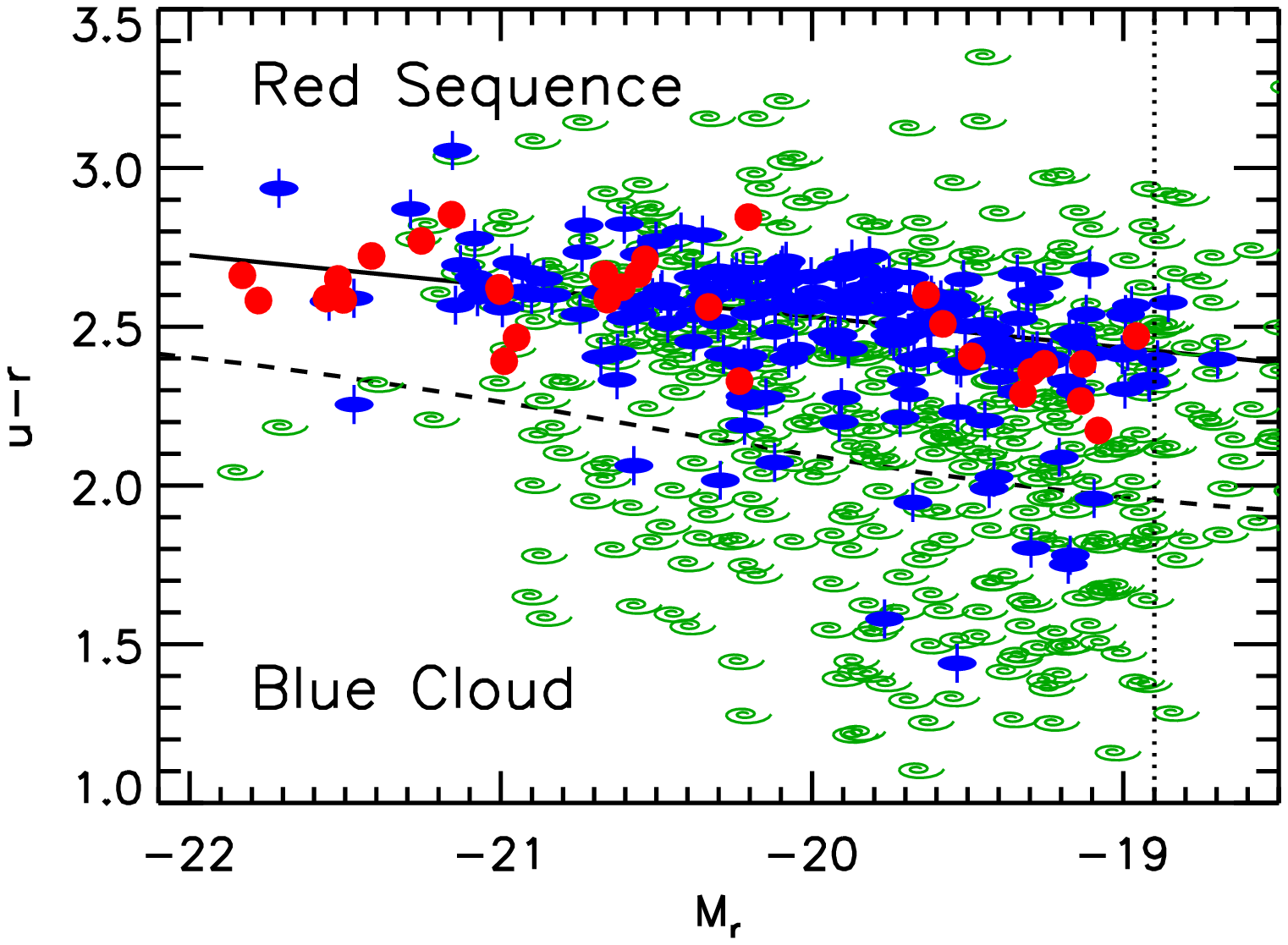}
\caption{$u$-$r$ versus $M_r$ colour-magnitude diagram for the morphologically-selected \atl\ parent sample with SDSS photometry. The blue ellipses with axis are fast-rotators ETGs, the red filled circles are slow rotators ETGs, while the green spirals are spiral galaxies. The dashed line indicates the separation between red sequence and blue cloud established by \citet{Baldry2004,Baldry2006}, from a sample of 151,642 galaxies. The vertical dotted line indicates our approximate survey completeness limit in $r$-band $M_r\la-18.9$ mag. The solid line is a linear robust fit to the ETGs only, minimizing the sum of the absolute residuals.}
\label{fig:colour_magnitude}
\end{figure}

The decision to select the \atl\ sample based on morphology instead of colour was based on (i) the broad similarity of the two criteria, (ii) the non-availability of reliable colours for the whole parent sample and (iii) the robustness of morphology, as opposed to colours, against dust extinction. Still, we expect the \atl\ sample to include mainly galaxies on the red sequence in a colour-magnitude diagram. This is verified in \reffig{fig:colour_magnitude}. The \atl\ sample indeed defines a narrow colour-magnitude sequence approximated, in SDSS magnitudes, by
\begin{equation}
    u-r = 2.53 - 0.097\times (M_r+20).
\end{equation}
As found by previous authors there is little scatter in the relation at the high-mass end, while at the lower mass-end some galaxies appear to be still in transition between the blue and red sequence \citep{Strateva2001,Conselice2006,vandenBergh2007,Bernardi2010}. The 31 ETGs with SDSS colour and defined as slow rotators in Paper III all lie close to the red sequence with an rms scatter of 0.13 mag from the best-fitting relation. All the deviants ETGs are classified as fast rotators in Paper III. The nature of these objects will be investigated in detail in subsequent papers of this series. Spiral galaxies span the full region of the diagram, both on the red sequence and the the blue clouds, as found by previous studies.

\section{\sauron\ data for the \atl\ survey}
\label{sec:sauron}

\subsection{Observing strategy}
\label{sec:observing}

The main aim of the \sauron\ \citep{Bacon2001} observations of the \atl\ sample is to obtain global galaxy quantities like the specific stellar angular momentum $\lambda_R$, the \vse\, the global kinematical misalignment, the luminosity-weighted second moment $\sigma_e$, the stellar and total mass-to-light ratio, the mean stellar population and the ionized gas emission. To be representative of the galaxies as a whole, these quantities need to be measured at least within one projected half-light radius \re. Moreover for a given observed area, more accurate values of the kinematical quantities are obtained when the quantities are measured within ellipses instead of circles, with ellipticity given by the galaxy photometry and the position angle defined by the stellar kinematics (see appendix~B of \citealt{Cappellari2007}). In addition, when galaxies are barred, the \sauron\ survey has shown that the kinematics are generally still aligned with the position angle of the galaxy photometry at large radii ${\rm PA_{phot}}$ \citep{Krajnovic08}, which defines the position of the line-of-nodes of the disk. These requirements, which derive from our experience with the \sauron\ survey \citep{deZeeuw2002}, lead to the following optimized observing strategy, which we systematically applied for the \sauron\ observation of the \atl\ sample:
\begin{enumerate}
\item When $\re\le30\arcsec$ take a single \sauron\ field and orient the \sauron\ major axis with the large radii ${\rm PA_{phot}}$;
\item When $\re>30\arcsec$ then we take a mosaic of two \sauron\ fields. Given the size of the \sauron\ field of $33''\times41''$, the criterion of maximizing the area of the largest isophote, of axial ratio $q'$, enclosed within the observed field-of-view, becomes:
    \begin{enumerate}
        \item If $q'<0.55$ the \sauron\ long axis is aligned with ${\rm PA_{phot}}$ and the mosaic is made by matching the two \sauron\ pointings along the short side;
        \item If $q'\ge0.55$ the \sauron\ short axis is aligned with ${\rm PA_{phot}}$ and the mosaic is made by matching the two \sauron\ pointings along the long side;
    \end{enumerate}
\end{enumerate}

\begin{figure}
\centering
\plotone{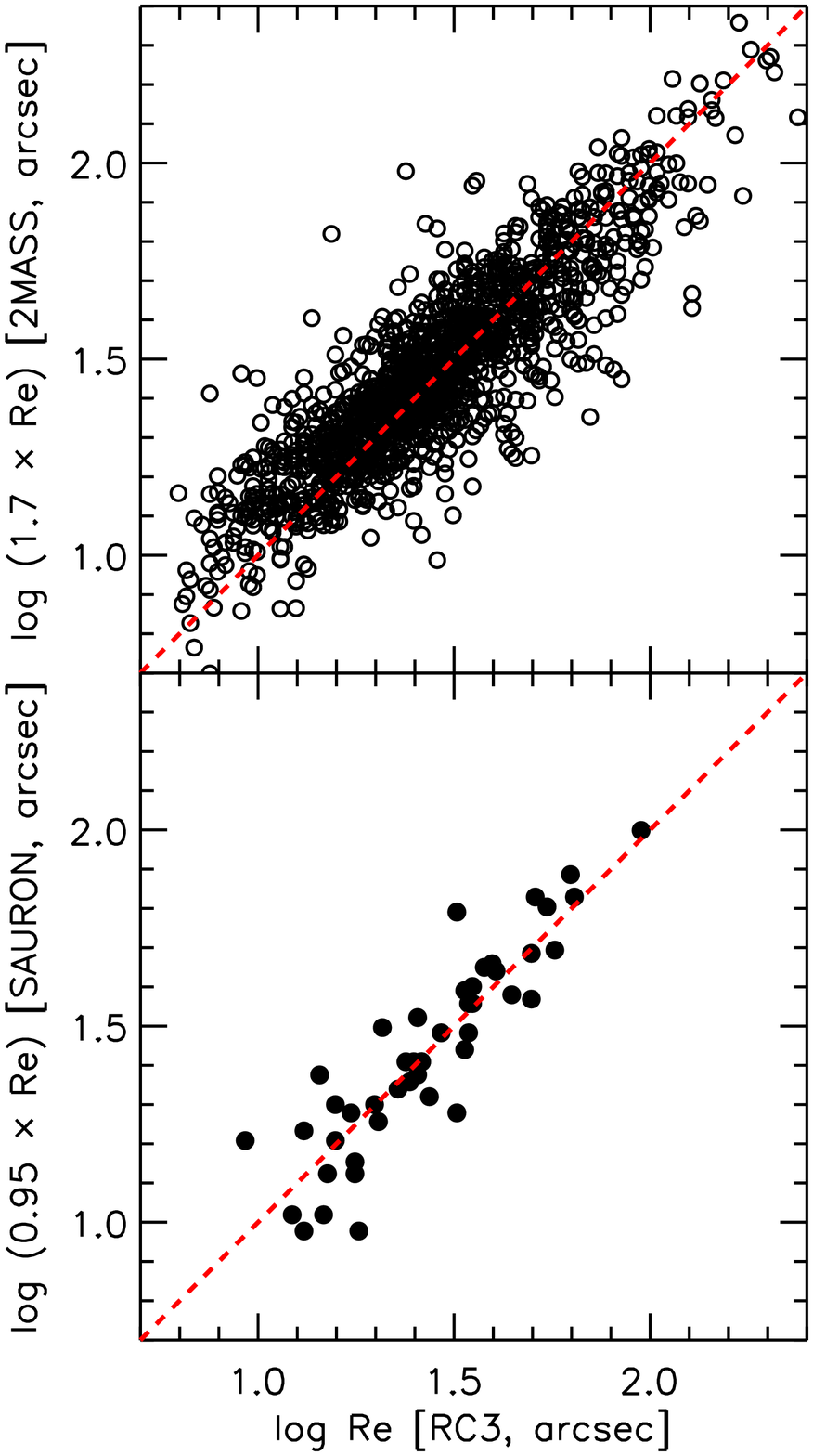}
\caption{Testing the accuracy of \re\ determinations. {\em Top Panel:} Comparison between 1353 values of \re\ given by RC3 and the ones given by 2MASS (computed as described in the text), scaled by the best-fitting factor of 1.7. {\em Bottom Panel:} Same as in the top panel for the RC3 \re\ and the ones for 46 ETGs of the \sauron\ survey. Once corrected for systematic differences, all three \re\ determinations are consistent with a similar error of $\approx22\%$.}
\label{fig:re_comparison}
\end{figure}

At the time of the \sauron\ observations the only photometry available to us for the whole sample was from 2MASS. We adopted the \re\ provided by the 2MASS XSC, which is determined via growth curves within elliptical apertures. Specifically, in terms of the XSC catalogs parameters, we defined
\begin{eqnarray}
\re^{\rm 2MASS} = \textrm{MEDIAN}(\textrm{j\_r\_eff, h\_r\_eff, k\_r\_eff})\times\sqrt{\textrm{k\_ba}},
\end{eqnarray}
as the median of the three 2MASS values in the $J$, $H$ and $K_s$-band, where the factor $\sqrt{\textrm{k\_ba}}$ takes into account the fact that the 2MASS values are the semi-major axes of the ellipses enclosing half of the galaxy light and we want the radius of the circle with the same area. This $\re^{\rm 2MASS}$ was compared to the $\re^{\rm RC3}$ provided by the RC3 catalogue and measured via growth curves within circular apertures. The two values correlate well, with an observed rms scatter of 0.12 dex in \re, which implies an error of about 22\% in each \re\ determination (assuming they have similar errors). However there is a general offset by a factor $\re^{\rm RC3}\approx1.7\times\re^{\rm 2MASS}$ between the two determinations (\reffig{fig:re_comparison}). The rms scatter in the $\re^{\rm RC3}-\re^{\rm 2MASS}$ correlation is close to the one (0.11 dex) we obtain when comparing $\re^{\rm RC3}$ to 46 values determined using growth curves in the $I$-band for the \sauron\ survey \citep{Cappellari2006,Kuntschner2006}. In that case however the offset in the values is negligible (factor 0.95). We conclude that the 2MASS \re\ determinations have comparable accuracy to the RC3 and \sauron\ determination, when they are increased by a factor 1.7 to account for the differences in the observed photometric band and in the depth of the photometry data. All three values are consistent with having a similar error of $\approx22\%$ in \re. This rather large error is consistent with the findings of \citet{Chen2010} from another extensive comparison of \re\ values. To further improve the accuracy we adopted $\re=(1.7\times\re^{\rm 2MASS}+\re^{\rm RC3})/2$ for the 412/871 galaxies with both 2MASS and RC3 determinations and $\re=1.7\times\re^{\rm 2MASS}$ when only 2MASS was available. The values of \re\ for the full parent sample, divided into ETGs and spirals, are given in Table~\ref{tab:atlas3d_sample} and Table~\ref{tab:atlas3d_spirals}.

\subsection{Integral-field spectroscopic observations}
\label{sec:ifu_obs}

The \sauron\ integral-field spectrograph was first mounted at the  William Herschel Telescope (WHT) at the Observatory of El Roque des Los Muchachos on La Palma in 1999. It has been used extensively in particular in the course of the \sauron\ survey, but also in separate smaller efforts \citep[e.g.][]{Bower2004,Allard2005,Dumas2007,Weijmans2010}. Given that the \atl\ selection criteria are by design very similar to the ones in the \sauron\ survey, a total of 64 ETGs had been observed before the beginning of the \atl\ observing campaign. 49 ETGs were part of the main survey \citep{deZeeuw2002}, out of which 47 were presented in the sub-sample of ETGs \citep{Emsellem2004} and two in the early-spirals sub-sample \citep{FalconBarroso06}. 14 `special' ETGs within the \atl\ volume had been observed with \sauron\ in the course of other projects \citep[table~3 of][]{Cappellari2007}. All these galaxies were observed before the volume phase holographic (VPH) grating came into use and were taken with an exposure time of 2 hours on source, in some cases with multiple spatial pointings to cover galaxies to roughly one effective (projected half-light \re) radius. All the observations were obtained in the low spatial resolution mode in which the instrument has a field-of-view of $33''\times41''$ sampled with 0\farcs94 square lenslets and with a spectral resolution of 4.2 \AA\ FWHM ($\sigma_{\rm instr}=105$ km s$^{-1}$), covering the wavelength range 4800--5380 \AA.

\begin{table}
\caption{\sauron\ observing runs for the \atl\ sample}
\centering
\begin{tabular}{ccc}
\hline
Run & Dates & Clear\\
\hline
1 & 2007 April 10--23 & 12/14 \\
2 & 2007 August 13--15 & 3/3 \\
3 & 2008 January 9--15 & 6/7 \\
4 & 2008 February 27 -- March 11 & 11/14 \\
\hline
\label{tab:observing_runs}
\end{tabular}
\end{table}

To observe the additional 196 galaxies we were allocated 38 observing nights comprising four observing runs spread over three consecutive semesters as part of a long-term project at the WHT (Table~\ref{tab:observing_runs}). The time allocation was split equally between Dutch and UK time. We had excellent weather with just 16\% of nights lost due to clouds, compared to a normal average for the period of around 30\%. This fact, combined with an efficient observing strategy allowed us to complete all observations of the \atl\ sample galaxies in the allocated time.

The optimal scheduling of the observations of the 196 galaxies, in some cases using multiple spatial pointings, was performed with a dedicated IDL script which gave higher priority to the intrinsically brightest galaxies, took into account the galaxy coordinates, the need for multiple pointings, the dates of the four observing runs, and the moon position and phase. The script could be easily re-run to modify the scheduling to account for time lost due to bad weather. The observations were performed with the VPH grating, which provides a resolution of 3.9 \AA\ FWHM ($\sigma_{\rm instr}=98$ \kms), about 7\% better than for the \sauron\ survey. The adopted on-source exposure time was 1 hour, split into two equal 30 min exposure dithered by a couple of arcseconds.

\subsection{Data reduction and stellar kinematics extraction}
\label{sec:kinematics}

\begin{figure}
\plotone{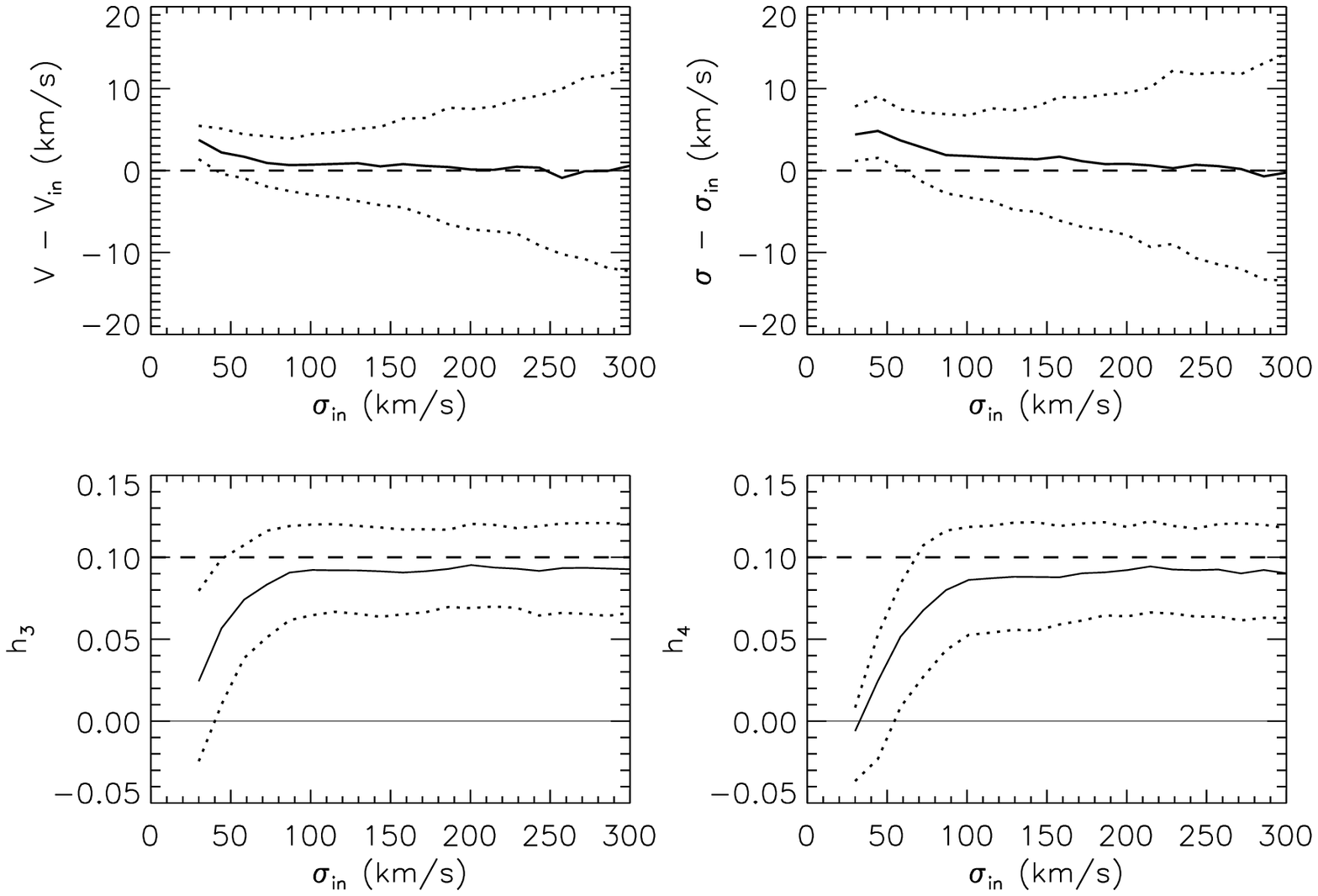}
\caption{Testing penalization in pPXF. We simulated spectra with the $S/N=40$ of our data and an LOSVD with $h_3=0.1$, $h_4=0.1$ and $\sigma$ in the range between 30 and 300 \kms. We extracted the kinematics with pPXF and a penalty $\lambda=0.5$. The lines in the top two panels show the 50th (median, solid line), 16th and 84th percentiles (1$\sigma$ errors, dotted lines) of the differences between the measured values and the input values of the mean velocity $V_{\rm in}$ and the velocity dispersion $\sigma_{\rm in}$. The bottom panels show the same lines for the recovered values of $h_3$ and $h_4$, compared to the input values (dashed line). The $h_3$ and $h_4$ parameters can only be recovered when $\sigma_{\rm in}\ga100$ \kms. Typical statistical errors in the kinematics parameter for $\sigma_{\rm in}\approx200$ \kms\ are 6 \kms, 7 \kms, 0.03 and 0.03, for $V$, $\sigma$, $h_3$ and $h_4$ respectively.}
\label{fig:ppxf_simulation}
\end{figure}

\begin{figure}
\plotone{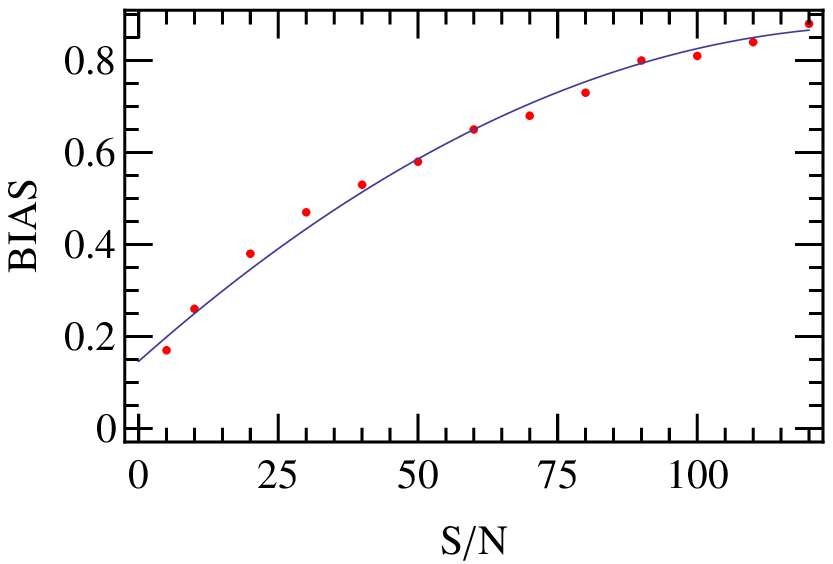}
\caption{Relation between the $S/N$ of the \sauron\ spectrum and the optimal penalty parameter $\lambda$ (keyword BIAS) in pPXF. The solid line is the polynomial ${\rm BIAS} = 0.15 + 0.0107\; S/N - 0.00004\; (S/N)^2$}
\label{fig:ppxf_bias_sn}
\end{figure}

The \sauron\ data reduction followed standard procedures and used the dedicated XSAURON software developed at Centre de Recherche Astronomique de Lyon (CRAL) and was already described in \citet{Bacon2001} and \citet{Emsellem2004}. However some details of our approach have been improved after the original publication of the data. For this reason we re-extracted all the stellar kinematics using the improved and completely homogeneous approach for the entire \atl\ dataset. We describe here the minor differences with respect to what was published before.

The data were spatially binned with the adaptive Voronoi method\footnote{Available from http://purl.org/cappellari/idl} of \citet{Cappellari2003}, which optimally solves the problem of preserving the maximum spatial resolution of two-dimensional data, given a constraint on the minimum signal-to-noise ratio ($S/N$). We adopted a target $S/N=40$ for all the data used in the \atl\ survey, including the previous \sauron\ observations, which were re-extracted adopting for consistency this lower $S/N$ instead of the $S/N=60$ as originally published in \citet{Emsellem2004}.

The stellar kinematics were extracted with the penalized pixel-fitting (pPXF) software\footnotemark[8] \citep{Cappellari2004}, which simultaneously fits the stellar kinematics and the optimal linear combination of spectral templates to the observed spectrum, using a maximum-likelihood approach to suppress noisy solutions. The line-of-sight velocity distribution (LOSVD) is described via the Gauss-Hermite parametrization up to $h_3-h_4$ \citep{vanDerMarel93,gerhard93}. We employed as stellar template an optimal linear combination of stars from the MILES library\footnote{Available from http://miles.iac.es/} \citep{SanchezBlazquez2006}, which was separately determined for every galaxy. We did not allow the template to change in every bin, to eliminate small artifacts in the kinematics due imperfections in the velocity alinement of the MILES stars. We adjusted the penalty in pPXF to a value $\lambda=0.5$, optimized for the adopted $S/N$. Following the pPXF documentation we determined the optimal $\lambda$ by requiring the maximum bias in the Gauss-Hermite parameters $h_3$ and $h_4$ to be equal to rms$/3$, where the rms is the scatter of the measurements obtained from a Monte Carlo simulations with the adopted $S/N=40$ and a well resolved stellar dispersion $\sigma\ga180$ \kms\ (\reffig{fig:ppxf_simulation}). In a handful of cases we could not reach the required $S/N$ without employing excessively large Voronoi bins. In those cases we further reduced the target $S/N$. For those galaxies we correspondingly adapted the penalty in pPXF according to the empirical relation of \reffig{fig:ppxf_bias_sn}. For usage in cases where we need to approximate the stellar velocity second moments and not the full LOSVD --- e.g. to fit models based on the \citet{Jeans1922} equations or measure $\lambda_R$ or $V/\sigma$ --- we separately extracted the kinematics assuming a simple Gaussian LOSVD. In that case the pPXF penalty is ignored. In all cases the errors on the kinematics were determined via bootstrapping \citep{Efron1993}, by randomly re-sampling the pPXF fit residuals of the best fit and repeating the kinematic fit for 100 realizations, with a zero penalty \citep[see Sec.~3.4 of][]{Cappellari2004}. The homogeneous set of integral-field \sauron\ kinematics introduced in this paper, together with the stellar population parameters, the characteristics of the ionized gas and the entire data cubes for the full \atl\ sample will be made available via the project Web page\footnote{http://purl.org/atlas3d} at the end of our project.

\begin{figure*}
\includegraphics[width=0.77\textwidth]{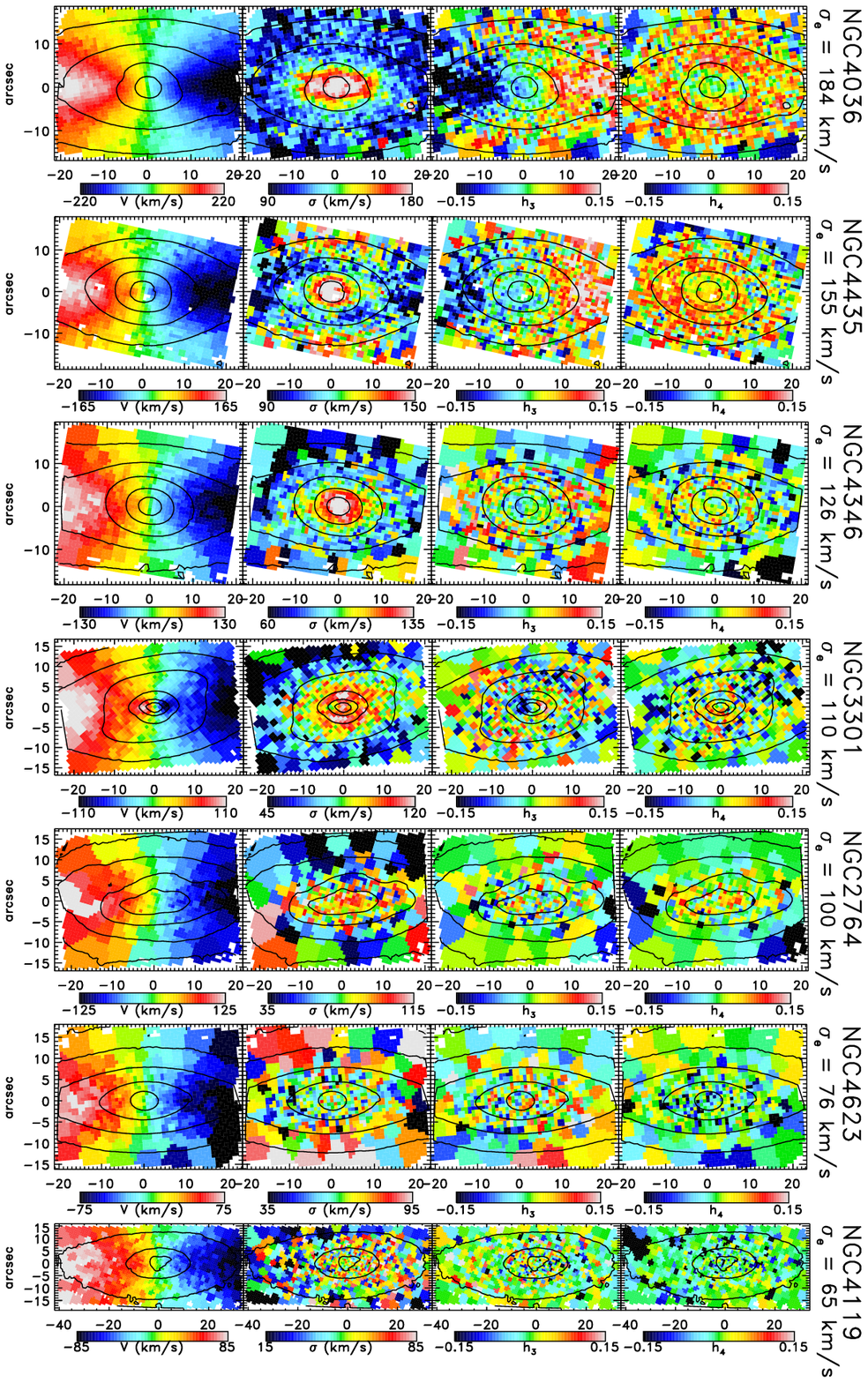}
\caption{Quality of the \sauron\ kinematics in the \atl\ survey. In each column from left to right shows the  Voronoi binned kinematic moments extracted via pPXF from the \sauron\ data: mean velocity, $V$, velocity dispersion, $\sigma$, and higher Gauss-Hermite moments, $h_3$ and $h_4$. From top to bottom the data for seven newly-observed fast-rotators in the \atl\ sample are sorted according to the luminosity-weighted dispersion \se\ within 1\re. Below $\se\la120$ \kms\ the data have insufficient information to constrain the full LOSVD and the Gauss-Hermite moments are automatically and gradually suppressed by pPXF towards zero to reduce the noise in $V$ and $\sigma$, which can still be reliably recovered. About 40\% of the galaxies in the sample have $\se\la120$ \kms.}
\label{fig:ppxf_kinematics}
\end{figure*}

As discussed in detail in \citet{Cappellari2004} and illustrated in \reffig{fig:ppxf_simulation}, the \sauron\ spectroscopic data allow the extraction of the full stellar Line-of-Sight Velocity-Distribution (LOSVD), including the Gauss-Hermite parameters, only for observed velocity dispersions $\sigma\ga120$ \kms. Below this value the $h_3$ and $h_4$ start becoming unconstrained by the data,  due to the spectral under-sampling and for this reason the pPXF method automatically and gradually penalizes them towards zero to keep the noise on the mean velocity $V$ and velocity dispersion $\sigma$ under control. As the minimum $S/N$ of our data is defined by our Voronoi binning criterion, the degree of penalization depends only on $\sigma$ as illustrated in the Monte Carlo simulation of \reffig{fig:ppxf_simulation}. This effect is also illustrated for \atl\ data in \reffig{fig:ppxf_kinematics}, for a small set of galaxies with a range of luminosity-weighed $\sigma_e$ within 1Re \citep[defined as in][]{Cappellari2006}. The figure shows the range of data quality for the stellar kinematical data for the \atl\ survey. At high $\sigma_e$ the data have a quality which is comparable to the one for the \sauron\ survey presented in \citet{Emsellem2004}. The shorter exposure time of the \atl\ survey is in part compensated by the use of the VPH grating and in part by the adoption of larger Voronoi bins. However at $\sigma_e\la120$ \kms\ there is not enough information in the data to constrain $h_3$ and $h_4$. Contrary to the \sauron\ survey the volume-limited \atl\ survey is dominated by the more common low-luminosity galaxies, which also tend to have low dispersion. In practice we find that $\approx40$\% of the galaxies in the sample have $\sigma_e\la120$ \kms\ and for those objects we can only recover reliable $V$ and $\sigma$ values. Although the kinematics are not sufficient to constrain general dynamical models for the full sample, they still provide very reliable global values of specific angular momenta and galaxy masses for all galaxies \citep[e.g.][]{Cappellari2010iau}.

We also used the \sauron\ stellar kinematic to measure extremely robust heliocentric recession velocities $V_{\rm hel}$ for all galaxies in the sample. Repeated determinations indicate an 1$\sigma$ accuracy $\Delta V_{\rm hel}\approx5$ \kms.  The values were derived from the integral-field stellar kinematics by finding the value which needs to be subtracted from the observed velocity field to best fit a bi-anti-symmetric version of the velocity field\footnote{This was performed with the IDL routine FIT\_KINEMATICS\_PA described in Appendix C of \citet{krajnovic06} and available here http://purl.org/cappellari/idl.}. This technique does not suffer from the uncertainties due to the slit centering, which affects spectroscopic surveys, or from the possibility of the gas not being associated to the galaxy, which affects \hi\ determinations. The observed velocities were converted into velocities $V_{\rm hel}$ relative to the barycenter of the Solar System via the IDL routine \texttt{baryvel}, which implements the algorithm by \citet{Stumpff1980} and is part of the IDL Astronomy User's Library \citep{Landsman1993}. The measured $V_{\rm hel}$ values are given in Table~\ref{tab:atlas3d_sample}.

\section{Additional \atl\ data and simulations}

\subsection{\hi, CO and optical observations}
\label{sec:data}

\begin{table*}
\caption{Multiwavelength \atl\ data}
\centering
\begin{tabular}{cccccc}
Instrument & Tracer/Band/Res.  & Number of Objects & Selection & Reference & Time Allocation \\
\hline
WSRT & \ion{H}{i} @ 21 cm & 170 & $\delta>10^\circ$ & Serra et al. in preparation & 1212 h\\
IRAM 30m & $^{12}$CO J$=$1-0 and 2-1 & 259 & All & Paper IV & 211 h\\
CARMA & $^{12}$CO J$=$1-0 & 32 & $f>19$ Jy \kms\ & Altalo et al. in preparation & 467 h\\
INT $+$ SDSS & $u, g, r, i, z$ & 35 $+$ 225 & All & Scott et al. in prep. $+$ \citet{Abazajian2009} & 5 n\\ SAURON & 480--538 nm; $R=1300$ & 260 & All & This paper & 38 n\\
\hline
\label{tab:data_sets}
\end{tabular}
\end{table*}

Apart from the \sauron\ integral-field spectroscopic observations presented in detail in \refsec{sec:sauron}, the \atl\ survey includes other multi-wavelength observations obtained with different instruments and facilities. These datasets will be presented in subsequent papers, however a summary of the main ones is presented in Table~\ref{tab:data_sets}, and more information are given below:
\begin{enumerate}
\item {\bf \hi\ interferometry:} We have observed the \hi\ properties of all galaxies in the \atl\ sample above $\delta=10^\circ$ (due to the telescope latitude). This sub-sample includes 170 galaxies --- 43 inside and 127 outside the Virgo cluster. We observed all galaxies outside Virgo, and galaxies inside Virgo detected by the Alfalfa survey \citep{Giovanelli2005}, with the Westerbork Synthesis Radio Telescope (WSRT). Some of the galaxies were observed with the WSRT as part of previous studies \citep{Morganti2006,Jozsa2009,Oosterloo2010}. The integration time for all galaxies observed with the WSRT is 12 h, providing \hi\ cubes at a resolution of $\sim 30$ arcsec and 16 km/s over a field of view of $\sim 1$ deg$^2$ and a velocity range of $\sim 4000$ km/s. We detect \hi\ gas down to a column density of a few times 10$^{19}$ cm$^{-2}$. The upper limits on \mhi\ derived from these data ranges between 10$^6$ and a few times 10$^7$ \msun\ depending on galaxy distance. This is typically $\sim 5$ times lower than upper limits derived from Alfalfa spectra. These observations and a discussion of the \hi\ properties will be presented in Serra et al. (in preparation). Interesting objects, like extended disks, have been followed-up for deeper \hi\ observations.

\item {\bf CO single-dish:} All of the \atl\ galaxies have been searched for $^{12}$CO J$=$1-0 and 2-1 emission with the IRAM 30m telescope, including 204 new observations with the remainder collected from the recent literature.  The data consist of a single pointing at the galaxy center, covering a bandwidth of 1300 \kms\ centered on the optical velocity.   The rms noise levels of the 1-0 spectra are 3 mK ($T_A^*$)  or better after binning to 31 \kms\ channels, so that the 3$\sigma$ upper limit for a sum over a 300 \kms\ linewidth corresponds to a $H_2$ mass $\sim 1\times 10^7$ M$_\odot$ for the most nearby sample members and $\sim 1\times 10^8$ M$_\odot$ for the most distant members. A detailed description of the observations and the corresponding measurements are presented in Paper IV.

\item {\bf CO interferometry:} The brighter CO detections have been, observed in the 1-0 line with the BIMA, Plateau de Bure, and CARMA millimeter interferometers in order to map the distribution and kinematics of the molecular gas.  These observations are designed to provide the molecular surface densities and angular momenta for 80\% of all of the molecular gas found in the \atl\ sample, typically at resolutions of 5\arcsec.  Some additional data at higher and lower resolutions are obtained as necessary to probe the structure of the gas and recover the bulk of the emission detected in the single dish data.  On-source integration times range from 4 to 20 hours and are also adjusted as necessary for high quality detections. A detailed description of the observations and the corresponding measurements are presented in Alatalo et al. (in preparation).

\item {\bf INT optical imaging:} Observations with the Wide-Field Camera (WFC) at the 2.5m Isaac Newton Telescope (INT) were carried out to obtain $u$, $g$, $r$, $i$ and $z$-band imaging for galaxies not observed by the SDSS. Images were taken through the 5 filters for 55 galaxies from the \atl\ sample. Integration times were typically 60 to 160 seconds reaching sensitivities comparable or deeper than the SDSS. Galaxies already observed by SDSS were observed in the runs as a cross-check in general and to bring the INT imaging onto the same photometric system as SDSS in particular. The images have been reduced and calibrated using the Astro-WISE system \citep{Valentijn2007}. A detailed description of the observations and the corresponding measurements are presented in Scott et al. (in preparation).

\item {\bf Targeted follow-ups:} We are also obtaining data for targeted subsets of the sample: Deep optical images of \atl\ galaxies were obtained with the MegaCam camera installed on the Canada-France-Hawaii Telescope. This imaging part of the project aims to reach surface-brightness limits as low as 28.5 mag arcsec$^2$ in the g-band. Reaching such  values allows to disclose very faint, diffuse structures in the outskirts of the ETGs, such as shells and tidal tails,  that  tell about their past mass accretion history. We have also started an observing campaign to obtain stellar kinematics and absorption line strengths out to large radii ($3-5 \re$) with IFUs (e.g. \sauron, VIRUS-P) for a number of \atl\ galaxies. Following the methods outlined in \citet{Weijmans2009} we will construct dynamical models to trace the halo mass profiles. We primarily target galaxies that have been detected in \hi\ to have a regularly rotating disc and ring, so that the \hi\ kinematics can be added to the dynamical modeling.
\end{enumerate}

\subsection{Numerical simulations}

The \atl\ project includes a theoretical effort to interpret the observations using models of galaxy formation. We are attacking the problem via three parallel approaches as described below:
\begin{enumerate}

\item {\bf Binary mergers:} An extensive set of ``idealized'' (i.e. without the cosmological context) numerical simulations is being conducted. These simulations have been made at an unmatched resolution (softening length of 58~pc and a total number of particles of $1.2 \times 10^7$) with the goal to better understand the role of mergers in the formation and evolution of galaxies of the red sequence and to understand in details the physical processes involved during a merger (e.g. the formation of the Kinematically Decoupled Components, the energy and angular momentum exchanges). These simulations are also a powerful tool to perform direct comparisons with observations such as e.g., the morphology via the ellipticity, the kinematics via $\lambda_R$, the redistribution of the gas at large scales, the metallicity gradients and the  Mg {\em b}~--~V$_{\mbox{esc}}$ relation \citep{Davies1993,Scott2009}. The detailed description of the simulations and their associated results are presented in \citet{Bois2010} and Paper~VI.
    Other idealized simulations of \atl\ galaxies are performed with the RAMSES high-resolution grid-based hydrodynamical code \citep{Teyssier2002}. Following the technique developed in \citet{Bournaud2010}, we model the dynamics of atomic and molecular gas disks in early-type galaxies with parsec-scale resolution, based on accurate mass models extracted from the \atl\ data. These models aim at understanding the dynamics, stability and star formation activity of the ISM in ETGs.

\item {\bf Semi-analytic modeling:} In a second strand of simulations we address the formation of elliptical galaxies within a large scale cosmological setting using a semi-analytic modeling (SAM) approach. Using the knowledge gained from idealized high-resolution simulations of mergers and  the formation  of a limited number of cosmologically simulated ETGs we test formation scenarios within our SAM making full use of the completeness of the \atl\ sample. The SAM we use is an extension of earlier work by \citet{Khochfar2005} and \citet{Khochfar2006}. Within the SAM we follow the individual history of a large statistical sample of galaxies to the present day, taking into account  their merging history and physical processes related to e.g. gas cooling or star formation. In addition the SAM is used to make predictions on the evolution of these classes of ETGs, that can be tested with future observations (Khochfar et al in prep.).

\item {\bf Cosmological simulations:} We will also make use of high resolution simulations of individual galaxies in a full cosmological context \citep[i.e.][]{Naab2007,Naab2009} to investigate the physical processes setting the present day kinematical properties of ETGs. We will use a new sample of simulations \citep{Oser2010} covering the full mass range of the \atl\ galaxies. From these simulated galaxies we will construct two-dimensional kinematical maps \citep{jesseit07} to compare directly to the \atl\ observations.  The use of cosmological simulations is advantageous as we can link the present day properties to the evolutionary history embedded in the favored cosmology. We will also be able to investigate the influence of the merging history, dark matter and various feedback mechanisms on kinematic properties and the stellar populations.

\end{enumerate}

\section{Summary}

We described the motivation and goals of the \atl\ project, which aims at constraining models of galaxy formation by obtaining two-dimensional observations of the distribution and kinematics of the atomic (\hi), molecular (CO) and ionized gas, together with the stellar population and kinematics, for a volume-limited nearly mass-selected ($K_s$-band) sample of ETGs.

We defined the selection criteria for the volume-limited ($1.16\times10^5$ Mpc$^3$) {\em parent} sample of 871 galaxies with $D<42$ Mpc and $M_K<-21.5$ mag, satisfying our observability criteria, and investigated possible selection biases, especially due to redshift incompleteness. We found that incompleteness cannot amount to more than a couple of percent, making the sample essentially complete and representative of the nearby population. We additionally tested the representativeness of the sample by comparing its $K_s$-band luminosity function with the one derived from a much larger sample \citep{Bell2003} and found a very good agreement.
We described the morphological selection used to extract the 260 ETGs of the \atl\ sample from the parent sample and showed that the ETGs define a narrow red-sequence, on a colour-magnitude diagram, with few objects in transition from the blue cloud. We presented the size-luminosity relation for the \atl\ sample and the full parent sample to illustrate the general characteristic of our galaxies.

We described the strategy for the \sauron\ integral-field observations, the data reduction, the extraction of the stellar kinematics and their typical errors. We gave an overview of the additional dataset already available for our sample, which include interferometric observations of the atomic gas as traced by \hi, single-dish and interferometric observations of molecular gas as traced by the CO lines, and multi-band optical photometry. We summarized the ongoing semi-analytic modeling and the cosmological and binary-merger N-body simulations we are performing to interpret our observations.

This is the first paper of a series devoted to our understanding of the formation of ETGs. Key additional elements are provided by the kinematics, ages and chemical composition of the stars in the galaxies, the presence of cold atomic or molecular gas, the photometric profiles and the dynamical masses, as a function of environment. The observations for the \atl\ sample will be compared against the model predictions, to test formation scenarios and to tune the model parameter. This will be the topic of future papers of this series. The \atl\ project aims to constitute a zero redshift baseline, against which one can investigate the evolution of galaxy global parameters with redshift, to trace galaxy evolution back time. Future studies should extend this effort to more dense environment than can be explored in the nearby universe, and to increasingly higher redshifts to explore the time evolution of the structure of ETGs.

\section*{acknowledgements}

We thank the anonymous referee for a useful report.
MC acknowledges support from a STFC Advanced Fellowship PP/D005574/1 and a Royal Society University Research Fellowship.
This work was supported by the rolling grants `Astrophysics at Oxford' PP/E001114/1 and ST/H002456/1 and visitors grants PPA/V/S/2002/00553, PP/E001564/1 and ST/H504862/1 from the UK Research Councils. RLD acknowledges travel and computer grants from Christ Church, Oxford and support from the Royal Society in the form of a Wolfson Merit Award 502011.K502/jd. RLD also acknowledges the support of the ESO Visitor Programme which funded a 3 month stay in 2010.
SK acknowledges support from the the Royal Society Joint Projects Grant JP0869822.
RMcD is supported by the Gemini Observatory, which is operated by the Association of Universities for Research in Astronomy, Inc., on behalf of the international Gemini partnership of Argentina, Australia, Brazil, Canada, Chile, the United Kingdom, and the United States of America.
TN and MBois acknowledge support from the DFG Cluster of Excellence `Origin and Structure of the Universe'.
MS acknowledges support from a STFC Advanced Fellowship ST/F009186/1.
NS and TD acknowledge support from an STFC studentship.
The authors acknowledge financial support from ESO. We acknowledge the usage in pPXF of the MPFIT routine by \citet{Markwardt2009}.
The SAURON observations were obtained at the William Herschel Telescope, operated by the Isaac Newton Group in the Spanish Observatorio del Roque de los Muchachos of the Instituto de Astrofisica de Canarias. We are grateful to the ING staff for their excellent support and creativity in quickly solving technical problem during the \sauron\ runs. We are grateful to Jes\'us Falcon-Barroso for useful discussions and help with the observations. MC is grateful to the NED staff for prompt support.
This research has made use of the NASA/IPAC Extragalactic Database (NED) which is operated by the Jet Propulsion Laboratory, California Institute of Technology, under contract with the National Aeronautics and Space Administration. We acknowledge the usage of the HyperLeda database (http://leda.univ-lyon1.fr). Funding for the SDSS and SDSS-II was provided by the Alfred P. Sloan Foundation, the Participating Institutions, the National Science Foundation, the U.S. Department of Energy, the National Aeronautics and Space Administration, the Japanese Monbukagakusho, the Max Planck Society, and the Higher Education Funding Council for England. The SDSS was managed by the Astrophysical Research Consortium for the Participating Institutions. This publication makes use of data products from the Two Micron All Sky Survey, which is a joint project of the University of Massachusetts and the Infrared Processing and Analysis Center/California Institute of Technology, funded by the National Aeronautics and Space Administration and the National Science Foundation.


\begin{thebibliography}{}

\bibitem[\protect\citeauthoryear{{Abazajian}, {Adelman-McCarthy} \& {et
  al.}}{{Abazajian} et~al.}{2009}]{Abazajian2009}
{Abazajian} K.~N.,  {et al.} 2009, \apjs, 182, 543

\bibitem[\protect\citeauthoryear{{Allard}, {Peletier} \& {Knapen}}{{Allard}
  et~al.}{2005}]{Allard2005}
{Allard} E.~L.,  {Peletier} R.~F.,    {Knapen} J.~H.,  2005, \apjl, 633, L25

\bibitem[\protect\citeauthoryear{{Bacon}, {Copin} \& {et al.}}{{Bacon}
  et~al.}{2001}]{Bacon2001}
{Bacon} R.,  {et al.} 2001, \mnras, 326, 23

\bibitem[\protect\citeauthoryear{{Baldry}, {Balogh}, {Bower}, {Glazebrook},
  {Nichol}, {Bamford} \& {Budavari}}{{Baldry} et~al.}{2006}]{Baldry2006}
{Baldry} I.~K.,  {Balogh} M.~L.,  {Bower} R.~G.,  {Glazebrook} K.,  {Nichol}
  R.~C.,  {Bamford} S.~P.,    {Budavari} T.,  2006, \mnras, 373, 469

\bibitem[\protect\citeauthoryear{{Baldry}, {Glazebrook}, {Brinkmann},
  {Ivezi{\'c}}, {Lupton}, {Nichol} \& {Szalay}}{{Baldry}
  et~al.}{2004}]{Baldry2004}
{Baldry} I.~K.,  {Glazebrook} K.,  {Brinkmann} J.,  {Ivezi{\'c}} {\v Z}.,
  {Lupton} R.~H.,  {Nichol} R.~C.,    {Szalay} A.~S.,  2004, \apj, 600, 681

\bibitem[\protect\citeauthoryear{{Barnes}}{{Barnes}}{1992}]{Barnes1992}
{Barnes} J.~E.,  1992, \apj, 393, 484

\bibitem[\protect\citeauthoryear{{Bell} \& {de Jong}}{{Bell} et~al.}{2001}]{Bell2001}
{Bell} E.~F.,  {de Jong} R.~S.,  2001, \apj, 550, 212

\bibitem[\protect\citeauthoryear{{Bell}, {Wolf}, {Meisenheimer} \& {et
  al.}}{{Bell} et~al.}{2003}]{Bell2003}
{Bell} E.~F.,  {et al.} 2003, \apjs, 149, 289

\bibitem[\protect\citeauthoryear{{Bell}, {Wolf}, {Meisenheimer} \& {et
  al.}}{{Bell} et~al.}{2004}]{Bell2004}
{Bell} E.~F.,  {et al.} 2004, \apj, 608, 752

\bibitem[\protect\citeauthoryear{{Bender}, {Burstein} \& {Faber}}{{Bender}
  et~al.}{1992}]{bender92}
{Bender} R.,  {Burstein} D.,    {Faber} S.~M.,  1992, \apj, 399, 462

\bibitem[\protect\citeauthoryear{{Bender}, {Surma}, {Doebereiner},
  {Moellenhoff} \& {Madejsky}}{{Bender} et~al.}{1989}]{Bender1989}
{Bender} R.,  {Surma} P.,  {Doebereiner} S.,  {Moellenhoff} C.,    {Madejsky}
  R.,  1989, \aap, 217, 35

\bibitem[\protect\citeauthoryear{{Bernardi}, {Shankar}, {Hyde}, {Mei},
  {Marulli} \& {Sheth}}{{Bernardi} et~al.}{2010}]{Bernardi2010}
{Bernardi} M.,  {Shankar} F.,  {Hyde} J.~B.,  {Mei} S.,  {Marulli} F.,
  {Sheth} R.~K.,  2010, \mnras, 404, 2087

\bibitem[\protect\citeauthoryear{{Binney}}{{Binney}}{1978}]{Binney1978}
{Binney} J.,  1978, \mnras, 183, 501

\bibitem[\protect\citeauthoryear{{Blakeslee}, {Cantiello} \& {et
  al.}}{{Blakeslee} et~al.}{2010}]{Blakeslee2010}
{Blakeslee} J.~P.,  {et al.} 2010, \apj, 724, 657

\bibitem[\protect\citeauthoryear{{Blakeslee}, {Jord{\'a}n} \& {et
  al.}}{{Blakeslee} et~al.}{2009}]{Blakeslee2009}
{Blakeslee} J.~P., {et al.} 2009, \apj, 694, 556

\bibitem[\protect\citeauthoryear{{Bois}, {Bournaud} \& {et al.}}{{Bois}
  et~al.}{2010}]{Bois2010}
{Bois} M.,  {et al.} 2010, \mnras, 406, 2405

\bibitem[\protect\citeauthoryear{{Bois}, {Emsellem} \& {et al.}}{{Bois}
  et~al.}{2011}]{Bois2011}
{Bois} M.,  {et al.} 2011, \mnras, submitted (Paper VI)

\bibitem[\protect\citeauthoryear{Bournaud et
al.}{2010}]{Bournaud2010} Bournaud F., Elmegreen B.~G., Teyssier R.,
Block D.~L., Puerari I., 2010, MNRAS, 409, 1088

\bibitem[\protect\citeauthoryear{{Bournaud}, {Jog} \& {Combes}}{{Bournaud}
  et~al.}{2007}]{Bournaud2007}
{Bournaud} F.,  {Jog} C.~J.,    {Combes} F.,  2007, \aap, 476, 1179

\bibitem[\protect\citeauthoryear{{Bower}, {Benson}, {Malbon}, {Helly}, {Frenk},
  {Baugh}, {Cole} \& {Lacey}}{{Bower} et~al.}{2006}]{Bower2006}
{Bower} R.~G.,  {Benson} A.~J.,  {Malbon} R.,  {Helly} J.~C.,  {Frenk} C.~S.,
  {Baugh} C.~M.,  {Cole} S.,    {Lacey} C.~G.,  2006, \mnras, 370, 645

\bibitem[\protect\citeauthoryear{{Bower}, {Morris} \& {et al.}}{{Bower}
  et~al.}{2004}]{Bower2004}
{Bower} R.~G.,  {et al.} 2004, \mnras, 351, 63

\bibitem[\protect\citeauthoryear{{Cappellari}, {Bacon} \& {et
  al.}}{{Cappellari} et~al.}{2006}]{Cappellari2006}
{Cappellari} M.,  {et al.} 2006, \mnras, 366, 1126

\bibitem[\protect\citeauthoryear{{Cappellari} \& {Copin}}{{Cappellari} \&
  {Copin}}{2003}]{Cappellari2003}
{Cappellari} M.,  {Copin} Y.,  2003, \mnras, 342, 345

\bibitem[\protect\citeauthoryear{{Cappellari} \& {Emsellem}}{{Cappellari} \&
  {Emsellem}}{2004}]{Cappellari2004}
{Cappellari} M.,  {Emsellem} E.,  2004, \pasp, 116, 138

\bibitem[\protect\citeauthoryear{{Cappellari}, {Emsellem} \& {et
  al.}}{{Cappellari} et~al.}{2007}]{Cappellari2007}
{Cappellari} M.,  {et al.} 2007, \mnras, 379, 418

\bibitem[\protect\citeauthoryear{{Cappellari}, {Emsellem} \& {et
  al.}}{{Cappellari} et~al.}{2011}]{Cappellari2011}
{Cappellari} M.,  {et al.} 2011, \mnras, submitted (Paper
  VII)

\bibitem[\protect\citeauthoryear{{Cappellari}, {Scott} \& {et
  al.}}{{Cappellari} et~al.}{2010}]{Cappellari2010iau}
{Cappellari} M.,  {et al.} 2010, Highlights of Astronomy, 15, 81

\bibitem[\protect\citeauthoryear{{Cattaneo}, {Dekel}, {Devriendt}, {Guiderdoni}
  \& {Blaizot}}{{Cattaneo} et~al.}{2006}]{Cattaneo2006}
{Cattaneo} A.,  {Dekel} A.,  {Devriendt} J.,  {Guiderdoni} B.,    {Blaizot} J.,
   2006, \mnras, 370, 1651

\bibitem[\protect\citeauthoryear{{Chen}, {C{\^o}t{\'e}}, {West}, {Peng} \&
  {Ferrarese}}{{Chen} et~al.}{2010}]{Chen2010}
{Chen} C.,  {C{\^o}t{\'e}} P.,  {West} A.~A.,  {Peng} E.~W.,    {Ferrarese} L.,
   2010, \apjs, 191, 1

\bibitem[\protect\citeauthoryear{Cimatti et
al.}{2008}]{Cimatti2008} Cimatti A., et al., 2008, A\&A, 482, 21

\bibitem[\protect\citeauthoryear{{Cole}, {Norberg} \& {et al.}}{{Cole}
  et~al.}{2001}]{Cole2001}
{Cole} S.,  {et al.} 2001, \mnras, 326, 255

\bibitem[\protect\citeauthoryear{{Conselice}}{{Conselice}}{2006}]{Conselice200%
6}
{Conselice} C.~J.,  2006, \mnras, 373, 1389

\bibitem[\protect\citeauthoryear{{Cox}, {Dutta}, {Di Matteo}, {Hernquist},
  {Hopkins}, {Robertson} \& {Springel}}{{Cox} et~al.}{2006}]{Cox2006}
{Cox} T.~J.,  {Dutta} S.~N.,  {Di Matteo} T.,  {Hernquist} L.,  {Hopkins}
  P.~F.,  {Robertson} B.,    {Springel} V.,  2006, \apj, 650, 791

\bibitem[\protect\citeauthoryear{{Crook} \& {et al.}}{{Crook}, {et
  al.}}{2007}]{Crook2007}
{Crook} J.~R.,  {et al.} 2007, \apj, 655, 790

\bibitem[\protect\citeauthoryear{{Croton}, {Springel} \& {et al.}}{{Croton}
  et~al.}{2006}]{Croton2006}
{Croton} D.~J., {et al.} 2006, \mnras, 365, 11

\bibitem[\protect\citeauthoryear{{Daddi}, {Renzini} \& {et al.}}{{Daddi}
  et~al.}{2005}]{Daddi2005}
{Daddi} E., {et al.} 2005, \apj, 626, 680

\bibitem[\protect\citeauthoryear{{Davies}, {Efstathiou}, {Fall}, {Illingworth}
  \& {Schechter}}{{Davies} et~al.}{1983}]{Davies1983}
{Davies} R.~L.,  {Efstathiou} G.,  {Fall} S.~M.,  {Illingworth} G.,
  {Schechter} P.~L.,  1983, \apj, 266, 41

\bibitem[\protect\citeauthoryear{{Davies}, {Sadler} \& {Peletier}}{{Davies}
  et~al.}{1993}]{Davies1993}
{Davies} R.~L.,  {Sadler} E.~M.,    {Peletier} R.~F.,  1993, \mnras, 262, 650

\bibitem[\protect\citeauthoryear{{Davis}, {Bureau} \& {et al.}}{{Davis}
  et~al.}{2011}]{Davis2010}
{Davis} T.~A.,  {et al.} 2011, \mnras, submitted (Paper V)

\bibitem[\protect\citeauthoryear{{de Vaucouleurs}}{{de
  Vaucouleurs}}{1959}]{deVaucouleurs1959}
{de Vaucouleurs} G.,  1959, Handbuch der Physik, 53, 311

\bibitem[\protect\citeauthoryear{{de Vaucouleurs}}{{de
  Vaucouleurs}}{1963}]{deVaucouleurs1963}
{de Vaucouleurs} G.,  1963, \apjs, 8, 31

\bibitem[\protect\citeauthoryear{{de Vaucouleurs}, {de Vaucouleurs}, {Corwin}
  Jr., {Buta}, {Paturel} \& {Fouque}}{{de Vaucouleurs}
  et~al.}{1991}]{deVaucouleurs1991}
{de Vaucouleurs} G.,  {de Vaucouleurs} A.,  {Corwin} Jr. H.~G.,  {Buta} R.~J.,
  {Paturel} G.,    {Fouque} P.,  1991, {Third Reference Catalogue of Bright
  Galaxies}.
Volume 1-3, XII, 2069 pp.~7 figs..~ Springer-Verlag Berlin Heidelberg New York

\bibitem[\protect\citeauthoryear{{de Vaucouleurs}, {de Vaucouleurs} \&
  {Corwin}}{{de Vaucouleurs} et~al.}{1976}]{deVaucouleurs1976}
{de Vaucouleurs} G.,  {de Vaucouleurs} A.,    {Corwin} J.~R.,  1976, in Second
  reference catalogue of bright galaxies, 1976, Austin: University of Texas
  Press. {Second reference catalogue of bright galaxies}.
pp~0--+

\bibitem[\protect\citeauthoryear{{de Zeeuw}, {Bureau} \& {et al.}}{{de Zeeuw}
  et~al.}{2002}]{deZeeuw2002}
{de Zeeuw} P.~T., {et al.} 2002, \mnras, 329, 513

\bibitem[\protect\citeauthoryear{{Dekel} \& {Birnboim}}{{Dekel} \&
  {Birnboim}}{2006}]{Dekel2006}
{Dekel} A.,  {Birnboim} Y.,  2006, \mnras, 368, 2

\bibitem[\protect\citeauthoryear{{Di Matteo}, {Springel} \& {Hernquist}}{{Di
  Matteo} et~al.}{2005}]{DiMatteo2005}
{Di Matteo} T.,  {Springel} V.,    {Hernquist} L.,  2005, \nat, 433, 604

\bibitem[\protect\citeauthoryear{{Djorgovski} \& {Davis}}{{Djorgovski} \&
  {Davis}}{1987}]{Djorgovski1987}
{Djorgovski} S.,  {Davis} M.,  1987, \apj, 313, 59

\bibitem[\protect\citeauthoryear{{Dressler}, {Lynden-Bell}, {Burstein},
  {Davies}, {Faber}, {Terlevich} \& {Wegner}}{{Dressler}
  et~al.}{1987}]{Dressler1987}
{Dressler} A.,  {Lynden-Bell} D.,  {Burstein} D.,  {Davies} R.~L.,  {Faber}
  S.~M.,  {Terlevich} R.,    {Wegner} G.,  1987, \apj, 313, 42

\bibitem[\protect\citeauthoryear{{Dumas}, {Mundell}, {Emsellem} \&
  {Nagar}}{{Dumas} et~al.}{2007}]{Dumas2007}
{Dumas} G.,  {Mundell} C.~G.,  {Emsellem} E.,    {Nagar} N.~M.,  2007, \mnras,
  379, 1249

\bibitem[\protect\citeauthoryear{{Dunkley}, {Komatsu} \& {et al.}}{{Dunkley}
  et~al.}{2009}]{Dunkley2009}
{Dunkley} J.,  {et al.} 2009, \apjs, 180, 306

\bibitem[\protect\citeauthoryear{Efron \& Tibshirani}{Efron \&
  Tibshirani}{1993}]{Efron1993}
Efron B.,  Tibshirani R.,  1993

\bibitem[\protect\citeauthoryear{{Emsellem}, {Cappellari} \& {et
  al.}}{{Emsellem} et~al.}{2004}]{Emsellem2004}
{Emsellem} E.,  {et al.} 2004, \mnras, 352, 721

\bibitem[\protect\citeauthoryear{{Emsellem}, {Cappellari} \& {et
  al.}}{{Emsellem} et~al.}{2007}]{Emsellem2007}
{Emsellem} E.,  {et al.} 2007, \mnras, 379, 401

\bibitem[\protect\citeauthoryear{{Emsellem}, {Cappellari} \& {et
  al.}}{{Emsellem} et~al.}{2011}]{Emsellem2010}
{Emsellem} E.,  {et al.} 2011, \mnras, submitted (Paper
  III)

\bibitem[\protect\citeauthoryear{{Faber} \& {et al.}}{{Faber} et~al.}{1997}]{faber97}
{Faber} S.~M.,  {et al.} 1997, \aj, 114, 1771

\bibitem[\protect\citeauthoryear{{Faber} \& {Jackson}}{{Faber} \&
  {Jackson}}{1976}]{faber76}
{Faber} S.~M.,  {Jackson} R.~E.,  1976, \apj, 204, 668

\bibitem[\protect\citeauthoryear{{Faber}, {Willmer} \& {et al.}}{{Faber}
  et~al.}{2007}]{Faber2007}
{Faber} S.~M.,  {et al.} 2007, \apj, 665, 265

\bibitem[\protect\citeauthoryear{{Falc{\'o}n-Barroso}, {Bacon} \& {et
  al.}}{{Falc{\'o}n-Barroso} et~al.}{2006}]{FalconBarroso06}
{Falc{\'o}n-Barroso} J.,  {et al.} 2006, \mnras, 369, 529

\bibitem[\protect\citeauthoryear{{Ferrarese}, {C{\^o}t{\'e}}, {Dalla
  Bont{\`a}}, {Peng}, {Merritt}, {Jord{\'a}n}, {Blakeslee}, {Ha{\c s}egan},
  {Mei}, {Piatek}, {Tonry} \& {West}}{{Ferrarese} et~al.}{2006}]{Ferrarese2006}
{Ferrarese} L.,  {et al.}  2006, \apjl, 644,
  L21

\bibitem[\protect\citeauthoryear{{Ferrarese}, {van den Bosch}, {Ford}, {Jaffe}
  \& {O'Connell}}{{Ferrarese} et~al.}{1994}]{Ferrarese1994}
{Ferrarese} L.,  {van den Bosch} F.~C.,  {Ford} H.~C.,  {Jaffe} W.,
  {O'Connell} R.~W.,  1994, \aj, 108, 1598

\bibitem[\protect\citeauthoryear{{Gerhard}}{{Gerhard}}{1993}]{gerhard93}
{Gerhard} O.~E.,  1993, \mnras, 265, 213

\bibitem[\protect\citeauthoryear{{Giovanelli}, {Haynes}, {Kent} \& {et
  al.}}{{Giovanelli} et~al.}{2005}]{Giovanelli2005}
{Giovanelli} R.,  {et al.} 2005, \aj, 130,
  2598

\bibitem[\protect\citeauthoryear{{Graham}}{{Graham}}{2004}]{Graham2004}
{Graham} A.~W.,  2004, \apjl, 613, L33

\bibitem[\protect\citeauthoryear{{Granato}, {De Zotti}, {Silva}, {Bressan} \&
  {Danese}}{{Granato} et~al.}{2004}]{Granato2004}
{Granato} G.~L.,  {De Zotti} G.,  {Silva} L.,  {Bressan} A.,    {Danese} L.,
  2004, \apj, 600, 580

\bibitem[\protect\citeauthoryear{{Hernquist}}{{Hernquist}}{1992}]{Hernquist199%
2}
{Hernquist} L.,  1992, \apj, 400, 460

\bibitem[\protect\citeauthoryear{{Hopkins}, {Hernquist}, {Cox}, {Keres} \&
  {Wuyts}}{{Hopkins} et~al.}{2009}]{Hopkins2009}
{Hopkins} P.~F.,  {Hernquist} L.,  {Cox} T.~J.,  {Keres} D.,    {Wuyts} S.,
  2009, \apj, 691, 1424

\bibitem[\protect\citeauthoryear{{Hubble}}{{Hubble}}{1936}]{Hubble1936}
{Hubble} E.~P.,  1936, {Realm of the Nebulae}.
Yale Univ. Press, New Haven

\bibitem[\protect\citeauthoryear{{Huchra}, {Davis}, {Latham} \&
  {Tonry}}{{Huchra} et~al.}{1983}]{Huchra1983}
{Huchra} J.,  {Davis} M.,  {Latham} D.,    {Tonry} J.,  1983, \apjs, 52, 89

\bibitem[\protect\citeauthoryear{{Huchra} \& {et al.}}{{Huchra} {et
  al.}}{2005}]{Huchra2005}
{Huchra} J.~P.,  {et al.} 2005, in {Colless} M.,  {Staveley-Smith} L.,
  {Stathakis} R.,  eds, Maps of the Cosmos Vol.~216 of IAU Symposium, {2MASS
  redshift survey}.
p.~170

\bibitem[\protect\citeauthoryear{{Huchra}, {Geller}, {Clemens}, {Tokarz} \&
  {Michel}}{{Huchra} et~al.}{1992}]{Huchra1992}
{Huchra} J.~P.,  {Geller} M.~J.,  {Clemens} C.~M.,  {Tokarz} S.~P.,    {Michel}
  A.,  1992, Bulletin d'Information du Centre de Donnees Stellaires, 41, 31

\bibitem[\protect\citeauthoryear{{Illingworth}}{{Illingworth}}{1977}]{Illingwo%
rth1977}
{Illingworth} G.,  1977, \apjl, 218, L43

\bibitem[\protect\citeauthoryear{{Jarrett}, {Chester}, {Cutri}, {Schneider},
  {Skrutskie} \& {Huchra}}{{Jarrett} et~al.}{2000}]{Jarrett2000}
{Jarrett} T.~H.,  {Chester} T.,  {Cutri} R.,  {Schneider} S.,  {Skrutskie} M.,
    {Huchra} J.~P.,  2000, \aj, 119, 2498

\bibitem[\protect\citeauthoryear{{Jeans}}{{Jeans}}{1922}]{Jeans1922}
{Jeans} J.~H.,  1922, \mnras, 82, 122

\bibitem[\protect\citeauthoryear{{Jeong}, {Yi} \& {et al.}}{{Jeong}
  et~al.}{2009}]{Jeong2009}
{Jeong} H.,  {Yi} S.~K.,    {et al.} 2009, \mnras, 398, 2028

\bibitem[\protect\citeauthoryear{{Jesseit}, {Cappellari}, {Naab}, {Emsellem} \&
  {Burkert}}{{Jesseit} et~al.}{2009}]{Jesseit2009}
{Jesseit} R.,  {Cappellari} M.,  {Naab} T.,  {Emsellem} E.,    {Burkert} A.,
  2009, \mnras, 397, 1202

\bibitem[\protect\citeauthoryear{{Jesseit}, {Naab}, {Peletier} \&
  {Burkert}}{{Jesseit} et~al.}{2007}]{jesseit07}
{Jesseit} R.,  {Naab} T.,  {Peletier} R.~F.,    {Burkert} A.,  2007, \mnras,
  376, 997

\bibitem[\protect\citeauthoryear{{Johansson}, {Naab} \& {Ostriker}}{{Johansson}
  et~al.}{2009}]{Johansson2009}
{Johansson} P.~H.,  {Naab} T.,    {Ostriker} J.~P.,  2009, \apjl, 697, L38

\bibitem[\protect\citeauthoryear{{J{\'o}zsa}, {Oosterloo}, {Morganti}, {Klein}
  \& {Erben}}{{J{\'o}zsa} et~al.}{2009}]{Jozsa2009}
{J{\'o}zsa} G.~I.~G.,  {Oosterloo} T.~A.,  {Morganti} R.,  {Klein} U.,
  {Erben} T.,  2009, \aap, 494, 489

\bibitem[\protect\citeauthoryear{{Karachentsev} \& {Nasonova}}{{Karachentsev}
  \& {Nasonova}}{2010}]{Karachentsev2010}
{Karachentsev} I.~D.,  {Nasonova} O.~G.,  2010, \mnras, 405, 1075

\bibitem[\protect\citeauthoryear{{Kere{\v s}}, {Katz}, {Weinberg} \&
  {Dav{\'e}}}{{Kere{\v s}} et~al.}{2005}]{Keres2005}
{Kere{\v s}} D.,  {Katz} N.,  {Weinberg} D.~H.,    {Dav{\'e}} R.,  2005,
  \mnras, 363, 2

\bibitem[\protect\citeauthoryear{{Khochfar} \& {Burkert}}{{Khochfar} \&
  {Burkert}}{2003}]{Khochfar2003}
{Khochfar} S.,  {Burkert} A.,  2003, \apjl, 597, L117

\bibitem[\protect\citeauthoryear{{Khochfar} \& {Burkert}}{{Khochfar} \&
  {Burkert}}{2005}]{Khochfar2005}
{Khochfar} S.,  {Burkert} A.,  2005, \mnras, 359, 1379

\bibitem[\protect\citeauthoryear{{Khochfar} \& {Ostriker}}{{Khochfar} \&
  {Ostriker}}{2008}]{Khochfar2008}
{Khochfar} S.,  {Ostriker} J.~P.,  2008, \apj, 680, 54

\bibitem[\protect\citeauthoryear{{Khochfar} \& {Silk}}{{Khochfar} \&
  {Silk}}{2006}]{Khochfar2006}
{Khochfar} S.,  {Silk} J.,  2006, \apjl, 648, L21

\bibitem[\protect\citeauthoryear{{Khochfar} \& {Silk}}{{Khochfar} \&
  {Silk}}{2009}]{Khochfar2009}
{Khochfar} S.,  {Silk} J.,  2009, \mnras, 397, 506

\bibitem[\protect\citeauthoryear{{Kochanek}, {Pahre}, {Falco} \& {et
  al.}}{{Kochanek} et~al.}{2001}]{Kochanek2001}
{Kochanek} C.~S.,  {et al.} 2001, \apj, 560,
  566

\bibitem[\protect\citeauthoryear{{Kormendy} \& {Bender}}{{Kormendy} \&
  {Bender}}{1996}]{Kormendy1996}
{Kormendy} J.,  {Bender} R.,  1996, \apjl, 464, L119+

\bibitem[\protect\citeauthoryear{{Kormendy}, {Fisher}, {Cornell} \&
  {Bender}}{{Kormendy} et~al.}{2009}]{Kormendy2009}
{Kormendy} J.,  {Fisher} D.~B.,  {Cornell} M.~E.,    {Bender} R.,  2009, \apjs,
  182, 216

\bibitem[\protect\citeauthoryear{{Kormendy} \& {Richstone}}{{Kormendy} \&
  {Richstone}}{1995}]{kormendy95}
{Kormendy} J.,  {Richstone} D.,  1995, \araa, 33, 581

\bibitem[\protect\citeauthoryear{{Krajnovi{\'c}}, {Bacon}, {Cappellari} \& {et
  al.}}{{Krajnovi{\'c}} et~al.}{2008}]{Krajnovic08}
{Krajnovi{\'c}} D., {et al.} 2008, \mnras,
  390, 93

\bibitem[\protect\citeauthoryear{{Krajnovi{\'c}}, {Cappellari}, {de Zeeuw} \&
  {Copin}}{{Krajnovi{\'c}} et~al.}{2006}]{krajnovic06}
{Krajnovi{\'c}} D.,  {Cappellari} M.,  {de Zeeuw} P.~T.,    {Copin} Y.,  2006,
  \mnras, 366, 787

\bibitem[\protect\citeauthoryear{{Krajnovi\'c}, {Emsellem} \& {et
  al.}}{{Krajnovi\'c} et~al.}{2011}]{Krajnovic2010}
{Krajnovi\'c} D.,  {et al.} 2011, \mnras, submitted (Paper
  II)

\bibitem[\protect\citeauthoryear{{Kuntschner}, {Emsellem}, {Bacon} \& {et
  al.}}{{Kuntschner} et~al.}{2006}]{Kuntschner2006}
{Kuntschner} H.,  {et al.} 2006, \mnras, 369,
  497

\bibitem[\protect\citeauthoryear{{Kuntschner}, {Emsellem}, {Bacon} \& {et
  al.}}{{Kuntschner} et~al.}{2010}]{Kuntschner2010}
{Kuntschner} H.,  {et al.} 2010, ArXiv e-prints

\bibitem[\protect\citeauthoryear{{Landsman}}{{Landsman}}{1993}]{Landsman1993}
{Landsman} W.~B.,  1993, in {R.~J.~Hanisch, R.~J.~V.~Brissenden, \& J.~Barnes}
  ed., Astronomical Data Analysis Software and Systems II Vol.~52 of
  Astronomical Society of the Pacific Conference Series, {The IDL Astronomy
  User's Library}.
pp 246--+

\bibitem[\protect\citeauthoryear{{Lauer}, {Ajhar}, {Byun} \& {et al.}}{{Lauer}
  et~al.}{1995}]{Lauer1995}
{Lauer} T.~R., {et al.} 1995, \aj, 110, 2622

\bibitem[\protect\citeauthoryear{{Lupton}, {Blanton}, {Fekete}, {Hogg},
  {O'Mullane}, {Szalay} \& {Wherry}}{{Lupton} et~al.}{2004}]{Lupton2004}
{Lupton} R.,  {Blanton} M.~R.,  {Fekete} G.,  {Hogg} D.~W.,  {O'Mullane} W.,
  {Szalay} A.,    {Wherry} N.,  2004, \pasp, 116, 133

\bibitem[\protect\citeauthoryear{{Maraston}}{{Maraston}}{2005}]{Maraston2005}
{Maraston} C.,  2005, \mnras, 362, 799

\bibitem[\protect\citeauthoryear{{Markwardt}}{{Markwardt}}{2009}]{Markwardt200%
9}
{Markwardt} C.~B.,  2009, in {D.~A.~Bohlender, D.~Durand, \& P.~Dowler} ed.,
  Astronomical Society of the Pacific Conference Series Vol.~411 of
  Astronomical Society of the Pacific Conference Series, {Non-linear
  Least-squares Fitting in IDL with MPFIT}.
pp 251--+

\bibitem[\protect\citeauthoryear{{McDermid}, {Emsellem}, {Shapiro} \& {et
  al.}}{{McDermid} et~al.}{2006}]{mcdermid06}
{McDermid} R.~M.,  {et al.} 2006, \mnras,
  373, 906

\bibitem[\protect\citeauthoryear{{Mei}, {Blakeslee}, {Cote} \& {et al.}}{{Mei}
  et~al.}{2007}]{Mei2007}
{Mei} S., {et al.} 2007, \apj, 655, 144

\bibitem[\protect\citeauthoryear{{Morganti}, {de Zeeuw}, {Oosterloo},
  {McDermid}, {Krajnovi{\'c}}, {Cappellari}, {Kenn}, {Weijmans} \&
  {Sarzi}}{{Morganti} et~al.}{2006}]{Morganti2006}
{Morganti} R.,  {et al.}  2006, \mnras, 371, 157

\bibitem[\protect\citeauthoryear{{Mould}, {Huchra} \& {et al.}}{{Mould}
  et~al.}{2000}]{Mould2000}
{Mould} J.~R.,  {et al.} 2000, \apj, 529, 786

\bibitem[\protect\citeauthoryear{{Naab} \& {Burkert}}{{Naab} \&
  {Burkert}}{2003}]{Naab2003}
{Naab} T.,  {Burkert} A.,  2003, \apj, 597, 893

\bibitem[\protect\citeauthoryear{{Naab}, {Burkert} \& {Hernquist}}{{Naab}
  et~al.}{1999}]{Naab1999}
{Naab} T.,  {Burkert} A.,    {Hernquist} L.,  1999, \apjl, 523, L133

\bibitem[\protect\citeauthoryear{{Naab}, {Jesseit} \& {Burkert}}{{Naab}
  et~al.}{2006}]{Naab2006}
{Naab} T.,  {Jesseit} R.,    {Burkert} A.,  2006, \mnras, 372, 839

\bibitem[\protect\citeauthoryear{{Naab}, {Johansson} \& {Ostriker}}{{Naab}
  et~al.}{2009}]{Naab2009}
{Naab} T.,  {Johansson} P.~H.,    {Ostriker} J.~P.,  2009, \apjl, 699, L178

\bibitem[\protect\citeauthoryear{{Naab}, {Johansson}, {Ostriker} \&
  {Efstathiou}}{{Naab} et~al.}{2007}]{Naab2007}
{Naab} T.,  {Johansson} P.~H.,  {Ostriker} J.~P.,    {Efstathiou} G.,  2007,
  \apj, 658, 710

\bibitem[\protect\citeauthoryear{{Naab}, {Khochfar} \& {Burkert}}{{Naab}
  et~al.}{2006}]{Naab2006kb}
{Naab} T.,  {Khochfar} S.,    {Burkert} A.,  2006, \apjl, 636, L81

\bibitem[\protect\citeauthoryear{{Oosterloo}, {Morganti} \& {et
  al.}}{{Oosterloo} et~al.}{2010}]{Oosterloo2010}
{Oosterloo} T.,  {et al.} 2010, \mnras, 409, 500

\bibitem[\protect\citeauthoryear{{Oser}, {Ostriker}, {Naab}, {Johansson} \&
  {Burkert}}{{Oser} et~al.}{2010}]{Oser2010}
{Oser} L.,  {Ostriker} J.~P.,  {Naab} T.,  {Johansson} P.~H.,    {Burkert} A.,
  2010, MNRAS, submitted (arXiv:1010.1381)

\bibitem[\protect\citeauthoryear{{Paturel}, {Petit}, {Prugniel}, {Theureau},
  {Rousseau}, {Brouty}, {Dubois} \& {Cambr{\'e}sy}}{{Paturel}
  et~al.}{2003}]{Paturel2003}
{Paturel} G.,  {Petit} C.,  {Prugniel} P.,  {Theureau} G.,  {Rousseau} J.,
  {Brouty} M.,  {Dubois} P.,    {Cambr{\'e}sy} L.,  2003, \aap, 412, 45

\bibitem[\protect\citeauthoryear{{Perlmutter}, {Aldering}, {Goldhaber} \& {et
  al.}}{{Perlmutter} et~al.}{1999}]{Perlmutter1999}
{Perlmutter} S.,   {et al.} 1999, \apj, 517,
  565

\bibitem[\protect\citeauthoryear{{Press}, {Teukolsky}, {Vetterling} \&
  {Flannery}}{{Press} et~al.}{1992}]{press92}
{Press} W.~H.,  {Teukolsky} S.~A.,  {Vetterling} W.~T.,    {Flannery} B.~P.,
  1992, {Numerical recipes in FORTRAN. The art of scientific computing}.
Cambridge: University Press, |c1992, 2nd ed.

\bibitem[\protect\citeauthoryear{{Riess}, {Filippenko}, {Challis} \& {et
  al.}}{{Riess} et~al.}{1998}]{Riess1998}
{Riess} A.~G., {et al.} 1998, \aj, 116,
  1009

\bibitem[\protect\citeauthoryear{{Robertson}, {Bullock}, {Cox}, {Di Matteo},
  {Hernquist}, {Springel} \& {Yoshida}}{{Robertson}
  et~al.}{2006}]{Robertson2006}
{Robertson} B.,  {Bullock} J.~S.,  {Cox} T.~J.,  {Di Matteo} T.,  {Hernquist}
  L.,  {Springel} V.,    {Yoshida} N.,  2006, \apj, 645, 986

\bibitem[\protect\citeauthoryear{{S{\'a}nchez-Bl{\'a}zquez}, {Peletier},
  {Jim{\'e}nez-Vicente}, {Cardiel}, {Cenarro}, {Falc{\'o}n-Barroso}, {Gorgas},
  {Selam} \& {Vazdekis}}{{S{\'a}nchez-Bl{\'a}zquez}
  et~al.}{2006}]{SanchezBlazquez2006}
{S{\'a}nchez-Bl{\'a}zquez} P.,  {et al.}  2006, \mnras, 371, 703

\bibitem[\protect\citeauthoryear{{Sandage}}{{Sandage}}{1961}]{Sandage1961}
{Sandage} A.,  1961, The Hubble Atlas.
Carnegie Institution, Washington

\bibitem[\protect\citeauthoryear{{Sandage}}{{Sandage}}{1975}]{Sandage1975}
{Sandage} A.,  1975, {Classification and Stellar Content of Galaxies Obtained
  from Direct Photography}.
Galaxies and the Universe, p.~170

\bibitem[\protect\citeauthoryear{{Sarzi}, {Falc{\'o}n-Barroso}, {Davies} \& {et
  al.}}{{Sarzi} et~al.}{2006}]{Sarzi2006}
{Sarzi} M.,  {et al.} 2006,
  \mnras, 366, 1151

\bibitem[\protect\citeauthoryear{{Sarzi}, {Shields}, {Schawinski} \& {et
  al.}}{{Sarzi} et~al.}{2010}]{Sarzi2010}
{Sarzi} M., {et al.} 2010, \mnras, 402,
  2187

\bibitem[\protect\citeauthoryear{{Schechter}}{{Schechter}}{1976}]{Schechter197%
6}
{Schechter} P.,  1976, \apj, 203, 297

\bibitem[\protect\citeauthoryear{{Schlegel}, {Finkbeiner} \&
  {Davis}}{{Schlegel} et~al.}{1998}]{schlegel98}
{Schlegel} D.~J.,  {Finkbeiner} D.~P.,    {Davis} M.,  1998, \apj, 500, 525

\bibitem[\protect\citeauthoryear{{Scott}, {Cappellari}, {Davies} \& {et
  al.}}{{Scott} et~al.}{2009}]{Scott2009}
{Scott} N.,   {et al.} 2009, \mnras, 398,
  1835

\bibitem[\protect\citeauthoryear{{Sersic}}{{Sersic}}{1968}]{sersic68}
{Sersic} J.~L.,  1968, {Atlas de galaxias australes}.
Cordoba, Argentina: Observatorio Astronomico, 1968

\bibitem[\protect\citeauthoryear{{Shankar} \& {Bernardi}}{{Shankar} \&
  {Bernardi}}{2009}]{Shankar2009}
{Shankar} F.,  {Bernardi} M.,  2009, \mnras, 396, L76

\bibitem[\protect\citeauthoryear{{Shapiro}, {Falc{\'o}n-Barroso} \& {et
  al.}}{{Shapiro} et~al.}{2010}]{Shapiro2010}
{Shapiro} K.~L.,   {et al.} 2010, \mnras, 402, 2140

\bibitem[\protect\citeauthoryear{{Shen}, {Mo}, {White}, {Blanton}, {Kauffmann},
  {Voges}, {Brinkmann} \& {Csabai}}{{Shen} et~al.}{2003}]{Shen2003}
{Shen} S.,  {Mo} H.~J.,  {White} S.~D.~M.,  {Blanton} M.~R.,  {Kauffmann} G.,
  {Voges} W.,  {Brinkmann} J.,    {Csabai} I.,  2003, \mnras, 343, 978

\bibitem[\protect\citeauthoryear{{Skrutskie} \& {et al.}}{{Skrutskie} \& {et
  al.}}{2006}]{Skrutskie2006}
{Skrutskie} M.~F.,  {et al.} 2006, \aj, 131, 1163

\bibitem[\protect\citeauthoryear{{Spergel}, {Bean}, {Dor{\'e}} \& {et
  al.}}{{Spergel} et~al.}{2007}]{Spergel2007}
{Spergel} D.~N.,  {Bean} R.,  {Dor{\'e}} O.,    {et al.} 2007, \apjs, 170, 377

\bibitem[\protect\citeauthoryear{{Springel}, {Di Matteo} \&
  {Hernquist}}{{Springel} et~al.}{2005}]{Springel2005}
{Springel} V.,  {Di Matteo} T.,    {Hernquist} L.,  2005, \apjl, 620, L79

\bibitem[\protect\citeauthoryear{{Springel}, {White}, {Jenkins} \& {et
  al.}}{{Springel} et~al.}{2005}]{Springel2005nat}
{Springel} V.,  {White} S.~D.~M.,  {Jenkins} A.,    {et al.} 2005, \nat, 435,
  629

\bibitem[\protect\citeauthoryear{{Strateva}, {Ivezi{\'c}}, {Knapp} \& {et
  al.}}{{Strateva} et~al.}{2001}]{Strateva2001}
{Strateva} I.,  {Ivezi{\'c}} {\v Z}.,  {Knapp} G.~R.,    {et al.} 2001, \aj,
  122, 1861

\bibitem[\protect\citeauthoryear{{Stumpff}}{{Stumpff}}{1980}]{Stumpff1980}
{Stumpff} P.,  1980, \aaps, 41, 1

\bibitem[\protect\citeauthoryear{Taylor et al.}{2010}]{Taylor2010}
Taylor E.~N., Franx M., Glazebrook K., Brinchmann J., van der Wel A., van
Dokkum P.~G., 2010, ApJ, 720, 723

\bibitem[\protect\citeauthoryear{{Teyssier}}{{Teyssier}}{2002}]{Teyssier2002}
{Teyssier} R.,  2002, \aap, 385, 337

\bibitem[\protect\citeauthoryear{{Tonry}, {Dressler}, {Blakeslee}, {Ajhar},
  {Fletcher}, {Luppino}, {Metzger} \& {Moore}}{{Tonry}
  et~al.}{2001}]{Tonry2001}
{Tonry} J.~L.,  {Dressler} A.,  {Blakeslee} J.~P.,  {Ajhar} E.~A.,  {Fletcher}
  A.~B.,  {Luppino} G.~A.,  {Metzger} M.~R.,    {Moore} C.~B.,  2001, \apj,
  546, 681

\bibitem[\protect\citeauthoryear{{Trujillo}, {Cenarro}, {de
  Lorenzo-C{\'a}ceres}, {Vazdekis}, {de la Rosa} \& {Cava}}{{Trujillo}
  et~al.}{2009}]{Trujillo2009}
{Trujillo} I.,  {Cenarro} A.~J.,  {de Lorenzo-C{\'a}ceres} A.,  {Vazdekis} A.,
  {de la Rosa} I.~G.,    {Cava} A.,  2009, \apjl, 692, L118

\bibitem[\protect\citeauthoryear{{Trujillo}, {Feulner}, {Goranova} \& {et
  al.}}{{Trujillo} et~al.}{2006}]{Trujillo2006}
{Trujillo} I.,  {Feulner} G.,  {Goranova} Y.,    {et al.} 2006, \mnras, 373,
  L36

\bibitem[\protect\citeauthoryear{{Tully} \& {Fisher}}{{Tully} \&
  {Fisher}}{1977}]{Tully1977}
{Tully} R.~B.,  {Fisher} J.~R.,  1977, \aap, 54, 661

\bibitem[\protect\citeauthoryear{{Tully}, {Rizzi}, {Shaya}, {Courtois},
  {Makarov} \& {Jacobs}}{{Tully} et~al.}{2009}]{Tully2009}
{Tully} R.~B.,  {Rizzi} L.,  {Shaya} E.~J.,  {Courtois} H.~M.,  {Makarov}
  D.~I.,    {Jacobs} B.~A.,  2009, \aj, 138, 323

\bibitem[\protect\citeauthoryear{{Tully}, {Shaya}, {Karachentsev}, {Courtois},
  {Kocevski}, {Rizzi} \& {Peel}}{{Tully} et~al.}{2008}]{Tully2008}
{Tully} R.~B.,  {Shaya} E.~J.,  {Karachentsev} I.~D.,  {Courtois} H.~M.,
  {Kocevski} D.~D.,  {Rizzi} L.,    {Peel} A.,  2008, \apj, 676, 184

\bibitem[\protect\citeauthoryear{{Valentijn}, {McFarland}, {Snigula} \& {et
  al.}}{{Valentijn} et~al.}{2007}]{Valentijn2007}
{Valentijn} E.~A.,  {McFarland} J.~P.,  {Snigula} J.,    {et al.} 2007, in
  {R.~A.~Shaw, F.~Hill, \& D.~J.~Bell} ed., Astronomical Data Analysis Software
  and Systems XVI Vol.~376 of Astronomical Society of the Pacific Conference
  Series, {Astro-WISE: Chaining to the Universe}.
pp 491

\bibitem[\protect\citeauthoryear{{Valentinuzzi}, {Fritz}, {Poggianti}, {Cava},
  {Bettoni}, {Fasano}, {D'Onofrio}, {Couch}, {Dressler}, {Moles}, {Moretti},
  {Omizzolo}, {Kj{\ae}rgaard}, {Vanzella} \& {Varela}}{{Valentinuzzi}
  et~al.}{2010}]{Valentinuzzi2010}
{Valentinuzzi} T.,  {et al.}  2010, \apj, 712, 226

\bibitem[\protect\citeauthoryear{{van den Bergh}}{{van den
  Bergh}}{2007}]{vandenBergh2007}
{van den Bergh} S.,  2007, \aj, 134, 1508

\bibitem[\protect\citeauthoryear{{van der Marel} \& {Franx}}{{van der Marel} \&
  {Franx}}{1993}]{vanDerMarel93}
{van der Marel} R.~P.,  {Franx} M.,  1993, \apj, 407, 525

\bibitem[\protect\citeauthoryear{van der Wel et
al.}{2009}]{vanderWel2009} van der Wel A., Bell E.~F., van den Bosch
F.~C., Gallazzi A., Rix H.-W., 2009, ApJ, 698, 1232

\bibitem[\protect\citeauthoryear{{van Dokkum}, {Franx}, {Kriek}, {Holden},
  {Illingworth}, {Magee}, {Bouwens}, {Marchesini}, {Quadri}, {Rudnick},
  {Taylor} \& {Toft}}{{van Dokkum} et~al.}{2008}]{vanDokkum2008}
{van Dokkum} P.~G.,  {et al.} 2008, \apjl, 677, L5

\bibitem[\protect\citeauthoryear{{Weijmans}, {Bower}, {Geach}, {Swinbank},
  {Wilman}, {de Zeeuw} \& {Morris}}{{Weijmans} et~al.}{2010}]{Weijmans2010}
{Weijmans} A.,  {Bower} R.~G.,  {Geach} J.~E.,  {Swinbank} A.~M.,  {Wilman}
  R.~J.,  {de Zeeuw} P.~T.,    {Morris} S.~L.,  2010, \mnras, 402, 2245

\bibitem[\protect\citeauthoryear{{Weijmans}, {Cappellari} \& {et
  al.}}{{Weijmans} et~al.}{2009}]{Weijmans2009}
{Weijmans} A.,  {et al.} 2009, \mnras, 398, 561

\bibitem[\protect\citeauthoryear{{York}, {Adelman}, {Anderson} Jr. \& {et
  al.}}{{York} et~al.}{2000}]{York2000}
{York} D.~G.,  {et al.} 2000, \aj, 120,
  1579

\bibitem[\protect\citeauthoryear{{Young}, {Bureau} \& {et al.}}{{Young}
  et~al.}{2011}]{Young2010}
{Young} L.~M.,  {et al.} 2011, \mnras, submitted (Paper IV)

\end{thebibliography}


\clearpage
\begin{deluxetable}{rrrcccrrrrrrrrrrrrr}
\tablewidth{0pt}
\tablecaption{The \atl\ sample or 260 early-type (E and S0) galaxies\label{tab:atlas3d_sample}}
\tablehead{
 \colhead{Galaxy} &
 \colhead{RA} &
 \colhead{DEC} &
 \colhead{SBF} &
 \colhead{NED-D} &
 \colhead{Virgo} &
  \colhead{$V_{\rm hel}$} &
 \colhead{$D$} &
 \colhead{$M_K$} &
 \colhead{$A_B$} &
 \colhead{$T$-type} &
 \colhead{$\log\re$} \\
 \colhead{} &
 \colhead{(deg)} &
 \colhead{(deg)} &
 \colhead{} &
 \colhead{} &
 \colhead{} &
 \colhead{(\kms)} &
 \colhead{(Mpc)} &
 \colhead{(mag)} &
 \colhead{(mag)} &
 \colhead{} &
 \colhead{(\arcsec)}  \\
 \colhead{(1)} &
 \colhead{(2)} &
 \colhead{(3)} &
 \colhead{(4)} &
 \colhead{(5)} &
 \colhead{(6)} &
 \colhead{(7)} &
 \colhead{(8)} &
 \colhead{(9)} &
 \colhead{(10)} &
 \colhead{(11)} &
 \colhead{(12)}
}
\startdata
     IC0560 &  146.472656 &   -0.268221 &  0 &  0 &  0 &  1853 &  27.2 &  -22.10 &  0.59 &  -0.7 &  1.11 \\
     IC0598 &  153.202423 &   43.145546 &  0 &  0 &  0 &  2256 &  35.3 &  -22.60 &  0.06 &  -0.1 &  1.02 \\
     IC0676 &  168.165909 &    9.055736 &  0 &  0 &  0 &  1429 &  24.6 &  -22.27 &  0.11 &  -1.3 &  1.35 \\
     IC0719 &  175.077042 &    9.009861 &  0 &  0 &  0 &  1833 &  29.4 &  -22.70 &  0.22 &  -2.0 &  1.10 \\
     IC0782 &  185.404053 &    5.765672 &  0 &  0 &  0 &  2424 &  36.3 &  -22.02 &  0.09 &   2.7 &  1.35 \\
     IC1024 &  217.863419 &    3.009107 &  0 &  0 &  0 &  1479 &  24.2 &  -21.85 &  0.13 &  -2.0 &  1.05 \\
     IC3631 &  189.950195 &   12.973927 &  0 &  0 &  0 &  2822 &  42.0 &  -22.01 &  0.17 &  -1.3 &  1.13 \\
    NGC0448 &   18.818876 &   -1.626105 &  1 &  1 &  0 &  1908 &  29.5 &  -23.02 &  0.26 &  -2.5 &  1.05 \\
    NGC0474 &   20.027901 &    3.415270 &  0 &  1 &  0 &  2315 &  30.9 &  -23.91 &  0.15 &  -2.0 &  1.52 \\
    NGC0502 &   20.731415 &    9.049169 &  0 &  0 &  0 &  2524 &  35.9 &  -23.05 &  0.17 &  -2.0 &  1.07 \\
    NGC0509 &   20.850327 &    9.433469 &  0 &  0 &  0 &  2261 &  32.3 &  -21.89 &  0.20 &  -1.3 &  1.37 \\
    NGC0516 &   21.033607 &    9.551668 &  0 &  0 &  0 &  2437 &  34.7 &  -22.21 &  0.29 &  -1.5 &  1.16 \\
    NGC0524 &   21.198778 &    9.538793 &  1 & 10 &  0 &  2403 &  23.3 &  -24.71 &  0.36 &  -1.2 &  1.64 \\
    NGC0525 &   21.220442 &    9.703240 &  0 &  0 &  0 &  2139 &  30.7 &  -21.86 &  0.38 &  -2.0 &  1.06 \\
    NGC0661 &   26.060976 &   28.705988 &  0 &  1 &  0 &  3815 &  30.6 &  -23.19 &  0.30 &  -4.4 &  1.12 \\
    NGC0680 &   27.447035 &   21.970827 &  0 &  1 &  0 &  2928 &  37.5 &  -24.17 &  0.34 &  -4.0 &  1.16 \\
    NGC0770 &   29.806850 &   18.954695 &  0 &  0 &  0 &  2543 &  36.7 &  -22.57 &  0.31 &  -4.2 &  0.94 \\
    NGC0821 &   32.088123 &   10.994870 &  1 &  4 &  0 &  1718 &  23.4 &  -23.99 &  0.48 &  -4.8 &  1.60 \\
    NGC0936 &   36.906090 &   -1.156280 &  1 &  4 &  0 &  1429 &  22.4 &  -24.85 &  0.15 &  -1.2 &  1.72 \\
    NGC1023 &   40.100052 &   39.063251 &  1 &  8 &  0 &   602 &  11.1 &  -24.01 &  0.26 &  -2.7 &  1.68 \\
    NGC1121 &   42.663387 &   -1.734040 &  0 &  0 &  0 &  2558 &  35.3 &  -22.70 &  0.29 &  -1.8 &  0.87 \\
    NGC1222 &   47.236446 &   -2.955212 &  0 &  0 &  0 &  2422 &  33.3 &  -22.71 &  0.26 &  -3.0 &  1.10 \\
    NGC1248 &   48.202328 &   -5.224674 &  0 &  0 &  0 &  2217 &  30.4 &  -22.90 &  0.27 &  -2.0 &  1.20 \\
    NGC1266 &   49.003120 &   -2.427370 &  0 &  0 &  0 &  2170 &  29.9 &  -22.93 &  0.43 &  -2.1 &  1.31 \\
    NGC1289 &   49.707592 &   -1.973354 &  0 &  0 &  0 &  2792 &  38.4 &  -23.46 &  0.37 &  -2.1 &  1.26 \\
    NGC1665 &   72.071098 &   -5.427655 &  0 &  0 &  0 &  2745 &  37.5 &  -23.63 &  0.26 &  -1.8 &  1.50 \\
    NGC2481 &  119.307182 &   23.767693 &  0 &  0 &  0 &  2157 &  32.0 &  -23.38 &  0.28 &   0.4 &  1.02 \\
    NGC2549 &  124.743111 &   57.803108 &  1 &  1 &  0 &  1051 &  12.3 &  -22.43 &  0.28 &  -2.0 &  1.28 \\
    NGC2577 &  125.681137 &   22.553040 &  0 &  0 &  0 &  2062 &  30.8 &  -23.41 &  0.23 &  -3.0 &  1.15 \\
    NGC2592 &  126.783669 &   25.970339 &  1 &  1 &  0 &  1979 &  25.0 &  -22.88 &  0.25 &  -4.8 &  1.09 \\
    NGC2594 &  126.821609 &   25.878935 &  0 &  0 &  0 &  2362 &  35.1 &  -22.36 &  0.24 &   0.0 &  0.82 \\
    NGC2679 &  132.887192 &   30.865419 &  0 &  0 &  0 &  2027 &  31.1 &  -22.81 &  0.14 &  -2.0 &  1.35 \\
    NGC2685 &  133.894791 &   58.734409 &  0 &  0 &  0 &   875 &  16.7 &  -22.78 &  0.27 &  -1.0 &  1.41 \\
    NGC2695 &  133.612778 &   -3.067101 &  1 &  3 &  0 &  1833 &  31.5 &  -23.64 &  0.08 &  -2.1 &  1.21 \\
    NGC2698 &  133.902222 &   -3.183882 &  0 &  0 &  0 &  1900 &  27.1 &  -23.32 &  0.08 &  -1.0 &  1.10 \\
    NGC2699 &  133.953415 &   -3.127507 &  1 &  3 &  0 &  1868 &  26.2 &  -22.72 &  0.08 &  -5.0 &  1.06 \\
    NGC2764 &  137.072983 &   21.443447 &  0 &  0 &  0 &  2706 &  39.6 &  -23.19 &  0.17 &  -2.0 &  1.09 \\
    NGC2768 &  137.906265 &   60.037209 &  1 &  5 &  0 &  1353 &  21.8 &  -24.71 &  0.20 &  -4.4 &  1.80 \\
    NGC2778 &  138.101639 &   35.027424 &  1 &  1 &  0 &  2025 &  22.3 &  -22.23 &  0.09 &  -4.8 &  1.20 \\
    NGC2824 &  139.759277 &   26.269999 &  0 &  0 &  0 &  2758 &  40.7 &  -22.93 &  0.14 &  -2.0 &  0.86 \\
    NGC2852 &  140.810684 &   40.163879 &  0 &  0 &  0 &  1781 &  28.5 &  -22.18 &  0.06 &   1.0 &  0.85 \\
    NGC2859 &  141.077286 &   34.513378 &  0 &  0 &  0 &  1690 &  27.0 &  -24.13 &  0.09 &  -1.2 &  1.43 \\
    NGC2880 &  142.394241 &   62.490620 &  1 &  1 &  0 &  1554 &  21.3 &  -22.98 &  0.14 &  -2.7 &  1.32 \\
    NGC2950 &  145.646317 &   58.851219 &  1 &  1 &  0 &  1322 &  14.5 &  -22.93 &  0.07 &  -2.0 &  1.19 \\
    NGC2962 &  145.224609 &    5.165820 &  0 &  3 &  0 &  1967 &  34.0 &  -24.01 &  0.25 &  -1.1 &  1.39 \\
    NGC2974 &  145.638611 &   -3.699116 &  1 &  3 &  0 &  1887 &  20.9 &  -23.62 &  0.23 &  -4.2 &  1.58 \\
    NGC3032 &  148.034119 &   29.236279 &  1 &  2 &  0 &  1562 &  21.4 &  -22.01 &  0.07 &  -1.9 &  1.12 \\
    NGC3073 &  150.216843 &   55.618935 &  1 &  1 &  0 &  1173 &  32.8 &  -21.78 &  0.05 &  -2.8 &  1.13 \\
    NGC3098 &  150.569458 &   24.711092 &  0 &  0 &  0 &  1397 &  23.0 &  -22.72 &  0.16 &  -1.5 &  1.12 \\
    NGC3156 &  153.171692 &    3.129320 &  1 &  3 &  0 &  1338 &  21.8 &  -22.15 &  0.15 &  -2.5 &  1.24 \\
    NGC3182 &  154.887558 &   58.205818 &  0 &  0 &  0 &  2118 &  34.0 &  -23.19 &  0.05 &   0.4 &  1.32 \\
    NGC3193 &  154.603683 &   21.893978 &  1 &  3 &  0 &  1381 &  33.1 &  -24.63 &  0.11 &  -4.8 &  1.42 \\
    NGC3226 &  155.862549 &   19.898439 &  1 &  3 &  0 &  1315 &  22.9 &  -23.24 &  0.10 &  -4.8 &  1.49 \\
    NGC3230 &  155.933090 &   12.567883 &  0 &  0 &  0 &  2795 &  40.8 &  -24.18 &  0.16 &  -1.8 &  1.26 \\
    NGC3245 &  156.826523 &   28.507435 &  1 &  1 &  0 &  1326 &  20.3 &  -23.69 &  0.11 &  -2.1 &  1.40 \\
    NGC3248 &  156.939270 &   22.847170 &  0 &  0 &  0 &  1481 &  24.6 &  -22.43 &  0.09 &  -2.0 &  1.20 \\
    NGC3301 &  159.233459 &   21.882166 &  0 &  0 &  0 &  1339 &  22.8 &  -23.28 &  0.10 &  -0.4 &  1.30 \\
    NGC3377 &  161.926666 &   13.985640 &  1 & 10 &  0 &   690 &  10.9 &  -22.76 &  0.15 &  -4.8 &  1.55 \\
    NGC3379 &  161.956665 &   12.581630 &  1 & 15 &  0 &   918 &  10.3 &  -23.80 &  0.11 &  -4.8 &  1.60 \\
    NGC3384 &  162.070404 &   12.629300 &  1 & 10 &  0 &   733 &  11.3 &  -23.52 &  0.12 &  -2.7 &  1.51 \\
    NGC3400 &  162.689590 &   28.468929 &  0 &  0 &  0 &  1441 &  24.7 &  -21.82 &  0.08 &   0.7 &  1.23 \\
    NGC3412 &  162.722137 &   13.412142 &  1 &  5 &  0 &   860 &  11.0 &  -22.55 &  0.12 &  -2.0 &  1.49 \\
    NGC3414 &  162.817673 &   27.974968 &  1 &  1 &  0 &  1470 &  24.5 &  -23.98 &  0.11 &  -2.0 &  1.38 \\
    NGC3457 &  163.702591 &   17.621157 &  1 &  1 &  0 &  1148 &  20.1 &  -21.89 &  0.13 &  -5.0 &  1.13 \\
    NGC3458 &  164.006042 &   57.116970 &  0 &  0 &  0 &  1877 &  30.9 &  -23.12 &  0.04 &  -2.0 &  1.06 \\
    NGC3489 &  165.077454 &   13.901258 &  1 &  5 &  0 &   695 &  11.7 &  -22.99 &  0.07 &  -1.2 &  1.35 \\
    NGC3499 &  165.796280 &   56.221664 &  0 &  0 &  0 &  1535 &  26.4 &  -21.88 &  0.04 &   0.0 &  0.94 \\
    NGC3522 &  166.668549 &   20.085621 &  0 &  1 &  0 &  1228 &  25.5 &  -21.67 &  0.10 &  -4.9 &  1.01 \\
    NGC3530 &  167.168411 &   57.230160 &  0 &  0 &  0 &  1894 &  31.2 &  -22.00 &  0.04 &   0.0 &  0.87 \\
    NGC3595 &  168.856461 &   47.447147 &  0 &  0 &  0 &  2177 &  34.7 &  -23.28 &  0.09 &  -3.3 &  1.15 \\
    NGC3599 &  168.862305 &   18.110369 &  1 &  3 &  0 &   839 &  19.8 &  -22.22 &  0.09 &  -2.0 &  1.37 \\
    NGC3605 &  169.194260 &   18.017141 &  1 &  3 &  0 &   661 &  20.1 &  -21.83 &  0.09 &  -4.5 &  1.23 \\
    NGC3607 &  169.227737 &   18.051809 &  1 &  4 &  0 &   942 &  22.2 &  -24.74 &  0.09 &  -3.1 &  1.59 \\
    NGC3608 &  169.245697 &   18.148531 &  1 &  4 &  0 &  1226 &  22.3 &  -23.65 &  0.09 &  -4.8 &  1.47 \\
    NGC3610 &  169.605316 &   58.786247 &  1 &  4 &  0 &  1707 &  20.8 &  -23.69 &  0.04 &  -4.2 &  1.20 \\
    NGC3613 &  169.650543 &   57.999924 &  1 &  3 &  0 &  2051 &  28.3 &  -24.26 &  0.05 &  -4.7 &  1.42 \\
    NGC3619 &  169.840088 &   57.757683 &  0 &  0 &  0 &  1560 &  26.8 &  -23.57 &  0.08 &  -0.9 &  1.42 \\
    NGC3626 &  170.015808 &   18.356791 &  1 &  1 &  0 &  1486 &  19.5 &  -23.30 &  0.08 &  -1.0 &  1.41 \\
    NGC3630 &  170.070786 &    2.964170 &  0 &  0 &  0 &  1499 &  25.0 &  -23.16 &  0.18 &  -1.5 &  1.10 \\
    NGC3640 &  170.278549 &    3.234764 &  1 &  4 &  0 &  1298 &  26.3 &  -24.60 &  0.19 &  -4.9 &  1.49 \\
    NGC3641 &  170.286621 &    3.194489 &  1 &  1 &  0 &  1780 &  25.9 &  -21.85 &  0.18 &  -4.9 &  0.97 \\
    NGC3648 &  170.631195 &   39.876972 &  0 &  0 &  0 &  1970 &  31.9 &  -23.06 &  0.09 &  -2.0 &  1.12 \\
    NGC3658 &  170.992706 &   38.562424 &  0 &  0 &  0 &  2039 &  32.7 &  -23.45 &  0.09 &  -2.2 &  1.28 \\
    NGC3665 &  171.181793 &   38.762791 &  0 &  0 &  0 &  2069 &  33.1 &  -24.92 &  0.08 &  -2.1 &  1.49 \\
    NGC3674 &  171.610870 &   57.048290 &  0 &  0 &  0 &  2055 &  33.4 &  -23.23 &  0.06 &  -1.9 &  1.05 \\
    NGC3694 &  172.225571 &   35.413857 &  0 &  0 &  0 &  2243 &  35.2 &  -22.35 &  0.10 &  -5.0 &  1.02 \\
    NGC3757 &  174.261765 &   58.415649 &  0 &  0 &  0 &  1245 &  22.6 &  -22.15 &  0.06 &  -2.0 &  0.95 \\
    NGC3796 &  175.129776 &   60.298958 &  0 &  0 &  0 &  1250 &  22.7 &  -21.84 &  0.06 &   0.0 &  1.06 \\
    NGC3838 &  176.057205 &   57.948101 &  0 &  0 &  0 &  1308 &  23.5 &  -22.52 &  0.05 &   0.0 &  1.04 \\
    NGC3941 &  178.230667 &   36.986378 &  1 &  2 &  0 &   930 &  11.9 &  -23.06 &  0.09 &  -2.0 &  1.40 \\
    NGC3945 &  178.307190 &   60.675560 &  0 &  0 &  0 &  1281 &  23.2 &  -24.31 &  0.12 &  -1.2 &  1.45 \\
    NGC3998 &  179.484039 &   55.453564 &  1 &  2 &  0 &  1048 &  13.7 &  -23.33 &  0.07 &  -2.1 &  1.30 \\
    NGC4026 &  179.854950 &   50.961689 &  1 &  2 &  0 &   985 &  13.2 &  -23.03 &  0.09 &  -1.8 &  1.31 \\
    NGC4036 &  180.362045 &   61.895699 &  0 &  0 &  0 &  1385 &  24.6 &  -24.40 &  0.10 &  -2.6 &  1.46 \\
    NGC4078 &  181.198456 &   10.595537 &  0 &  0 &  0 &  2546 &  38.1 &  -22.99 &  0.11 &  -2.0 &  0.92 \\
    NGC4111 &  181.763031 &   43.065392 &  1 &  1 &  0 &   792 &  14.6 &  -23.27 &  0.06 &  -1.4 &  1.08 \\
    NGC4119 &  182.040176 &   10.378720 &  0 &  0 &  1 &  1656 &  16.5 &  -22.60 &  0.12 &  -1.3 &  1.60 \\
    NGC4143 &  182.400360 &   42.534218 &  1 &  2 &  0 &   946 &  15.5 &  -23.10 &  0.05 &  -1.9 &  1.39 \\
    NGC4150 &  182.640228 &   30.401487 &  1 &  4 &  0 &   208 &  13.4 &  -21.65 &  0.08 &  -2.1 &  1.20 \\
    NGC4168 &  183.071808 &   13.205354 &  0 &  1 &  0 &  2286 &  30.9 &  -24.03 &  0.16 &  -4.8 &  1.48 \\
    NGC4179 &  183.217087 &    1.299673 &  0 &  0 &  1 &  1300 &  16.5 &  -23.18 &  0.14 &  -1.9 &  1.30 \\
    NGC4191 &  183.459915 &    7.200842 &  0 &  0 &  0 &  2646 &  39.2 &  -23.10 &  0.09 &  -1.8 &  1.06 \\
    NGC4203 &  183.770935 &   33.197243 &  1 &  2 &  0 &  1087 &  14.7 &  -23.44 &  0.05 &  -2.7 &  1.47 \\
    NGC4215 &  183.977142 &    6.401132 &  0 &  0 &  0 &  2011 &  31.5 &  -23.43 &  0.07 &  -0.9 &  1.18 \\
    NGC4233 &  184.282043 &    7.624434 &  0 &  1 &  0 &  2306 &  33.9 &  -23.88 &  0.10 &  -2.0 &  1.19 \\
    NGC4249 &  184.497650 &    5.598720 &  0 &  0 &  0 &  2618 &  38.7 &  -21.98 &  0.09 &  -1.3 &  1.10 \\
    NGC4251 &  184.534607 &   28.175299 &  1 &  1 &  0 &  1066 &  19.1 &  -23.68 &  0.10 &  -1.9 &  1.29 \\
    NGC4255 &  184.734100 &    4.785923 &  0 &  0 &  0 &  1995 &  31.2 &  -22.99 &  0.09 &  -1.9 &  1.10 \\
    NGC4259 &  184.842468 &    5.376242 &  0 &  0 &  0 &  2497 &  37.2 &  -22.19 &  0.08 &  -2.0 &  0.89 \\
    NGC4261 &  184.846924 &    5.824710 &  1 &  5 &  0 &  2212 &  30.8 &  -25.18 &  0.08 &  -4.8 &  1.58 \\
    NGC4262 &  184.877426 &   14.877717 &  2 &  1 &  1 &  1375 &  15.4 &  -22.60 &  0.16 &  -2.7 &  1.10 \\
    NGC4264 &  184.899078 &    5.846804 &  0 &  0 &  0 &  2518 &  37.5 &  -23.00 &  0.08 &  -1.1 &  1.14 \\
    NGC4267 &  184.938675 &   12.798356 &  2 &  1 &  1 &  1021 &  15.8 &  -23.18 &  0.20 &  -2.7 &  1.58 \\
    NGC4268 &  184.946762 &    5.283650 &  0 &  0 &  0 &  2034 &  31.7 &  -23.05 &  0.08 &  -0.3 &  1.20 \\
    NGC4270 &  184.955978 &    5.463371 &  0 &  0 &  0 &  2331 &  35.2 &  -23.69 &  0.08 &  -2.0 &  1.21 \\
    NGC4278 &  185.028320 &   29.280619 &  1 &  9 &  0 &   620 &  15.6 &  -23.80 &  0.13 &  -4.8 &  1.50 \\
    NGC4281 &  185.089691 &    5.386430 &  0 &  1 &  0 &  2671 &  24.4 &  -24.01 &  0.09 &  -1.5 &  1.34 \\
    NGC4283 &  185.086609 &   29.310898 &  1 &  3 &  0 &  1056 &  15.3 &  -21.80 &  0.11 &  -4.8 &  1.09 \\
    NGC4324 &  185.775726 &    5.250488 &  0 &  0 &  1 &  1665 &  16.5 &  -22.61 &  0.10 &  -0.9 &  1.30 \\
    NGC4339 &  185.895599 &    6.081713 &  1 &  3 &  1 &  1266 &  16.0 &  -22.49 &  0.11 &  -4.7 &  1.48 \\
    NGC4340 &  185.897141 &   16.722195 &  0 &  2 &  1 &   933 &  18.4 &  -23.01 &  0.11 &  -1.2 &  1.57 \\
    NGC4342 &  185.912598 &    7.053936 &  0 &  0 &  1 &   761 &  16.5 &  -22.07 &  0.09 &  -3.4 &  0.82 \\
    NGC4346 &  185.866425 &   46.993881 &  1 &  1 &  0 &   832 &  13.9 &  -22.55 &  0.05 &  -2.0 &  1.29 \\
    NGC4350 &  185.990891 &   16.693356 &  0 &  2 &  1 &  1210 &  15.4 &  -23.13 &  0.12 &  -1.8 &  1.23 \\
    NGC4365 &  186.117615 &    7.317520 &  2 & 13 &  0 &  1243 &  23.3 &  -25.21 &  0.09 &  -4.8 &  1.72 \\
    NGC4371 &  186.230957 &   11.704288 &  2 &  1 &  1 &   933 &  17.0 &  -23.45 &  0.16 &  -1.3 &  1.47 \\
    NGC4374 &  186.265747 &   12.886960 &  2 & 13 &  1 &  1017 &  18.5 &  -25.12 &  0.18 &  -4.3 &  1.72 \\
    NGC4377 &  186.301285 &   14.762218 &  2 &  1 &  1 &  1338 &  17.8 &  -22.43 &  0.17 &  -2.6 &  1.13 \\
    NGC4379 &  186.311386 &   15.607498 &  2 &  2 &  1 &  1074 &  15.8 &  -22.24 &  0.10 &  -2.8 &  1.27 \\
    NGC4382 &  186.350220 &   18.191080 &  2 &  7 &  1 &   746 &  17.9 &  -25.13 &  0.13 &  -1.3 &  1.82 \\
    NGC4387 &  186.423813 &   12.810359 &  2 &  3 &  1 &   565 &  17.9 &  -22.13 &  0.14 &  -4.9 &  1.20 \\
    NGC4406 &  186.549225 &   12.945970 &  2 & 15 &  1 &  -224 &  16.8 &  -25.04 &  0.13 &  -4.8 &  1.97 \\
    NGC4417 &  186.710938 &    9.584117 &  2 &  1 &  1 &   828 &  16.0 &  -22.86 &  0.11 &  -1.9 &  1.25 \\
    NGC4425 &  186.805664 &   12.734803 &  0 &  0 &  1 &  1908 &  16.5 &  -22.09 &  0.13 &  -0.6 &  1.38 \\
    NGC4429 &  186.860657 &   11.107540 &  0 &  0 &  1 &  1104 &  16.5 &  -24.32 &  0.14 &  -1.1 &  1.62 \\
    NGC4434 &  186.902832 &    8.154311 &  2 &  5 &  0 &  1070 &  22.4 &  -22.55 &  0.10 &  -4.8 &  1.16 \\
    NGC4435 &  186.918762 &   13.079021 &  2 &  0 &  1 &   791 &  16.7 &  -23.83 &  0.13 &  -2.1 &  1.49 \\
    NGC4442 &  187.016220 &    9.803620 &  2 &  1 &  1 &   547 &  15.3 &  -23.63 &  0.09 &  -1.9 &  1.44 \\
    NGC4452 &  187.180417 &   11.755000 &  0 &  2 &  1 &   188 &  15.6 &  -21.88 &  0.13 &  -2.0 &  1.30 \\
    NGC4458 &  187.239716 &   13.241916 &  2 &  8 &  1 &   677 &  16.4 &  -21.76 &  0.10 &  -4.8 &  1.37 \\
    NGC4459 &  187.250107 &   13.978580 &  2 &  3 &  1 &  1192 &  16.1 &  -23.89 &  0.19 &  -1.4 &  1.56 \\
    NGC4461 &  187.262543 &   13.183857 &  0 &  0 &  1 &  1924 &  16.5 &  -23.08 &  0.10 &  -0.8 &  1.40 \\
    NGC4472 &  187.444992 &    8.000410 &  2 & 18 &  1 &   981 &  17.1 &  -25.78 &  0.10 &  -4.8 &  1.98 \\
    NGC4473 &  187.453659 &   13.429320 &  2 &  8 &  1 &  2260 &  15.3 &  -23.77 &  0.12 &  -4.7 &  1.43 \\
    NGC4474 &  187.473099 &   14.068673 &  2 &  1 &  1 &  1611 &  15.6 &  -22.28 &  0.18 &  -2.0 &  1.30 \\
    NGC4476 &  187.496170 &   12.348669 &  2 &  3 &  1 &  1968 &  17.6 &  -21.78 &  0.12 &  -3.0 &  1.20 \\
    NGC4477 &  187.509048 &   13.636443 &  0 &  0 &  1 &  1338 &  16.5 &  -23.75 &  0.14 &  -1.9 &  1.59 \\
    NGC4478 &  187.572662 &   12.328578 &  2 &  9 &  1 &  1375 &  17.0 &  -22.80 &  0.10 &  -4.8 &  1.20 \\
    NGC4483 &  187.669250 &    9.015665 &  2 &  1 &  1 &   906 &  16.7 &  -21.84 &  0.09 &  -1.3 &  1.24 \\
    NGC4486 &  187.705933 &   12.391100 &  2 & 15 &  1 &  1284 &  17.2 &  -25.38 &  0.10 &  -4.3 &  1.91 \\
   NGC4486A &  187.740540 &   12.270361 &  2 &  0 &  1 &   758 &  18.3 &  -21.82 &  0.10 &  -5.0 &  0.94 \\
    NGC4489 &  187.717667 &   16.758696 &  2 &  5 &  1 &   961 &  15.4 &  -21.59 &  0.12 &  -4.8 &  1.42 \\
    NGC4494 &  187.850143 &   25.774981 &  1 &  9 &  0 &  1342 &  16.6 &  -24.11 &  0.09 &  -4.8 &  1.69 \\
    NGC4503 &  188.025803 &   11.176434 &  0 &  0 &  1 &  1334 &  16.5 &  -23.22 &  0.22 &  -1.8 &  1.45 \\
    NGC4521 &  188.198853 &   63.939293 &  0 &  0 &  0 &  2511 &  39.7 &  -23.92 &  0.08 &  -0.1 &  1.21 \\
    NGC4526 &  188.512619 &    7.699140 &  1 &  5 &  1 &   617 &  16.4 &  -24.62 &  0.10 &  -1.9 &  1.65 \\
    NGC4528 &  188.525269 &   11.321266 &  2 &  0 &  1 &  1378 &  15.8 &  -22.05 &  0.20 &  -2.0 &  1.15 \\
    NGC4546 &  188.872940 &   -3.793227 &  1 &  2 &  0 &  1057 &  13.7 &  -23.30 &  0.15 &  -2.7 &  1.40 \\
    NGC4550 &  188.877548 &   12.220955 &  2 &  3 &  1 &   459 &  15.5 &  -22.27 &  0.17 &  -2.1 &  1.19 \\
    NGC4551 &  188.908249 &   12.264010 &  2 &  5 &  1 &  1176 &  16.1 &  -22.18 &  0.17 &  -4.9 &  1.22 \\
    NGC4552 &  188.916183 &   12.556040 &  2 & 12 &  1 &   344 &  15.8 &  -24.29 &  0.18 &  -4.6 &  1.53 \\
    NGC4564 &  189.112473 &   11.439320 &  2 &  5 &  1 &  1155 &  15.8 &  -23.08 &  0.14 &  -4.8 &  1.31 \\
    NGC4570 &  189.222504 &    7.246663 &  2 &  1 &  1 &  1787 &  17.1 &  -23.48 &  0.09 &  -2.0 &  1.30 \\
    NGC4578 &  189.377274 &    9.555121 &  2 &  3 &  1 &  2292 &  16.3 &  -22.66 &  0.09 &  -2.0 &  1.51 \\
    NGC4596 &  189.983063 &   10.176031 &  0 &  0 &  1 &  1892 &  16.5 &  -23.63 &  0.09 &  -0.9 &  1.59 \\
    NGC4608 &  190.305374 &   10.155793 &  0 &  0 &  1 &  1850 &  16.5 &  -22.94 &  0.07 &  -1.9 &  1.42 \\
    NGC4612 &  190.386490 &    7.314782 &  2 &  1 &  1 &  1775 &  16.6 &  -22.55 &  0.11 &  -2.0 &  1.40 \\
    NGC4621 &  190.509674 &   11.646930 &  2 &  9 &  1 &   467 &  14.9 &  -24.14 &  0.14 &  -4.8 &  1.63 \\
    NGC4623 &  190.544601 &    7.676934 &  2 &  1 &  1 &  1807 &  17.4 &  -21.74 &  0.10 &  -1.5 &  1.31 \\
    NGC4624 &  191.274826 &    3.055684 &  0 &  0 &  1 &   912 &  16.5 &  -23.67 &  0.10 &  -0.6 &  1.64 \\
    NGC4636 &  190.707779 &    2.687780 &  1 & 11 &  0 &   930 &  14.3 &  -24.36 &  0.12 &  -4.8 &  1.95 \\
    NGC4638 &  190.697632 &   11.442459 &  2 &  2 &  1 &  1152 &  17.5 &  -23.01 &  0.11 &  -2.7 &  1.22 \\
    NGC4643 &  190.833893 &    1.978399 &  0 &  0 &  1 &  1333 &  16.5 &  -23.69 &  0.13 &  -0.6 &  1.38 \\
    NGC4649 &  190.916702 &   11.552610 &  2 & 11 &  1 &  1110 &  17.3 &  -25.46 &  0.11 &  -4.6 &  1.82 \\
    NGC4660 &  191.133209 &   11.190533 &  2 &  5 &  1 &  1087 &  15.0 &  -22.69 &  0.15 &  -4.7 &  1.09 \\
    NGC4684 &  191.822861 &   -2.727538 &  1 &  1 &  0 &  1560 &  13.1 &  -22.21 &  0.12 &  -1.2 &  1.32 \\
    NGC4690 &  191.981323 &   -1.655975 &  0 &  0 &  0 &  2765 &  40.2 &  -22.96 &  0.13 &  -3.0 &  1.25 \\
    NGC4694 &  192.062881 &   10.983624 &  0 &  0 &  1 &  1171 &  16.5 &  -22.15 &  0.17 &  -2.0 &  1.47 \\
    NGC4697 &  192.149612 &   -5.800850 &  1 &  8 &  0 &  1252 &  11.4 &  -23.93 &  0.13 &  -4.4 &  1.79 \\
    NGC4710 &  192.412323 &   15.165490 &  0 &  0 &  1 &  1102 &  16.5 &  -23.53 &  0.13 &  -0.9 &  1.48 \\
    NGC4733 &  192.778259 &   10.912103 &  1 &  1 &  1 &   925 &  14.5 &  -21.80 &  0.09 &  -3.8 &  1.52 \\
    NGC4753 &  193.092133 &   -1.199690 &  1 &  3 &  0 &  1163 &  22.9 &  -25.09 &  0.14 &  -1.4 &  1.69 \\
    NGC4754 &  193.073181 &   11.313660 &  2 &  3 &  1 &  1351 &  16.1 &  -23.64 &  0.14 &  -2.5 &  1.50 \\
    NGC4762 &  193.233536 &   11.230800 &  0 &  2 &  0 &   986 &  22.6 &  -24.48 &  0.09 &  -1.8 &  1.64 \\
    NGC4803 &  193.890289 &    8.240547 &  0 &  0 &  0 &  2645 &  39.4 &  -22.28 &  0.13 &   0.0 &  0.94 \\
    NGC5103 &  200.125229 &   43.084015 &  0 &  0 &  0 &  1273 &  23.4 &  -22.36 &  0.08 &   0.0 &  1.02 \\
    NGC5173 &  202.105301 &   46.591572 &  0 &  0 &  0 &  2424 &  38.4 &  -22.88 &  0.12 &  -4.9 &  1.01 \\
    NGC5198 &  202.547546 &   46.670830 &  0 &  0 &  0 &  2519 &  39.6 &  -24.10 &  0.10 &  -4.8 &  1.38 \\
    NGC5273 &  205.534943 &   35.654240 &  1 &  2 &  0 &  1085 &  16.1 &  -22.37 &  0.04 &  -1.9 &  1.57 \\
    NGC5308 &  206.751633 &   60.973038 &  0 &  6 &  0 &  1998 &  31.5 &  -24.13 &  0.08 &  -2.1 &  1.25 \\
    NGC5322 &  207.313339 &   60.190411 &  1 &  5 &  0 &  1780 &  30.3 &  -25.26 &  0.06 &  -4.8 &  1.60 \\
    NGC5342 &  207.857910 &   59.863014 &  0 &  0 &  0 &  2189 &  35.5 &  -22.61 &  0.05 &  -2.0 &  0.97 \\
    NGC5353 &  208.361420 &   40.283123 &  0 &  0 &  0 &  2198 &  35.2 &  -25.11 &  0.05 &  -2.1 &  1.30 \\
    NGC5355 &  208.439850 &   40.338795 &  0 &  0 &  0 &  2344 &  37.1 &  -22.40 &  0.05 &  -2.1 &  1.06 \\
    NGC5358 &  208.501801 &   40.277420 &  0 &  0 &  0 &  2412 &  38.0 &  -22.01 &  0.04 &  -0.2 &  1.05 \\
    NGC5379 &  208.893112 &   59.742825 &  0 &  0 &  0 &  1774 &  30.0 &  -22.08 &  0.08 &  -2.0 &  1.32 \\
    NGC5422 &  210.175262 &   55.164478 &  0 &  0 &  0 &  1838 &  30.8 &  -23.69 &  0.06 &  -1.5 &  1.33 \\
    NGC5473 &  211.180176 &   54.892620 &  0 &  0 &  0 &  2022 &  33.2 &  -24.25 &  0.05 &  -2.7 &  1.32 \\
    NGC5475 &  211.301437 &   55.741802 &  0 &  0 &  0 &  1671 &  28.6 &  -22.88 &  0.05 &   1.0 &  1.22 \\
    NGC5481 &  211.671722 &   50.723320 &  0 &  2 &  0 &  1989 &  25.8 &  -22.68 &  0.08 &  -3.9 &  1.35 \\
    NGC5485 &  211.797134 &   55.001518 &  1 &  3 &  0 &  1927 &  25.2 &  -23.61 &  0.07 &  -2.0 &  1.45 \\
    NGC5493 &  212.872421 &   -5.043663 &  0 &  0 &  0 &  2665 &  38.8 &  -24.49 &  0.15 &  -2.1 &  1.14 \\
    NGC5500 &  212.563522 &   48.546066 &  0 &  0 &  0 &  1914 &  31.7 &  -21.93 &  0.09 &  -4.9 &  1.18 \\
    NGC5507 &  213.332825 &   -3.148860 &  0 &  0 &  0 &  1851 &  28.5 &  -23.19 &  0.26 &  -2.3 &  1.09 \\
    NGC5557 &  214.607117 &   36.493690 &  0 &  3 &  0 &  3219 &  38.8 &  -24.87 &  0.03 &  -4.8 &  1.46 \\
    NGC5574 &  215.233109 &    3.237995 &  1 &  1 &  0 &  1589 &  23.2 &  -22.30 &  0.13 &  -2.8 &  1.13 \\
    NGC5576 &  215.265381 &    3.271049 &  1 &  3 &  0 &  1506 &  24.8 &  -24.15 &  0.13 &  -4.8 &  1.34 \\
    NGC5582 &  215.179703 &   39.693584 &  1 &  3 &  0 &  1430 &  27.7 &  -23.28 &  0.06 &  -4.9 &  1.44 \\
    NGC5611 &  216.019897 &   33.047501 &  1 &  1 &  0 &  1968 &  24.5 &  -22.20 &  0.05 &  -1.9 &  1.00 \\
    NGC5631 &  216.638687 &   56.582664 &  1 &  1 &  0 &  1944 &  27.0 &  -23.70 &  0.09 &  -1.9 &  1.32 \\
    NGC5638 &  217.418289 &    3.233443 &  1 &  3 &  0 &  1652 &  25.6 &  -23.80 &  0.14 &  -4.8 &  1.45 \\
    NGC5687 &  218.718201 &   54.475685 &  1 &  1 &  0 &  2143 &  27.2 &  -23.22 &  0.05 &  -3.0 &  1.36 \\
    NGC5770 &  223.312653 &    3.959721 &  1 &  1 &  0 &  1471 &  18.5 &  -22.15 &  0.17 &  -2.0 &  1.23 \\
    NGC5813 &  225.296936 &    1.701970 &  1 &  4 &  0 &  1956 &  31.3 &  -25.09 &  0.25 &  -4.8 &  1.76 \\
    NGC5831 &  226.029266 &    1.219917 &  1 &  3 &  0 &  1645 &  26.4 &  -23.69 &  0.25 &  -4.8 &  1.40 \\
    NGC5838 &  226.359467 &    2.099356 &  0 &  0 &  0 &  1341 &  21.8 &  -24.13 &  0.23 &  -2.6 &  1.40 \\
    NGC5839 &  226.364471 &    1.634633 &  1 &  1 &  0 &  1220 &  22.0 &  -22.53 &  0.23 &  -2.0 &  1.22 \\
    NGC5845 &  226.503281 &    1.633824 &  1 &  2 &  0 &  1472 &  25.2 &  -22.92 &  0.23 &  -4.9 &  0.80 \\
    NGC5846 &  226.621887 &    1.605637 &  1 &  6 &  0 &  1712 &  24.2 &  -25.01 &  0.24 &  -4.7 &  1.77 \\
    NGC5854 &  226.948853 &    2.568560 &  0 &  0 &  0 &  1663 &  26.2 &  -23.30 &  0.23 &  -1.1 &  1.26 \\
    NGC5864 &  227.389786 &    3.052741 &  0 &  0 &  0 &  1874 &  29.0 &  -23.62 &  0.19 &  -1.7 &  1.35 \\
    NGC5866 &  226.623169 &   55.763309 &  1 &  3 &  0 &   755 &  14.9 &  -24.00 &  0.06 &  -1.3 &  1.56 \\
    NGC5869 &  227.456055 &    0.469967 &  0 &  1 &  0 &  2065 &  24.9 &  -23.27 &  0.24 &  -2.3 &  1.31 \\
    NGC6010 &  238.579773 &    0.543033 &  0 &  0 &  0 &  2022 &  30.6 &  -23.53 &  0.45 &   0.4 &  1.16 \\
    NGC6014 &  238.989105 &    5.931838 &  0 &  0 &  0 &  2381 &  35.8 &  -22.99 &  0.22 &  -1.9 &  1.35 \\
    NGC6017 &  239.314529 &    5.998364 &  1 &  1 &  0 &  1788 &  29.0 &  -22.52 &  0.23 &  -5.0 &  0.85 \\
    NGC6149 &  246.851151 &   19.597290 &  0 &  0 &  0 &  2427 &  37.2 &  -22.60 &  0.30 &  -1.9 &  1.03 \\
    NGC6278 &  255.209763 &   23.010956 &  0 &  0 &  0 &  2832 &  42.9 &  -24.19 &  0.27 &  -1.9 &  1.22 \\
    NGC6547 &  271.291748 &   25.232645 &  0 &  0 &  0 &  2677 &  40.8 &  -23.60 &  0.50 &  -1.3 &  1.06 \\
    NGC6548 &  271.496826 &   18.587217 &  1 &  1 &  0 &  2208 &  22.4 &  -23.19 &  0.35 &  -1.9 &  1.35 \\
    NGC6703 &  281.828522 &   45.550648 &  1 &  4 &  0 &  2373 &  25.9 &  -23.85 &  0.37 &  -2.8 &  1.34 \\
    NGC6798 &  291.013306 &   53.624752 &  0 &  0 &  0 &  2360 &  37.5 &  -23.52 &  0.57 &  -2.0 &  1.23 \\
    NGC7280 &  336.614899 &   16.148266 &  1 &  2 &  0 &  1845 &  23.7 &  -22.83 &  0.24 &  -1.3 &  1.33 \\
    NGC7332 &  339.352173 &   23.798351 &  1 &  1 &  0 &  1197 &  22.4 &  -23.75 &  0.16 &  -1.9 &  1.24 \\
    NGC7454 &  345.277130 &   16.388371 &  1 &  3 &  0 &  2020 &  23.2 &  -23.00 &  0.33 &  -4.7 &  1.41 \\
    NGC7457 &  345.249725 &   30.144892 &  1 &  3 &  0 &   844 &  12.9 &  -22.38 &  0.23 &  -2.6 &  1.56 \\
    NGC7465 &  345.503967 &   15.964876 &  0 &  0 &  0 &  1960 &  29.3 &  -22.82 &  0.33 &  -1.9 &  0.90 \\
    NGC7693 &  353.293671 &   -1.292010 &  0 &  0 &  0 &  2502 &  35.4 &  -21.58 &  0.15 &  -1.0 &  1.11 \\
    NGC7710 &  353.942261 &   -2.880941 &  0 &  0 &  0 &  2407 &  34.0 &  -21.99 &  0.15 &  -1.9 &  0.92 \\
  PGC016060 &   72.143387 &   -3.867104 &  0 &  0 &  0 &  2764 &  37.8 &  -22.64 &  0.19 &  -0.6 &  1.01 \\
  PGC028887 &  149.931290 &   11.660812 &  0 &  0 &  0 &  2833 &  41.0 &  -22.26 &  0.17 &   0.0 &  1.08 \\
  PGC029321 &  151.463226 &   12.961213 &  0 &  0 &  0 &  2816 &  40.9 &  -21.66 &  0.16 &   0.0 &  0.89 \\
  PGC035754 &  173.614716 &   33.178913 &  0 &  0 &  0 &  2534 &  39.0 &  -21.90 &  0.11 &  -3.0 &  0.83 \\
  PGC042549 &  190.316513 &   -5.009177 &  0 &  0 &  0 &  2822 &  40.7 &  -22.71 &  0.11 &  -5.0 &  1.06 \\
  PGC044433 &  194.578110 &   13.391409 &  0 &  0 &  0 &  2675 &  40.1 &  -22.25 &  0.13 &   0.0 &  0.71 \\
  PGC050395 &  211.913544 &   54.794575 &  0 &  0 &  0 &  2322 &  37.2 &  -21.92 &  0.05 &   0.0 &  1.04 \\
  PGC051753 &  217.310318 &   44.699104 &  0 &  0 &  0 &  2418 &  38.3 &  -21.92 &  0.05 &   0.0 &  1.01 \\
  PGC054452 &  228.894180 &    2.248187 &  0 &  0 &  0 &  1918 &  29.5 &  -21.59 &  0.18 &  -2.0 &  1.14 \\
  PGC056772 &  240.548340 &    7.085953 &  0 &  0 &  0 &  2655 &  39.5 &  -22.06 &  0.19 &  -2.0 &  0.93 \\
  PGC058114 &  246.517838 &    2.906550 &  0 &  0 &  0 &  1507 &  23.8 &  -21.57 &  0.29 &  -2.0 &  0.97 \\
  PGC061468 &  272.360748 &   19.117682 &  0 &  0 &  0 &  2371 &  36.2 &  -21.68 &  0.35 &   0.0 &  1.06 \\
  PGC071531 &  352.121338 &   19.863962 &  0 &  0 &  0 &  2030 &  30.4 &  -21.74 &  0.53 &  -4.0 &  0.88 \\
  PGC170172 &  176.731720 &   -5.187745 &  0 &  0 &  0 &  2562 &  37.1 &  -21.89 &  0.08 &  -5.0 &  0.89 \\
   UGC03960 &  115.094856 &   23.275089 &  0 &  0 &  0 &  2255 &  33.2 &  -21.89 &  0.20 &  -4.9 &  1.24 \\
   UGC04551 &  131.024582 &   49.793968 &  0 &  0 &  0 &  1728 &  28.0 &  -22.92 &  0.10 &  -2.0 &  1.03 \\
   UGC05408 &  150.966095 &   59.436138 &  0 &  0 &  0 &  2998 &  45.8 &  -22.03 &  0.06 &  -3.3 &  0.84 \\
   UGC06062 &  164.656662 &    9.050468 &  0 &  0 &  0 &  2634 &  38.7 &  -22.82 &  0.13 &  -2.0 &  1.05 \\
   UGC06176 &  166.852753 &   21.657185 &  0 &  0 &  0 &  2677 &  40.1 &  -22.66 &  0.08 &  -2.0 &  1.03 \\
   UGC08876 &  209.241943 &   45.973179 &  0 &  0 &  0 &  2085 &  33.9 &  -22.37 &  0.04 &  -0.1 &  0.93 \\
   UGC09519 &  221.588028 &   34.370651 &  0 &  0 &  0 &  1631 &  27.6 &  -21.98 &  0.09 &  -1.9 &  0.87 \\
\enddata
\tablecomments{
Column (1): The Name is the principal designation from LEDA, which is used as standard designation for our project.
Column (2): Right Ascension in degrees and decimal (J2000.0).
Column (3): Declination in degrees and decimals (J2000.0). As the galaxy names may not be always consistent between different catalogues, these coordinate {\em define} the galaxies of the sample.
Column (4): ${\rm SBF} = 1$ if the galaxy is in \citet{Tonry2001} and ${\rm SBF} = 2$ if it is in \citet{Mei2007} or both.
Column (5): number of redshift-independent distance determinations listed in the NED-D catalogue, excluding the ones based on kinematical scaling relations. Column (6): ${\rm Virgo}=1$ if the galaxies is contained within a sphere of radius $R=3.5$ Mpc from the center of the cluster assumed at coordinates RA$=$12h28m19s and DEC$=$+12$^\circ$40$'$ \citep{Mould2000} and distance $D=16.5$ Mpc \citep{Mei2007}.
Column (7): Heliocentric velocity measured from the \sauron\ integral-field stellar kinematics (1$\sigma$ error $\Delta V_{\rm hel}=5$ \kms).
Column (8): distance in Mpc. When ${\rm SBF} = 1$ the distance comes from \citet{Tonry2001}, corrected by {\em subtracting} 0.06 mag to the distance modulus \citep{Mei2007}; When ${\rm SBF} = 2$ the distance comes from \citet{Mei2007}; When ${\rm SBF} = 0$ and NED-D $>0$ the distance is the median of the NED-D values, excluding the ones based on kinematical scaling relations;
When ${\rm SBF}=$ NED-D $=0$ and the galaxy is in Virgo (${\rm Virgo}=1$) then it is assigned the cluster distance $D=16.5$ Mpc \citep{Mei2007}; Otherwise $D = V_{\rm cosmic}/H_0$, with $H_0=72$ \kms\ Mpc$^{-1}$, where $V_{\rm cosmic}$ is the velocity derived from $V_{\rm hel}$ via the local flow field model of \citet{Mould2000} using only the Virgo attractor.
Column (9): total galaxy absolute magnitude derived from the apparent magnitude $K_T$ (2MASS keyword \texttt{k\_m\_ext}) at the adopted distance $D$ and corrected for the foreground galactic extinction: $M_K=K_T-5\log_{10} D - 25-A_B/11.8$, which assumes $A_B/A_K=11.8$.
Column (10): $B$-band foreground galactic extinction from \citet{schlegel98}.
Column (11): Morphological $T$-type from HyperLeda. E: $T\le-3.5$, S0: $-3.5<T\le-0.5$. This morphology was {\em not} the one used for the sample selection, but is printed in \reffig{fig:postage_stamps_sr} and \ref{fig:postage_stamps_fr}.
Column (12): Projected half-light effective radius. It was derived from a combination of RC3 and 2MASS determinations, which both use growth curves, as described in \refsec{sec:observing}, but it was normalized to agree on average with RC3.
}
\end{deluxetable}

\clearpage
\begin{deluxetable}{rrrcccrrrrrrr}
\tablewidth{0pt}
\tablecaption{The 611 spiral galaxies in the \atl\ {\em parent} sample.\label{tab:atlas3d_spirals}}
\tablehead{
 \colhead{Galaxy} &
 \colhead{RA} &
 \colhead{DEC} &
 \colhead{SBF} &
 \colhead{NED-D} &
 \colhead{Virgo} &
  \colhead{$V_{\rm hel}$} &
 \colhead{$D$} &
 \colhead{$M_K$} &
 \colhead{$A_B$} &
 \colhead{$T$-type} &
 \colhead{$\log\re$} \\
 \colhead{} &
 \colhead{(deg)} &
 \colhead{(deg)} &
 \colhead{} &
 \colhead{} &
 \colhead{} &
 \colhead{(\kms)} &
 \colhead{(Mpc)} &
 \colhead{(mag)} &
 \colhead{(mag)} &
 \colhead{} &
 \colhead{(\arcsec)}  \\
 \colhead{(1)} &
 \colhead{(2)} &
 \colhead{(3)} &
 \colhead{(4)} &
 \colhead{(5)} &
 \colhead{(6)} &
 \colhead{(7)} &
 \colhead{(8)} &
 \colhead{(9)} &
 \colhead{(10)} &
 \colhead{(11)} &
 \colhead{(12)}
}
\startdata
     IC0065 &   15.230966 &   47.681984 &  0 &  0 &  0 &  2614 &  39.7 &  -23.54 &  0.64 &   4.0 &  1.38 \\
     IC0163 &   27.312431 &   20.711317 &  0 &  0 &  0 &  2749 &  39.7 &  -22.38 &  0.36 &   8.0 &  1.35 \\
     IC0239 &   39.116250 &   38.970000 &  0 &  0 &  0 &   903 &  15.7 &  -22.23 &  0.31 &   6.0 &  1.90 \\
     IC0540 &  142.542755 &    7.902529 &  0 &  0 &  0 &  2035 &  30.0 &  -21.89 &  0.26 &   3.5 &  1.08 \\
     IC0591 &  151.865479 &   12.274520 &  0 &  0 &  0 &  2839 &  41.2 &  -21.82 &  0.15 &   6.0 &  1.19 \\
     IC0610 &  156.618179 &   20.228252 &  0 &  0 &  0 &  1170 &  19.6 &  -21.53 &  0.09 &   3.9 &  1.02 \\
     IC0750 &  179.717606 &   42.722404 &  0 &  1 &  0 &   701 &  36.8 &  -24.71 &  0.09 &   2.1 &  1.24 \\
     IC0777 &  184.849228 &   28.309881 &  0 &  0 &  0 &  2541 &  39.0 &  -21.92 &  0.10 &   2.6 &  0.95 \\
     IC0800 &  188.486313 &   15.354542 &  0 &  0 &  0 &  2326 &  35.8 &  -22.20 &  0.16 &   5.2 &  1.56 \\
     IC0851 &  197.143127 &   21.049742 &  0 &  0 &  0 &  2615 &  39.8 &  -21.82 &  0.15 &   3.1 &  1.31 \\
\enddata
\tablecomments{
The meaning of the columns is the same as in Table~\ref{tab:atlas3d_sample}, except for Column (7), which contain here the heliocentric velocity taken from NED. Only the first 10 rows are shown while the full 611 will be published electronically. Both Table~\ref{tab:atlas3d_sample} and \ref{tab:atlas3d_spirals} are also available from our project website http://purl.com/atlas3d.
}
\end{deluxetable}

\label{lastpage}

\end{document}